\documentclass[12pt,a4paper]{article}
\pdfoutput=1
\usepackage{cite,color,graphics,amsmath,epsfig,rotating}
\numberwithin{equation}{section}
\usepackage{latexsym}
\usepackage{amssymb}
\usepackage{bbm}
\usepackage{graphicx}
\usepackage{subfig}
\usepackage{wrapfig}
\usepackage{psfrag}
\usepackage{mathrsfs}
\usepackage{url}
\usepackage{multicol}
\usepackage{color}
\usepackage[toc,page]{appendix}
\ifx\pdfoutput\undefined
\usepackage[bookmarks]{hyperref}	
\else
\usepackage{hyperref}	
\fi
\hypersetup{colorlinks=true,linktocpage,bookmarksopen,bookmarksnumbered,citecolor=azur,
linkcolor=green1,pdfstartview=FitH,urlcolor=darkred,breaklinks=true,
pdftitle={Probing U(1) extensions of the MSSM  at the LHC Run I and in dark matter searches},
pdfauthor={G. B\'elanger, J. Da Silva, U. Laa, A. Pukhov},
pdfsubject={paper},
pdfkeywords ={Supersymmetry, Higgs sector, dark matter, low energy observables, LHC}}

\definecolor{azur}{rgb}{0.118,0.498,0.796}
\definecolor{darkred}{cmyk}{0,1,1,0.4}
\definecolor{green1}{rgb}{0.21,0.6,0.32}
\def\mhref#1{\href{mailto:#1}{#1}}		

\usepackage{marginnote}

\usepackage{eurosym}
\usepackage{feynmp}
\begin{document}

\setlength{\unitlength}{1mm}
\renewcommand{\arraystretch}{1.4}


\def\NTools{{\tt NMSSMTools}}
\def\UTools{{\tt UMSSMTools}}
\def\HB{{\tt HiggsBounds\;}}
\def\micro{{\tt micrOMEGAs\;}}
\def\chep{{\tt CalcHEP\;}}
\def\lhep{{\tt LanHEP\;}}

\def\ra{\rightarrow}
\def\snr{\tilde{\nu}_R}
\def\lsp{\tilde{\nu}_R}
\def\mlsp{m_{\tilde{\nu}_R}}
\def\snl{\tilde{\nu}_L}
\def\mneut{m_{\tilde{\chi}^0_1}}
\def\mchi{m_{\tilde{\chi}^0_i}}
\def\mneutt{m_{\tilde{\chi}^0_2}}
\def\mneuth{m_{\tilde{\chi}^0_3}}
\def\mneutf{m_{\tilde{\chi}^0_4}}
\def\mchar{m_{\tilde{\chi}^+_1}}
\def\mchart{m_{\tilde{\chi}^+_2}}
\def\msel{m_{\tilde{e}_L}}
\def\mser{m_{\tilde{e}_R}}
\def\mslo{m_{\tilde{\tau}_1}}
\def\mslt{m_{\tilde{\tau}_2}}
\def\msul{m_{\tilde{u}_L}}
\def\msur{m_{\tilde{u}_R}}
\def\msdl{m_{\tilde{d}_L}}
\def\msdr{m_{\tilde{d}_R}}
\def\msto{m_{\tilde{t}_1}}
\def\mstt{m_{\tilde{t}_2}}
\def\msbo{m_{\tilde{b}_1}}
\def\msbt{m_{\tilde{b}_2}}
\def\sw{s_W}
\def\cw{c_W}
\def\tb{\tan\beta}
\def\cb{c_\beta}
\def\sb{s_\beta}

\def\bsg{\mathscr{B}(\bar{B}^0 \to X_s\gamma)}
\def\bXsl{\mathscr{B}(\bar{B}^0 \to X_s\ell^+ \ell^-)}
\def\bXsmu{\mathscr{B}(\bar{B}^0 \to X_s\mu^+ \mu^-)}
\def\bsmu{\mathscr{B}(B^0_s \to \mu^+ \mu^-)}
\def\btau{\mathscr{B}(B^\pm \to \tau^\pm \nu_\tau)}
\def\Omg{\Omega h^2}
\def\sip{\sigma^{SI}_{\chi p}}
\def\amu{\Delta a_\mu}
\def\neut1{\chi^0_1}
\def\neuti{\chi^0_i}
\def\neutj{\chi^0_j}
\def\neut2{\chi^0_2}
\def\neut3{\chi^0_3}
\def\neut4{\chi^0_4}
\def\chargi{\chi^\pm_i}
\def\charg1{\chi^\pm_1}
\def\charg2{\chi^\pm_2}
\def\gluino{\tilde{g}}
\def\ul{\tilde{u}_L}
\def\ur{\tilde{u}_R}
\def\stau{\tilde{\tau}}
\def\sl{\tilde{l}}
\def\sq{\tilde{q}}
\def\sneutrino{\tilde\nu}
\def\msnu{m_{\tilde\nu_R}}
\def\anu{A_{\tilde\nu}}
\def\mzp{M_{Z_2}}
\def\azz{\alpha_{Z}}
\def\te6{\theta_{E_6}}
\def\beq{\begin{equation}}
\def\eeq{\end{equation}}
\def\wino{\tilde{W}}
\def\bino{\tilde{B}}
\def\binop{\tilde{B'}}
\def\cw{c_W}
\def\sw{s_W}
\def\Vub{\left|V_{ub}\right|}
\def\Vtbts{\left|V_{ts}^*V_{tb}\right|}
\def\Vtbtd{\left|V_{td}^*V_{tb}\right|}
\def\VtsbVcb{\left|\frac{V_{ts}^*V_{tb}}{V_{cb}}\right|}
\def\l{\lambda}
\def\e{\epsilon}
\newcommand{\scs}{\scriptscriptstyle}
\def\simleq{\stackrel{<}{\scs \sim}}
\def\simgeq{\stackrel{>}{\scs \sim}}
\newcommand{\gb}{\textcolor{green}}
\newcommand{\jd}{\textcolor{blue}}
\newcommand{\uschi}{\textcolor{magenta}}
\def\smodelsnn  {{\sc SModelS}}
\def\smodels  {{\sc SModelS}\,v1.0.1}

\newcommand{\ablabels}[3]{
  \begin{picture}(100,0)\setlength{\unitlength}{1mm}
    \put(#1,#3){\bf (a)}
    \put(#2,#3){\bf (b)}
  \end{picture}\\[-8mm]
} 

\begin{titlepage}
\begin{center}
\vspace*{-1cm}
\begin{flushright}
LAPTH-029/15\\
MAN/HEP/2015/10\\
MCnet-15-10\\
LPSC15130
\end{flushright}

\vspace*{1.6cm}
{\Large\bf Probing U(1) extensions of the MSSM  at the LHC Run I and in dark matter searches} 

\vspace*{1cm}\renewcommand{\thefootnote}{\fnsymbol{footnote}}

{\large 
G.~B\'elanger$^{1}$\footnote[1]{Email: \mhref{belanger@lapth.cnrs.fr}},
J.~ Da Silva$^{2}$\footnote[2]{Email: \mhref{dasilva@lapth.cnrs.fr}},
U.~Laa$^{1,3}$\footnote[3]{Email: \mhref{ursula.laa@lpsc.in2p3.fr}},
A.~Pukhov$^{4}$\footnote[4]{Email: \mhref{pukhov@lapth.cnrs.fr}},
} 

\renewcommand{\thefootnote}{\arabic{footnote}}

\vspace*{1cm} 
{\normalsize \it 
$^1\,$\href{http://lapth.cnrs.fr}{LAPTH}, Universit\'e Savoie Mont Blanc, CNRS, \\ B.P.110, F-74941 Annecy-le-Vieux Cedex, France\\[2mm]
$^2\,$\href{http://www.hep.man.ac.uk}{Consortium for Fundamental Physics, School of Physics and Astronomy}, 
\\ University of Manchester, Oxford Road, Manchester, M13 9PL, United Kingdom\\[2mm]
$^3\,$\href{http://lpsc.in2p3.fr}{Laboratoire de Physique Subatomique et de Cosmologie}, Universit\'e Grenoble-Alpes, CNRS/IN2P3, 53 Avenue des Martyrs, 38026 Grenoble, France\\[2mm]
$^4\,$\href{http://theory.sinp.msu.ru}{Skobeltsyn Institute of Nuclear Physics (SINP MSU)}, Lomonosov Moscow State University, 1(2) Leninskie gory, GSP-1, Moscow 119991, Russia\\[2mm]
}

\vspace{1cm}

\begin{abstract}
The  U(1) extended supersymmetric standard model  (UMSSM) can accommodate a Higgs boson at 125 GeV without relying on large corrections from the top/stop sector. 
After imposing LHC results on the Higgs sector, on $B$-physics and on new particle searches as well as dark matter constraints, we show that this model offers two  viable dark matter candidates, the right-handed (RH) sneutrino or the neutralino.  
Limits on supersymmetric partners from LHC simplified model searches are imposed using \smodelsnn~and allow for light squarks and gluinos. Moreover the upper limit on the relic abundance
often favours scenarios with long-lived particles.  Searches for a $Z'$ at the LHC remain the most unambiguous probes of this model. 
Interestingly, the  $D$-term contributions to the sfermion masses allow to explain  the anomalous magnetic moment of the muon in specific corners of the parameter space with  light smuons or left-handed (LH) sneutrinos. 
We finally emphasize  the interplay between direct searches for dark matter and LHC simplified model searches.
\end{abstract}

\end{center}

\end{titlepage}

\tableofcontents

\section{Introduction}

The discovery  by the  ATLAS and CMS collaborations~\cite{Chatrchyan:2012ufa,Aad:2012tfa,Chatrchyan:2013lba} of a 125 GeV Higgs boson whose properties are compatible with the standard model (SM) predictions coupled with the fruitless searches for new particles at Run I of the LHC~\cite{Khachatryan:2014doa,Khachatryan:2014qwa,Aad:2014wea,Aad:2014kra} has left the community with little guidance for which direction to search for new physics at the TeV scale. The dark matter (DM) problem remains a strong motivation for considering extensions of the SM, in particular supersymmetry.

In the minimal supersymmetric standard model (MSSM), the Higgs couplings are to a large extent SM-like, especially when the mass scales of the second Higgs doublet and/or of other new particles that enter the loop-induced Higgs couplings are well above the electroweak scale.  This is to be expected in any model where the Higgs is responsible for electroweak symmetry breaking. The main challenge for the MSSM  is however to explain a  Higgs mass of 125 GeV. To achieve such a high mass  requires large contributions from one-loop diagrams involving top squarks --- 
in fact the loop contribution has to be of the same order as the tree-level contribution --- thus introducing a large amount of fine-tuning~\cite{Hall:2011aa,Baer:2014ica}.
The fine-tuning is reduced in extensions of the minimal model containing an additional singlet scalar field~\cite{Ellwanger:2014dfa,Kaminska:2014wia,Farina:2013fsa,Athron:2013ipa}.
For example in the next-to-minimal supersymmetric standard model (NMSSM), terms in the superpotential give an extra tree-level contribution to the light Higgs mass, thus reducing the amount of fine-tuning required to reach $m_h=125~{\rm GeV}$. A doublet-singlet mixing  can also modify significantly the tree-level couplings of the light Higgs. 
In the UMSSM, where the gauge group contains an extra U(1) symmetry, contributions from U(1) $D$-terms in addition to those from the superpotential present in the NMSSM, can further increase the light Higgs mass~\cite{Cvetic:1997ky,Barger:2006dh} reaching easily 125 GeV without a very large contribution from the stop sector. Furthermore, because the singlet mass is driven by the mass of the new gauge boson which is strongly constrained by LHC searches to be above the TeV scale~\cite{Aad:2014cka,Khachatryan:2014fba}\footnote{In this paper we concentrate on a $Z'$ above the electroweak scale, for scenarios with light $Z'$ see \cite{Frank:2014bma}.}, the tree-level couplings of the light Higgs are expected to be SM-like, in agreement with the latest results of ATLAS and CMS~\cite{ATLAS-CONF-2015-007,Khachatryan:2014jba}. This heavy $Z'$ was also found to increase the fine-tuning of supersymmetric models with U(1) extended gauge symmetry \cite{Athron:2015tsa}.
Another nice feature of the UMSSM (as the NMSSM) is that the $\mu$ parameter, generated from the vacuum expectation value (vev) of the singlet field responsible for the breaking of the U(1) symmetry, is naturally at the weak scale. Finally, this model is well motivated within the context of superstring models~\cite{Cvetic:1995rj,Cvetic:1996mf,Cleaver:1997jb,Cleaver:1998gc,Cleaver:1998sm} and grand unified theories~\cite{Langacker:1980js,London:1986dk}.

The range of masses for the Higgs scalars and pseudoscalars were examined in a variety of singlet extension of the MSSM~\cite{Barger:2006dh}.   
The parameter space of a similar model with a new U(1) symmetry, the constrained $E_6$SSM, compatible with the Higgs at 125 GeV as well as limits on the Higgs sector and providing a dark matter candidate was examined in~\cite{Athron:2012sq}.
In this model the RH sneutrino does not have a U(1) charge and is expected to be very massive.
The $h\rightarrow \gamma\gamma$ signal in a U(1) extended MSSM model was discussed in~\cite{Cheung:2012pq,Basso:2012tr} with emphasis on the region that leads to an increase in the two-photon signal. Additional non-standard decays of Higgs particles were found in~\cite{Frank:2013yta,Athron:2014pua}. However, since the mass of the additional singlet Higgs is expected to be very large due to strong limits on the $Z'$ boson mass, it does not affect the property of the lightest Higgs which is hence expected to be SM-like.  
In the UMSSM model considered here RH sneutrinos can be charged under the additional U(1) symmetry, hence this model gives a new viable dark matter candidate in addition to the lightest neutralino as  observed in~\cite{Belanger:2011rs}. 
The properties of a RH sneutrino DM were also examined in the $U(1)_{B-L}$~\cite{Basso:2012gz}  and $U(1)_{B-L}\times U(1)_R$ extensions of the MSSM~\cite{Hirsch:2012kv}. Note that in such models the  sneutrino vev's were found to play an important role in the vacuum stability \cite{CamargoMolina:2012hv}. Furthermore the $Z'$ can contribute to the stabilization of the Higgs potential~\cite{DiChiara:2014wha}.

In this paper we explore the parameter space of the UMSSM (derived from $E_6$) that is compatible with 
both collider and dark matter observables.
We include in particular the Higgs mass and signal strengths in all channels, LHC constraints on $Z'$ and on supersymmetric particles, new results from $B$-physics, as well as the relic density and direct detection of dark matter.
Specifically we take into account the most recent LHC results for supersymmetric particle searches based on simplified models using \smodelsnn~\cite{Kraml:2013mwa,Kraml:2014sna}.
This allows us to also highlight the signatures not well constrained by current searches despite a spectrum well below the TeV scale. One salient feature of the model is that large $D$-term contributions can significantly reduce the mass of RH squarks thus splitting the u-type and d-type squarks and weakening  the constraints on first generation squarks. 
Another feature, which is also found in the MSSM, is that the relic density upper limit favors a neutralino with a large higgsino or wino component as the lightest supersymmetric particle (LSP). Scenarios with a  higgsino LSP can easily escape current search limits.
For example simplified model limits from top squark searches rely on the assumption that one decay channel is dominant, while for higgsino LSP branching ratios  into $t\tilde{\chi}_i^0$ and $b\tilde\chi_i^+$ can both be large, thus the mixed channels where each stop decay into a different final state are important.
Since a higgsino or wino LSP  may be associated with a chargino which is stable at the collider scale, we also impose the  D0 and ATLAS limits originating from searches for long-lived particles.
On the remaining parameter space, we then discuss  the expected spectra of SUSY particles, the expectations for the signal strengths for the Higgses as well as dark matter observables in direct and indirect detection. 

In general we do not attempt to explain the observed discrepancy with the standard model expectations  in the muon anomalous magnetic moment.
However, we highlight the region where the model can explain this discrepancy and investigate how  it may escape simplified model limits from the  LHC. The interplay between the muon anomalous magnetic moment constraint, LHC and DM limits was recently studied in the MSSM~\cite{Ajaib:2015yma}.

In contrast to previous studies~\cite{Kalinowski:2008iq,Belanger:2011rs} we explore the impact of LHC8TeV results on Higgs and new particle searches from the 8~TeV run on scenarios with arbitrary U(1) originating from $E_6$. Moreover we consider  both the cases of a neutralino and a RH sneutrino dark matter.
We further examine the implications of dark matter searches in these scenarios.
An attractive feature of the model is the possibility to obtain $m_h=125$~GeV despite small values of $\tan\beta$. The phenomenology of Higgs and SUSY searches could thus differ  from that of the much-studied MSSM.

This paper is organised as follows. In section~\ref{sec:model} we describe the model. Section~\ref{sec:Higgs} is devoted to a detailed view of the Higgs sector. Section~\ref{sec:constraints} presents the different constraints used in our study. Section~\ref{sec:results} contains the results for several sectors of the model after applying a basic set of constraints mostly related to Higgs and $B$-physics observables and after applying the DM relic abundance limits. Section~\ref{sec:LHC_SMS_ll} is dedicated to the application of the LHC simplified models searches on the remaining allowed parameter space of the UMSSM. Section~\ref{sec:afterLHC} contains a summary of LHC constraints after Run I and suggestions on how to extend simplified models searches to further probe the model. Section~\ref{sec:Higgs_coup_sig} shows prospects for probing the Higgs sector and section~\ref{sec:DM_ID-DD} prospects from astroparticle searches. Our conclusions are presented  in section~\ref{sec:conclusion}.

\section{The model}
\label{sec:model}

\begin{figure}[h!]\centering
\includegraphics[width=0.6\textwidth]{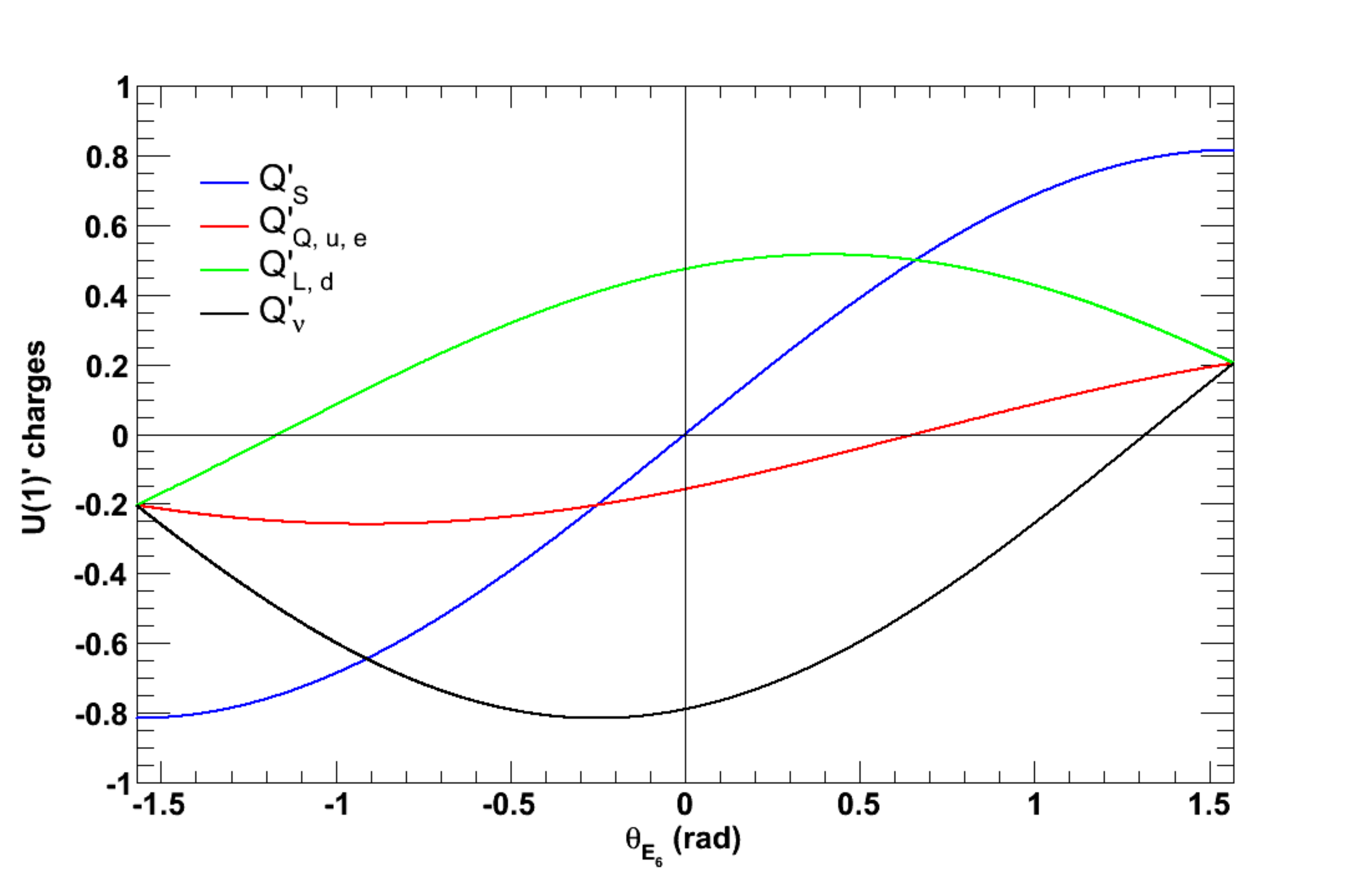}
\caption{$U(1)'$ charges of some matter fields in the UMSSM as a function of $\te6$.
\label{fig:Uprimecharge_te6} }
\end{figure}

The symmetry group of the model is $SU(3)_c  \otimes  SU(2)_L  \otimes  U(1)_Y \otimes U(1)'$ and 
we assume that this model is derived from an underlying $E_6$ model.  In this case the 
$U(1)'$ charges of each field $F$ of the model are parameterized by an angle $\te6$ as
\beq \mathcal{Q}'_F = \cos \te6 \mathcal{Q}'_{\chi} + \sin \te6 \mathcal{Q}'_{\psi},\eeq
where   $\te6 \in [-\pi/2, \pi/2]$  and the charges $\mathcal{Q}'_{\chi}$ and $\mathcal{Q}'_{\psi}$ are given in Table~\ref{tab:Ucharge} for all fermionic fields that we will consider \cite{Langacker:1998tc,Barger:2007nv}. The dependence on $\te6$ of the $U(1)'$ charge of some matter fields is shown in~figure~\ref{fig:Uprimecharge_te6}.

The matter sector of the $E_6$ model contains, in addition to the  chiral supermultiplets of the  SM fermions,  
three families of new particles,  each family containing : a RH neutrino, two  Higgs doublets ($H_u, H_d$), a singlet, and a colour $SU(3)_c$ (anti)triplet. While the complete matter sector is needed for anomaly cancellations, for simplicity we will assume that all exotic fields, with the exception of three RH neutrinos, two Higgs doublets and one singlet, are above a few TeV's and  can be  neglected. Similarly in addition to the MSSM chiral multiplets we will only consider the chiral multiplets corresponding to these fields,   that is the multiplet with a singlet  $S$ and the singlino $\tilde S$ and another multiplet with RH neutrinos $\nu_{iR}$ ($i \in \{e,\mu,\tau\}$) and their supersymmetric partners, the sneutrinos, $\tilde\nu_{iR}$.

\begin{table}[!htb]
\begin{center}
\begin{tabular}{cccccccccc}
       \hline \hline
       & $\mathcal{Q}'_Q$ & $\mathcal{Q}'_u$ & $\mathcal{Q}'_d$ & $\mathcal{Q}'_L$ & $\mathcal{Q}'_\nu$ & $\mathcal{Q}'_e$ & $\mathcal{Q}'_{H_u}$ & $\mathcal{Q}'_{H_d}$ & $\mathcal{Q}'_S$ \\ \hline \hline
      $\sqrt{40}\mathcal{Q}'_{\chi}$ & $-1$ & $-1$ & $3$ & $3$ & $-5$ & $-1$ & $2$ & $-2$ & $0$ \\ 
      $\sqrt{24}\mathcal{Q}'_{\psi}$ & $1$ & $1$ & $1$ & $1$ & $1$ & $1$ & $-2$ & $-2$ & $4$ \\
 \end{tabular}
\caption{\label{tab:Ucharge} $U(1)'$ charges of all matter fields considered.}
\end{center}
\end{table}

Finally the UMSSM model contains  a new vector multiplet, with a new boson $B'$ and the corresponding gaugino $\tilde{B}'$.   
The superpotential is the same as in the MSSM with $\mu=0$ but has additional terms involving the singlet,
\beq
\mathcal{W}_{\mathrm{UMSSM}} = \mathcal{W}_{\mathrm{MSSM}}|_{\mu = 0} + \l S H_u H_d + \tilde{\nu}^*_R \mathbf{y_{\nu}} \tilde{L}H_u + \cal{O}(\mathrm{TeVs})
\eeq
where $\mathbf{y_{\nu}}$ is the neutrino Yukawa matrix. The vev of $S$, $\langle S \rangle = \frac{v_s}{\sqrt{2}}$ breaks the $U(1)'$ symmetry and induces a $\mu$ term
\beq \mu= \l \frac{v_s}{\sqrt{2}}.\label{eq:mueff}\eeq
Note that for $\te6=0$ the $U(1)'$ symmetry cannot be broken by the singlet field since $\mathcal{Q}'_S=0$. Note also that the invariance of the superpotential under $U(1)'$ imposes a condition on the Higgs sector, namely $\mathcal{Q}'_{H_u} + \mathcal{Q}'_{H_d} + \mathcal{Q}'_S = 0$. The soft-breaking Lagrangian of the UMSSM is
\beq \begin{split}
\mathscr{L}_{\mathrm{UMSSM}}^{\mathrm{soft}} = \, & \mathscr{L}_{\mathrm{MSSM}}^{\mathrm{soft}}|_{b = 0} - \left( \frac{1}{2} M'_1\tilde{B'}\tilde{B'} + \tilde{\nu}^*_R \mathbf{a_{\nu}}\tilde{L}H_u + \textrm{h.c.} \right) - \tilde{\nu}^*_R \mathbf{m^2_{\tilde{\nu}_R}}\tilde{\nu}_R \\
& - m^2_S |S|^2 - (\l A_\l S H_uH_d + \textrm{h.c.}) + \cal{O}(\mathrm{TeVs}),
\end{split} \eeq
with the trilinear coupling $A_\l$, the $\tilde{B'}$ mass term $M'_1$, the singlet mass term $m_S$. The soft sneutrino mass term matrices $\mathbf{a_{\nu}}$ and $\mathbf{m^2_{\tilde{\nu}_R}}$  are taken  to be diagonal in the family space. Note that our study is based on the UMSSM model with parameters defined at the electroweak scale, we make no attempt to check the validity of the model at a high scale. We now describe briefly the sectors of the model 
that will play a role in the considered observables.

\subsection{Gauge bosons}

The two neutral massive gauge bosons, $Z^0$ and $Z' = B'$ can mix both through mass and kinetic
mixing~\cite{Langacker:2008yv,Kalinowski:2008iq}.  
In the following we will neglect the kinetic mixing\footnote{The impact of the kinetic mixing on the Higgs boson mass and on the $Z'$ and DM phenomenology was examined in the $U(1)_{B-L}$  extension of the MSSM in~\cite{O'Leary:2011yq,Basso:2012gz,Krauss:2012ku}.}.
The electroweak and $U(1)'$ symmetries are broken respectively by the vev's of the doublets, 
$v_u/\sqrt{2}=\langle H_u \rangle$, $v_d/\sqrt{2}=\langle H_d \rangle$ and singlet, $v_s/\sqrt{2}=\langle S \rangle$.
The mass matrix reads
\beq
  M^2_{Z} =
  \left( \begin{array}{cc}
   M^2_{Z^0} &   \Delta_Z^2\\
   \Delta_Z^2 & M^2_{Z'}
  \end{array}\right) \,,
\label{eq:mz}
\eeq
where 
\begin{eqnarray}
\label{eq:ZZ'}
M^2_{Z^0} &= &\frac{1}{4} \frac{g_2^2}{c_W^2} (v_u^2+v_d^2)\nonumber\\
M^2_{Z'} &= &{g'_1}^2 (\mathcal{Q}'^2_{H_d} v_d^2+ \mathcal{Q}'^2_{H_u} v_u^2+\mathcal{Q}'^2_S v_s^2)
\label{eq:mzz'}
\end{eqnarray}
\beq
\label{eq:deltaZZ'}
\Delta_Z^2 = \frac{g_2 g_1'}{2 c_W} (\mathcal{Q}'_{H_u} v_u^2 - \mathcal{Q}'_{H_d} v_d^2)
\eeq
where $g_2=e/s_W$, $g_1'=\sqrt{5/3} g_1$, $g_1 = e/c_W$ and $c_W$ ($s_W$) is the cosinus (sinus) of the Weinberg angle.  Diagonalisation of the mass matrix leads to two eigenstates
\begin{eqnarray}
Z_1&=&\cos\azz Z^0 + \sin\azz Z'\nonumber\\
Z_2&=& - \sin\azz Z^0 + \cos\azz Z'
\end{eqnarray}
where the mixing angle is defined as
\beq
\label{eq:Zangle}
\sin 2 \azz =\frac{2 \Delta_Z^2}{M^2_{Z_2} - M^2_{Z_1}}
\eeq
and  the masses of the physical fields are
\beq
\label{eq:Zmixing}
M^2_{Z_1,Z_2}=\frac{1}{2} \left( M^2_{Z^0}+M^2_{Z'} \mp \sqrt{\left(M^2_{Z^0}-M^2_{Z'}\right)^2+4\Delta_Z^4}  \right).
\eeq

Precision measurements at the $Z^0$-pole and from low energy neutral currents provide stringent constraints on the $Z^0-Z'$ mixing angle. Depending on the model parameters the constraints are below a few $10^{-3}$~\cite{Leike:1991if,Erler:2009jh}. The new gauge boson $Z_2$ will therefore have approximately the same properties as the $Z'$. 
As input parameters we choose the physical masses, $M_{Z_1}=91.187$~GeV, $M_{Z_2}$ and the mixing angle, 
$\azz$. From these together with the coupling constants, 
we extract both the value of  $\tb=v_u/v_d$ and the  value of $v_s$. Note that as in~\cite{Cerdeno:2004xw} we adopt the convention where both $\l$ and $\tb$ are positive while $\mu$ (and then $v_s$) and $A_\l$ can have both signs.
 From eqs.~\eqref{eq:deltaZZ'} and \eqref{eq:Zangle},
\beq 
\label{eq:7.c2b} 
\cos^2\beta = \frac{1}{\mathcal{Q}'_{H_d} + \mathcal{Q}'_{H_u}}\left( \frac{\sin 2 \azz(M^2_{Z_1} - M^2_{Z_2}) c_W}{ v^2 g'_1 g_2} + \mathcal{Q}'_{H_u}\right),
\eeq
where $v^2 = v^2_u + v^2_d$.

For each $U(1)'$ model the value of $\tan\beta$ can be strongly constrained as a consequence of the requirement $0\leq\cos^2\beta\leq1$. For example for the $U(1)_\psi$ case with $\sin \azz>0$ and $M_{Z_2} \gg M_{Z_1}$ the value of $\tan\beta$ has to be below 1. The reason is that for this choice of $\te6$ we have
\beq \Delta_Z^2  =  \frac{g_2 g'_1}{c_W \sqrt{24}} (\tan^2\beta-1) v^2_d < 0.\eeq
For other choices of parameters the value of $\tan\beta$ can be very large, $\cal{O}$(100). 
Another interesting relation is found for the case of small mass mixing between $Z^0$ and $Z'$ namely  $\azz \ll \frac{v^2}{M^{2}_{Z_2}}$. In this limit $\beta$ is determined from the $U(1)'$ charges only,
\beq \label{eq:c2b_azz0} 
\cos^2\beta \simeq \frac{\mathcal{Q}'_{H_u}}{\mathcal{Q}'_{H_d} + \mathcal{Q}'_{H_u}}.
\eeq

One might think that small values of $\tan\beta$ are problematic for the Higgs boson mass since the MSSM-type tree-level contribution becomes very small.
However, as we will see below, additional terms to the light Higgs mass and especially their dependence on $\azz$ can help raise its value to 125 GeV.

\subsection{Sfermions}
\label{subsec:sfer}

The important new feature in the sfermion sector is that the $U(1)'$ symmetry induces new $D$-term contributions to the sfermion masses. These are added to the diagonal part of the usual MSSM sfermion matrix,  and read
\beq 
\label{eq:7.sferU}
 \Delta_F= \frac{1}{2} {g'_1}^2 \mathcal{Q}'_F \left( \mathcal{Q}'_{H_d} v_d^2 + \mathcal{Q}'_{H_u} v_u^2 + \mathcal{Q}'_S v_s^2 \right),
 \eeq 
where $F \in \{Q, u, d, L, e, \nu\}$.

For large values of $v_s$ the new $D$-term contribution can completely dominate the sfermion mass. Moreover this term can  induce negative mass corrections, even driving the charged sfermion to be the LSP.  Thus the requirement that the LSP be neutral (either the lightest neutralino or RH sneutrino) constrains the values of $\te6$ (unless one allows large soft masses for the sfermions).
For example, for $-\tan^{-1}(3\sqrt{3/5}) < \te6 < 0$, the corrections to the d-squark and to LH slepton masses are negative, while for $0 < \te6 < \tan^{-1}(\sqrt{3/5})$ the corrections to the u-squark and RH slepton masses are negative. The latter implies that the u-type squarks (and in particular the lightest top squark) and the RH sleptons can be the Next-to-LSP (NLSP).  Interestingly for $\te6 = -\tan^{-1}(3\sqrt{3/5}) \approx -1.16$ the LH smuon/sneutrino can be sufficiently light to contribute significantly to the the anomalous magnetic moment of the muon and bring it in agreement with the data~\cite{Bennett:2006fi,Roberts:2010cj}.

\subsection{Neutralinos}

In the UMSSM the neutralino mass matrix in the basis $(\tilde{B},\tilde{W}^3,\tilde{H}_d,\tilde{H_u},\tilde{S},\tilde{B'})$ reads ($\cb = \cos \beta$ and $\sb = \sin \beta$)
\beq
\mathbf{M_{\tilde{\chi}^0}}=
  \left( \begin{array}{cccccc}
   M_1 	&   0    & -M_{Z^0} \cb \sw  &    M_{Z^0} \sb \sw  &   0   &   0\\  
   0 	&   M_2    & M_{Z^0} \cb \cw  & -M_{Z^0} \sb \cw  &    0   &   0\\  
   -M_{Z^0} \cb \sw &  M_{Z^0} \cb \cw  & 0 &  -\mu  & -\l \frac{v_u}{\sqrt{2}} & \mathcal{Q}'_{H_d} g'_1 v_d   \\  
    M_{Z^0} \sb \sw & -M_{Z^0} \sb \cw  & -\mu &   0 & -\l \frac{v_d}{\sqrt{2}} & \mathcal{Q}'_{H_u} g'_1 v_u \\  
   0 &  0 & -\l \frac{v_u}{\sqrt{2}} & -\l \frac{v_d}{\sqrt{2}} & 0 & \mathcal{Q}'_S g'_1 v_s\\  
   0 &  0 & \mathcal{Q}'_{H_d} g'_1 v_d  & \mathcal{Q}'_{H_u} g'_1 v_u  & \mathcal{Q}'_S g'_1 v_s  &  M'_1\\  
  \end{array}\right) \,.
\label{eq:neutralino}
\eeq
Diagonalisation by a 6$\times$6 unitary matrix $\mathbf{Z_n}$ leads to the neutralino mass eigenstates :
\beq \tilde{\chi}^0_i = Z_{n ij} \psi^0_j\textrm{,} \qquad i\textrm{,}j \in \{1,2,3,4,5,6\}.\eeq
The chargino sector is identical to that of the MSSM.

Several studies have  analysed the properties of the neutralino sector in the UMSSM \cite{Hesselbach:2001ri,Choi:2006fz}, in particular as concerns 
the neutralino LSP as a viable DM candidate~\cite{deCarlos:1997yv,Barger:2004bz}. In the weak scale model, the LSP can be any combination of bino/Higgsino/wino/singlino and bino'.
However, as we will show, the LSP is never pure bino', the pure bino and singlino tend to be overabundant while pure higgsino and wino lead to under abundance of DM.

\section{The Higgs sector}
\label{sec:Higgs}

The Higgs sector of the UMSSM consists of three CP-even Higgs bosons $h_i, i \in \{1,2,3\}$, two charged Higgs bosons $H^\pm$ and one CP-odd Higgs boson $A^0$.

 The Higgs potential  is a sum of F-, D- and soft supersymmetry breaking-terms belonging to the UMSSM Lagrangian : $V^{\mathrm{U}}_{\mathrm{tree}} = V_{\mathrm{F}}+ V_{\mathrm{D}} + V_{\mathrm{soft}}$, where
\beq \begin{split}
V_F & = |\l H_u\cdot H_d|^2 + |\l S|^2 \left(|H_d|^2+|H_u|^2 \right), \\
V_D & = \frac{(g_1^2 + g^2_2)^2}{8}\left( |H_d|^2-|H_u|^2 \right)^2+ \frac{g_{2}^2}{2} \left( |H_d|^2|H_u|^2-|H_u \cdot H_d|^2 \right)\\
 & \quad + {g'^2_1\over2}\left(\mathcal{Q}'_{H_d} |H_d|^2+ \mathcal{Q}'_{H_u} |H_u|^2+ \mathcal{Q}'_S |S|^2\right)^2,\\
V_{\rm soft} & = m_{H_d}^2|H_d|^2 + m_{H_u}^2|H_u|^2+ m_s^2|S|^2 + \left( A_\l \l S H_u\cdot H_d + h.c. \right).
\end{split} \eeq
At the minimum of the potential $V^{\mathrm{U}}_{\mathrm{tree}}$, the neutral Higgs fields are expanded as  
\beq
H_d^0 = \frac{1}{\sqrt{2}} \left( v_d + \phi_d + i \varphi_d \right), \quad
H_u^0 = \frac{1}{\sqrt{2}} \left( v_u + \phi_u + i \varphi_u \right), \quad
S     = \frac{1}{\sqrt{2}} \left( v_s + \sigma + i \xi \right),
\eeq
while the charged Higgs  :
\beq
H_d^- = -\cos\beta G_{W^-} + \sin\beta H^-, \quad
H_u^+ = \sin\beta G_{W^+} + \cos\beta H^+,
\eeq
with $G_{W}$ the Goldstone boson.\\

The minimization conditions of $V^{\mathrm{U}}_{\mathrm{tree}}$ are~\cite{Barger:2006dh}
\beq \begin{split} \label{eq:minLO}
\left(m^\mathrm{tree}_{H_d}\right)^2 = & - \frac{1}{2}\left[\frac{g_1^2 + g^2_2}{4} + \mathcal{Q}'^2_{H_d} g'^2_1\right]v_d^2 + \frac{1}{2}\left[\frac{g_1^2 + g^2_2}{4} - \l^2- \mathcal{Q}'_{H_d} \mathcal{Q}'_{H_u} g'^2_1\right]v_u^2 \\
& - \frac{1}{2}\left[\l^2 + \mathcal{Q}'_{H_d} \mathcal{Q}'_S g'^2_1\right]v^2_s + \frac{\l A_\l v_s v_u}{\sqrt{2} v_d}\\
\left(m^\mathrm{tree}_{H_u}\right)^2 = & \quad \, \frac{1}{2}\left[\frac{g_1^2 + g^2_2}{4} - \l^2- \mathcal{Q}'_{H_d} \mathcal{Q}'_{H_u} g'^2_1\right]v_d^2 - \frac{1}{2}\left[\frac{g_1^2 + g^2_2}{4} + \mathcal{Q}'^2_{H_u} g'^2_1\right]v_u^2 \\
& - \frac{1}{2}\left[\l^2 + \mathcal{Q}'_{H_u} \mathcal{Q}'_S g'^2_1\right]v^2_s + \frac{\l A_\l v_s v_d}{\sqrt{2} v_u}\\
\left(m^\mathrm{tree}_S\right)^2 = & - \frac{1}{2}\left[\l^2 + \mathcal{Q}'_{H_d} \mathcal{Q}'_S g'^2_1\right]v_d^2  - \frac{1}{2}\left[\l^2 + \mathcal{Q}'_{H_u} \mathcal{Q}'_S g'^2_1\right]v_u^2  - \frac{1}{2} \mathcal{Q}'^2_S g'^2_1 v^2_s + \frac{\l A_\l v_u v_d}{v_s \sqrt{2}}.
\end{split} \eeq \\
The tree-level mass-squared matrices for the CP-even $(\mathcal{M}_+^0)$ and CP-odd $(\mathcal{M}_-^0)$ Higgs bosons can be written in the basis $\{H^0_d, H^0_u, S\}$ using the relations
\beq
\left( {\mathcal{M}}_{+}^{0}\right)_{ij}= \left. \frac{\partial ^{2}V^{\mathrm{U}}_{\mathrm{tree}}}{\partial \phi _{i}\partial \phi_{j}} \right|_0 ,\qquad \left( {\mathcal{M}}_{-}^{0}\right) _{ij}= \left. \frac{\partial ^{2}V^{\mathrm{U}}_{\mathrm{tree}}}{\partial \varphi_{i}\partial  \varphi _{j}} \right|_0,
\eeq
where $(\phi_1,\phi_2,\phi_3) \equiv (\phi_d,\phi_u,\sigma)$ and $(\varphi_1,\varphi_2,\varphi_3) \equiv (\varphi_d,\varphi_u,\xi)$.
For the neutral CP-even Higgs bosons the relations are
\beq \begin{split}
\left({\mathcal{M}_{+}^0}\right)_{11} & =  \left[\frac{g_1^2 + g^2_2}{4} +  \mathcal{Q}'^2_{H_d} g'^2_1\right] v_d^2 + \frac{\l A_\l v_s v_u}{\sqrt{2} v_d} \\
\left({\mathcal{M}_{+}^0}\right)_{12} & = -\left[\frac{g_1^2 + g^2_2}{4} - \l^{2} - \mathcal{Q}'_{H_d} \mathcal{Q}'_{H_u} g'^2_1\right] v_u v_d - \frac{\l A_\l v_s}{\sqrt{2}}\\
\left({\mathcal{M}_{+}^0}\right)_{13} & =  \left[\l^{2} + \mathcal{Q}'_{H_d} \mathcal{Q}'_{S} g'^2_1\right] v_s v_d - \frac{\l A_\l v_u}{\sqrt{2}}\\
\left({\mathcal{M}_{+}^0}\right)_{22} & =  \left[\frac{g_1^2 + g^2_2}{4} + \mathcal{Q}'^2_{H_u} g'^2_1\right] v_u^2 + \frac{\l A_\l v_s v_d}{\sqrt{2} v_u} \\
\left({\mathcal{M}_{+}^0}\right)_{23} & =  \left[\l^{2} + \mathcal{Q}'_{H_u} \mathcal{Q}'_{S} g'^2_1\right] v_s v_u - \frac{\l A_\l v_d}{\sqrt{2}} \\
\left({\mathcal{M}_{+}^0}\right)_{33} & = \mathcal{Q}'^2_{S} g'^2_1 v^2_s + \frac{\l A_\l v_u v_d}{v_s \sqrt{2}}.\\
\end{split} \eeq
For the CP-odd sector the mass matrix
\beq \begin{split}
{\mathcal{M}_{-}^0} & = \frac{\l A_\l}{\sqrt{2}} 
  \begin{pmatrix}
    \frac{v_s v_u}{v_d} & v_s & v_u\\
    v_s & \frac{v_s v_d}{v_u} & v_d\\
    v_u & v_d & \frac{v_u v_d}{v_s}
  \end{pmatrix}, 
\end{split} \eeq
leads to
\beq
\left(m^\mathrm{tree}_{A^0}\right)^2 =  \frac{\l A_\l \sqrt{2}}{\sin 2 \beta} v_s \left( 1+ \frac{v^2}{4 v^2_s}\sin^2 2\beta\right).
\eeq
The charged Higgs mass at tree-level reads
\beq
\left(m^\mathrm{tree}_{H^\pm}\right)^2 = M^2_W + \frac{\l A_\l \sqrt{2}}{\sin 2 \beta} v_s - \frac{\l ^2}{2} v^2.  
\eeq
The radiative corrections to the Higgs sector are given in appendix~\ref{sec:app}.

\begin{figure}[h!]\centering
\includegraphics[width=0.65\textwidth]{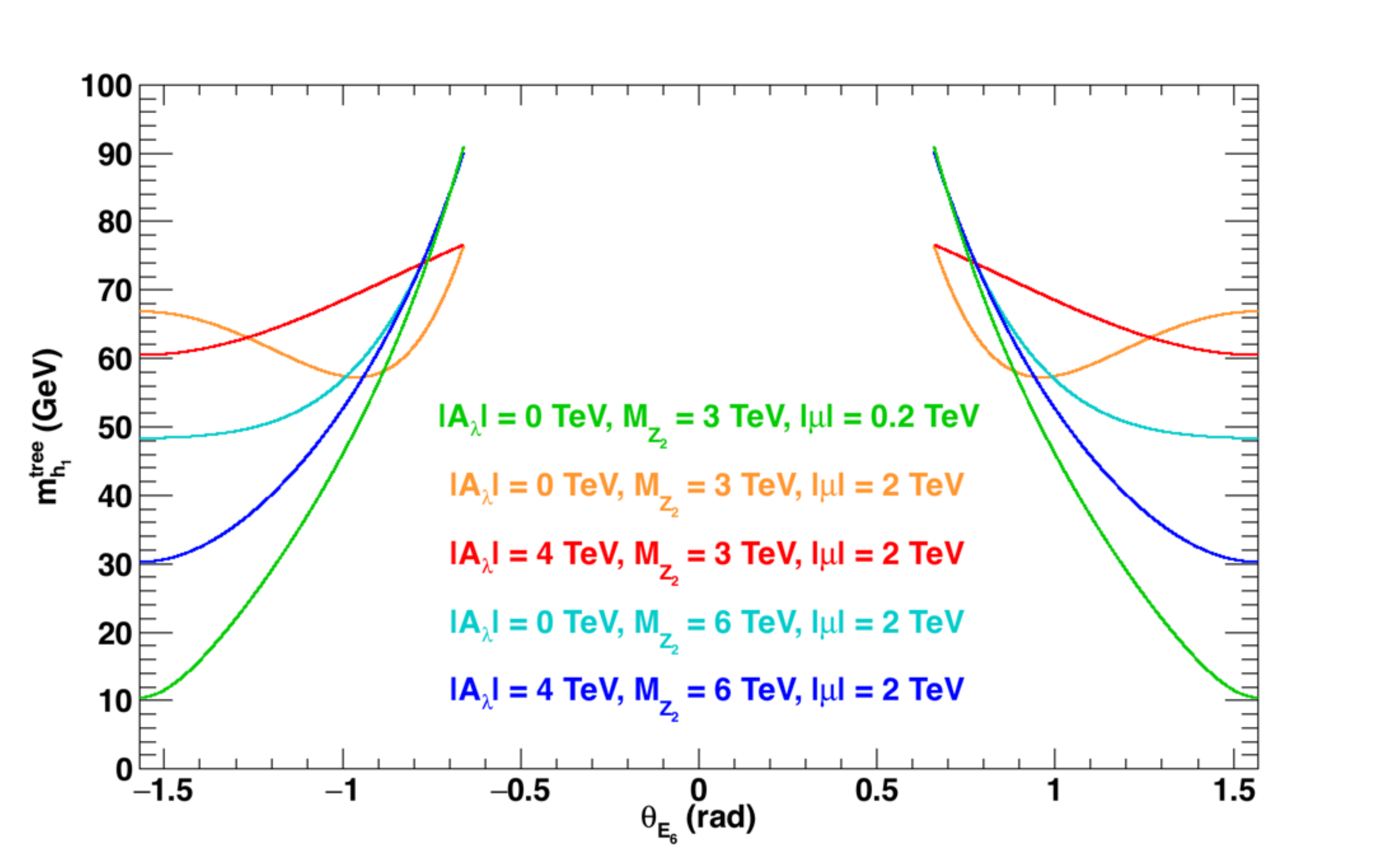}
\caption{The tree-level light Higgs mass in the approximation $\azz \ll \frac{v^2}{M^{2}_{Z_2}}$ as a function of $\te6$ for different values of $|A_\lambda|$, $M_{Z_2}$ and $|\mu|$.
\label{fig:thmh1caZ0} }
\end{figure}

The lightest Higgs is usually SM like but can be heavier than in the MSSM. Indeed the tree-level lightest Higgs boson mass squared, which can be approximated by \cite{King:2005jy}
\beq 
\begin{split}
 \label{eq:mhmax}
m^2_{h_1, \: \textrm{tree}} \simeq & \; M^2_{Z^0} \cos^2 2\beta + \frac{1}{2} \lambda^2 v^2 \sin^2 2\beta + g'^2_1 v^2 \left(\mathcal{Q}'_{H_d} \cos^2 \beta + \mathcal{Q}'_{H_u} \sin^2 \beta \right)^2\\
& -\frac{\l^4 v^2}{g'^2_1 \mathcal{Q}'^2_{S}}\left(1-\frac{A_\l \sin^2 2\beta}{2 \mu}+\frac{g'^2_1}{\l^2}\left(\mathcal{Q}'_{H_d} \cos^2 \beta + \mathcal{Q}'_{H_u} \sin^2 \beta \right)\mathcal{Q}'_{S}\right)^2\,,
\end{split} 
\eeq
receives three types of additional contributions as compared to the MSSM. The first one  proportional to $\lambda$  is also found in the NMSSM, the second one comes from the additional U(1) gauge coupling $g'_1$ and the last arises from a combination of pure UMSSM and NMSSM terms. The first term is not expected to play as important a  role as in the NMSSM since $\lambda$ is small. This is because $\lambda$ is inversely proportional to the vev of the singlet Higgs field which is in turn  related to the mass of new gauge boson, see eqs.~\eqref{eq:mueff} and \eqref{eq:mzz'}. The strong dependence of the latter two terms on the $U(1)'$ charges means that the size of the tree-level contribution to the Higgs mass will mostly depend on the value of $\te6$. 
We illustrate this taking the limit of small mass mixing between the two $Z$ bosons as given in eq.~\eqref{eq:c2b_azz0}. Figure~\ref{fig:thmh1caZ0}  shows that in  this limit   $m^\textrm{tree}_{h_1}$ does not exceed the MSSM upper bound and that its value depends strongly on $\te6$. The maximum is reached for $|\te6| = \tan^{-1}(\sqrt{3/5}) \approx 0.66$ which corresponds to $\cos^2 \beta$ or $\sin^2 \beta$ = 1.   A smaller value for the maximum  $m^\textrm{tree}_{h_1}$ is found for lower values of $M_{Z_2}$ since  
the last term  in eq.~\eqref{eq:mhmax} then gives a larger negative contribution to the tree-level mass. Furthermore, the tree-level mass tends to be suppressed for  small $\mu$ values since these are linked to small values of $\lambda$ and thus to a small contribution from the NMSSM term.
This behaviour shown in figure~\ref{fig:thmh1caZ0} is mostly observed for $|\mu| \leq 2$~TeV and $|A_\l| \leq 4$~TeV.
In the general case with non-zero $Z^0 - Z'$ mixing, a large tree-level contribution can be obtained for a wider range of parameters and a mass of 125 GeV for the lightest Higgs boson can easily be reached even with a small contribution from one-loop corrections that comes predominantly from the stop/top sector, as in the MSSM.

 Typically the Higgs spectrum will consist of a standard model like light Higgs, a heavy mostly doublet scalar which is almost degenerate with the pseudoscalar and the charged Higgs, and a predominantly singlet Higgs boson.  The latter can be either $h_2$ or $h_3$, depending on the values of the free parameters of the model,  in particular  $M_{Z_2}$ and $A_\l$. The singlet  Higgs is never  $h_{1}$  
because its mass depends on $v_s$ which is large due to the lower bound on $M_{Z_2}$.

\section{Constraints on the model}
\label{sec:constraints}

\subsection{Higgs physics}
\label{subsec:H-constraints}

For the Higgs sector we require that the light\footnote{ Strictly speaking it is also possible that the Higgs at 125 GeV corresponds to $h_2$, however we did not find such points in the scan.} Higgs mass lies in the range  $m_{h_1}=125.1 \pm 3 {\rm~GeV}$ allowing for a theoretical uncertainty around 2 GeV. We impose  constraints on the Higgs sector keeping only points allowed by \texttt{HiggsBounds-4.1.3}~\cite{Bechtle:2013wla} and  by \texttt{HiggsSignals-1.2.0}~\cite{Bechtle:2013xfa} at 95\% C.L  ($p$-value above 0.05). We also use constraints contained in \NTools\cite{Ellwanger:2005dv}, in particular the one on the heavy Higgs search in the $\tau^+\tau^-$ decay mode that rules out some of the large $\tan\beta$ region.  

Note that the Yukawa couplings evaluated at the SUSY scale which enter the computation of the Higgs boson masses must remain perturbative. We require that all  Yukawa couplings stay below
$\sqrt{4\pi}$ at the SUSY scale. This condition will impose restrictions on both  the  small  and the very large $\tb$ values (recall that  $\tb$ is not a free parameter of the model).  Yukawa couplings within the perturbative limit can nevertheless induce a 
very large width for some  of  the Higgs states, since we work in the context of elementary Higgs particles we impose the condition $\Gamma(h_i)/m_{h_i}<1$.

\subsection{Collider searches for $Z'$}
\label{subsec:Zp-constraints}

One of the main constraint on this model comes from the direct collider searches for a $Z'$ boson in the two-lepton decay channel. The best limits have been obtained at the LHC by the ATLAS~\cite{Aad:2014cka} and CMS~\cite{Khachatryan:2014fba} collaborations for $pp$ collisions at a center-of-mass energy of 8~TeV. In~\cite{Khachatryan:2014fba} limits were obtained with an integrated luminosity of 19.7~fb$^{-1}$ (20.6~fb$^{-1}$) in the dielectron (dimuon) channel and lead to $M_{Z_2} > 2.57$~TeV for $\te6=\theta_\psi$, assuming only SM decay modes. 
Such limits however depend on  the couplings of the $Z_2$, hence on $\te6$. To reinterpret this limit for any value of $\te6$, 
 we first simulate Monte Carlo signals for $Z'$ production using the same Monte Carlo generator and PDF set as in~\cite{Aad:2014cka}, respectively \textsc{Pythia} 8.165~\cite{Sjostrand:2007gs} and MSTW2008LO~\cite{Martin:2009iq}, for a large set of $\te6$ values.  We get results compatible with  with the ones derived in~\cite{Accomando:2015cfa} as well as the one obtained by the CMS collaboration~\cite{Khachatryan:2014fba}. Then we interpolate our limits for any possible choice of $\te6$. Note that  the coupling of $Z_2$ to the standard model fermions also weakly  depends on $\azz$. We have checked that this dependence does not modify significantly the $Z_2$ limits and are well below PDF uncertainties~\cite{Aad:2014cka}.
Furthermore in the UMSSM the $Z_2$ can  decay into supersymmetric particles,  RH neutrinos and Higgs bosons, thus reducing the branching ratio into leptons. The limits on the $Z_2$ mass are therefore weakened~\cite{Gunion:1986ky,Gherghetta:1996yr,Chang:2011be,Corcella:2014lha}. To take this effect into account we determine in a second step the modified leptonic branching ratio for each point in our scan, and re-derive the corresponding limit.

For any value of $\te6$ we restrict the scan to $|\alpha_Z|<10^{-3}$~\cite{Erler:2009jh}. In addition, the mixing between $Z^{0}$ and $Z'$ can be constrained by the $\Delta \rho$ parameter~\cite{Babu:1996vt}.
This observable, which measures the deviation of the $\rho$-parameter of the standard model from unity, receives a specific UMSSM tree-level contribution because $Z_1$ is no longer purely the $Z^0$ boson. In the limit where $M_{Z'}^2 \gg M_{Z^0}^2, \Delta_Z^2$, which is the case for the TeV scale $Z'$, this new contribution reads~\cite{Babu:1996vt}
\beq \Delta \rho_Z = \alpha_Z^2 \frac{M_{Z_2}^2}{M_{Z_1}^2}.\label{eq:rhoZZ'}\eeq
We compute $\Delta \rho$ for each point in the parameter space using a \micro routine which also contains leading one-loop third generation sfermions and leading two-loop QCD contributions. We impose the upper bound $\Delta \rho < 8.8\times10^{-4}$~\cite{Agashe:2014kda}.

\subsection{Collider searches for SUSY particles }
\label{subsec:susy-constraints}
 First we impose generic  constraints from LEP on neutralinos, charginos, sleptons and squarks. For the latter we ignore the possibility of very compressed spectra and use the generic limit at 103~GeV. 
 Lighter compressed squarks can in any case be  constrained from LHC monophoton searches~\cite{Belanger:2012mk,Aad:2014tda} and monojet analyses~\cite{Dreiner:2012gx}. 

The most powerful and comprehensive  constraints on supersymmetric partners  have been obtained by ATLAS and CMS using the data collected at 7 and 8 TeV. 
Searches were performed  for a wide variety of channels and results were presented both in the framework of specific models, such as the MSSM, and in the context  of  simplified model spectra (SMS). 
Here we use the SMS results to find the main constraints on the UMSSM.
We base our analysis on \smodels~\cite{Kraml:2013mwa,Kraml:2014sna}, a tool designed to decompose the signal of an arbitrary BSM model into simplified model topologies and to test it against LHC bounds. The version used includes more than 60 SMS results from both ATLAS and CMS.

The input to \smodelsnn~are SLHA files that contain the full mass spectrum, decay tables as well as production cross sections. 
The input files, including tree-level production cross sections, are generated using {\tt micrOMEGAs\_4.1.5} \cite{Belanger:2014vza}, for strongly produced particles, \smodelsnn~then calls \texttt{NLL-fast}~\cite{Beenakker:1996ch,Beenakker:1997ut,Kulesza:2008jb,Kulesza:2009kq,Beenakker:2009ha,Beenakker:2010nq,Beenakker:2011fu} to compute the k-factor at NLO+NLL order.
Subsequently the code decomposes the full model into simplified model components,
and calculates the weight (production cross section times branching ratio, $\sigma\times {\cal B}$) for each topology.
To limit computing time, topologies with small weights are not considered in the
decomposition.
As a minimum weight we have used a cutoff $\sigma_{cut}$ of $0.03$ fb.
The resulting list is then confronted with the \smodelsnn~database, for any matching
result $\sigma\times {\cal B}$ is compared against the experimental upper limit. If no matching results are found the point is labeled as not tested.
This  may happen for several reasons, either all cross sections are below $\sigma_{cut}$ in which case the decomposition
will not return any entries, there is no matching simplified model result
in the database, or the mass vector of the new particles lies outside the
experimental grids for all applicable SMS results.
Since soft decay products cannot be detected,  in this analysis we disregard  vertices where the mass splitting between the mother and daughter SUSY particles is less than 5 GeV. A fully invisible decay at the end of a decay chain is compressed, the corresponding mother SUSY particle is then treated as an effective LSP.

Note that topologies that contain long-lived charged particles corresponding to $c\tau>10$~mm are not tested against SMS results within \smodelsnn.  
However searches for long-lived particles leaving charged tracks in the detector have been performed at the  Tevatron~\cite{Abazov:2012ina} and the LHC~\cite{ATLAS:2014fka,Chatrchyan:2013oca} and were interpreted in the context of long-lived  charginos or in the context of the pMSSM~\cite{Khachatryan:2015lla}.
When the neutralino LSP is dominantly wino, typically, the NLSP chargino will be stable at the collider scale.
We have therefore considered the D0  and ATLAS  upper limits for points with charginos in the mass range $100 - 300$~GeV and $450 - 800$~GeV, and  decay lengths $c\tau > 10$~m and $21$~m respectively. We have not included the limits from CMS as these cannot be simply reinterpreted for direct production of chargino pairs~\cite{Chatrchyan:2013oca}. Long-lived gluinos or squarks are also possible, we have not considered these cases since the interpretation of a given experimental analysis relies on 
the modeling of R-hadrons, thus   introducing large uncertainties. Moreover we have not implemented current limits on long-lived staus as these rarely occur in the parameter space considered.

\subsection{Flavour physics}
\label{subsec:flav-constraints}

Indirect constraints coming from the flavour sector, especially those involving $B$-Mesons, play an important role in defining the allowed parameter space of supersymmetric models, \textit{e.g.} \cite{Bediaga:2012py,Bernal:2011pj,Arbey:2012ax,Domingo:2007dx}. The constraints imposed on the model are listed in Table~\ref{tab:flav_limits}, though we do not in general require agreement with the measured value of $\Delta a_\mu$.  
We do however highlight the specific regions consistent with the measured value of  the muon anomalous magnetic moment. These mostly correspond to regions with a light LH smuon/sneutrino as mentioned in section~\ref{subsec:sfer}.
 To compute these observables, we have adapted the \NTools~routine to the UMSSM, for more details see~\cite{DaSilva:2013jga}. 
The most powerful constraints are $\Delta M_s$ and $\Delta M_d$ for small values of $\tan\beta$ while  $\bsg$ and $\bsmu$ are also important to constrain some large values of $\tan\beta$. 
We also compute $\bXsmu$ but this channel does not give additional constraints.
Uncertainties coming from CKM matrix elements, rare decays, hadronic parameters and theory are taken into account when computing the observables listed in Table~\ref{tab:flav_limits}, see~\cite{DaSilva:2013jga}. The most important uncertainties in our computation of flavour observables are theoretical (10\%) and from the CKM element $|V_{ts}| = (42.9 \pm 2.6) \times 10^{-3}$ \cite{Beringer:1900zz}.

\begin{table}[!htb]
\begin{center}
\begin{tabular}{cc}\hline \hline
\textbf{Constraint} & \textbf{Range} \\ \hline \hline
$\btau$ & [0.70,~1.58]$\times$10$^{-4}$ \cite{HFAG_btau}\\ 
$\bsg$ & [2.99,~3.87]$\times$10$^{-4}$ \cite{HFAG_bsg}\\ 
$\bsmu$ & [1.6,~4.2]$\times$10$^{-9}$ \cite{CMS:2014xfa}\\
$\Delta M_s$ & [17.805,~17.717] ps$^{-1}$ \cite{HFAG_dms}\\
$\Delta M_d$ & [0.504,~0.516] ps$^{-1}$ \cite{HFAG_dmd}\\
$\amu$ & [7.73,~42.14]$\times$10$^{-10}$ \cite{Bennett:2006fi,Roberts:2010cj,Aoyama:2012wk}\\ \hline \hline
\end{tabular}
\caption{Flavour constraints used and their allowed ranges which correspond to the experimental results (or to the difference between the experimental value and the standard model expectation for $\amu$) $\pm$ 2$\sigma$.}
\label{tab:flav_limits}
\end{center}
\end{table}

\subsection{Dark matter}

The value of the dark matter relic density has recently been measured precisely by the \textit{Planck} collaboration and a combination of \textit{Planck} power spectra, \textit{Planck} lensing and other external data leads to~\cite{Planck:2015xua}
\begin{equation}
\Omega h^2 = 0.1188 \pm 0.0010.
\label{eq:omega}
\end{equation}
We will impose only the 2$\sigma$ upper bound from eq.~\eqref{eq:omega} on the value of the relic density. That is we assume that either there is another component of dark matter or that there exists some regeneration mechanism that can bring the dark matter within the range favoured by \textit{Planck}~\cite{Hall:2009bx,Chu:2011be}. 

This measurement puts a strong constraint on the parameter space of the UMSSM whether the dark matter candidate is 
the lightest neutralino or the supersymmetric partner of the right-handed neutrino. Since the three RH sneutrinos have the same coupling to all other particles in the model we assume for simplicity that the third generation sneutrino is the lightest. 
In previous studies it was shown that the favoured mass for the RH sneutrino LSP was near $M_{Z_2}/2$, although much lighter sneutrinos could also be found, especially near $m_{h_1}/2$ or when coannihilation was present~\cite{Belanger:2011rs}.
As in the MSSM the lightest neutralino covers a large range of mass,  the main new features being the possibility of a singlino LSP~\cite{Franke:2001nx,Suematsu:2005bc,Nakamura:2006ht} 
and  the possibility for this singlino to  have a non-negligeable  bino' component. Typical MSSM features can also be observed as the example of wino LSP annihilating efficiently into $W$'s and strongly degenerate in mass with chargino NLSP. However sometimes the mass degeneracy between the NLSP and the LSP can be sufficiently small to give an absolutely stable charged NLSP. When focusing on relic density constraints we will systematically discard these configurations.

One of the strongest constraint on DM arises from direct detection. We implement the upper limit from the LUX collaboration~\cite{Akerib:2013tjd} taking \micro default values for the quark coefficients in the nucleons.  This upper limit strongly constrains the scenarios where the LSP is ${\cal O}(100$~GeV).
Another relevant constraint is the one from FermiLAT searches for DM annihilation from the dwarf spheroidal satellite galaxies of the Milky Way where limits obtained for DM annihilation into $b\bar{b}$ and $\tau^+\tau^-$ can constrain scenarios with DM masses below 100~GeV~\cite{Ackermann:2015zua}.

\section{Results}
\label{sec:results}

\begin{table}[!htb]
\begin{center}
\begin{tabular}{cc|cc}\hline \hline
\textbf{Parameter} & \textbf{Range} & \textbf{Parameter} & \textbf{Range} \\ \hline \hline
$m_{\tilde{\nu}_{\tau R}}$ & [0, 2] TeV & $\mu, M_1$ & [-2, 2] TeV \\ 
$M_{Z_2}$ & [2.2, 7] TeV & $M_2, A_\l, A_t, A_b, A_l$ & [-4, 4] TeV \\ 
$M'_1$ & [-20, 20] TeV & $M_3$ & [0.4, 12] TeV\\
$\te6$ & [-$\pi$/2, $\pi$/2] rad & $m_{\tilde{F_i}}, m_{\tilde{\nu}_j}$ & [0, 4] TeV \\
$\azz$ & [-$10^{-3}$, $10^{-3}$] rad & $m_{t}$ & 173.34 $\pm$ 1 GeV \cite{ATLAS:2014wva} \\ \hline \hline
\end{tabular}
\caption{Range of the free parameters where concerning the soft mass terms we define $F \in \{Q, u, d, L, e\}$, $i \in \{1, 2, 3\}$ and $j \in \{1, 2\}$ and where $m_{\tilde{F_2}} = m_{\tilde{F_1}}, m_{\tilde{\nu}_2} = m_{\tilde{\nu}_1}$.}
\label{tab:free_param}
\end{center}
\end{table}

After imposing universality for the sfermion masses of the first and second generation and fixing the trilinear coupling of the first two generation sfermions to 0 GeV, the UMSSM features 24 free parameters. The range used for these parameters in the scans are listed in Table~\ref{tab:free_param}. In addition  we have allowed the top mass to vary.
We perform a random scan over the free parameters and impose first the set of basic constraints described from section~\ref{subsec:H-constraints} to section~\ref{subsec:susy-constraints} :  the Higgs mass and couplings allowed by \texttt{HiggsBounds}, \texttt{HiggsSignals} and our modified \NTools~routines, perturbative Yukawas for top and bottom quarks, agreement with LEP limits on sparticles and LHC limits on the $Z'$ and finally a neutral LSP. We then include the constraints from $B$-physics mentioned in section~\ref{subsec:flav-constraints}. Another scan is done to highlight the regions of  parameter space which give sufficient New Physics contribution to $\amu$. For this we restrict the soft masses of the second generation of sleptons to [0, 2]~TeV and we impose all constraints given in section~\ref{subsec:flav-constraints}. 

\begin{figure}[!htb]
\begin{center}
\centering
\subfloat[]{\label{fig:tree-level-h1-a}\includegraphics[width=8.cm,height=5.5cm]{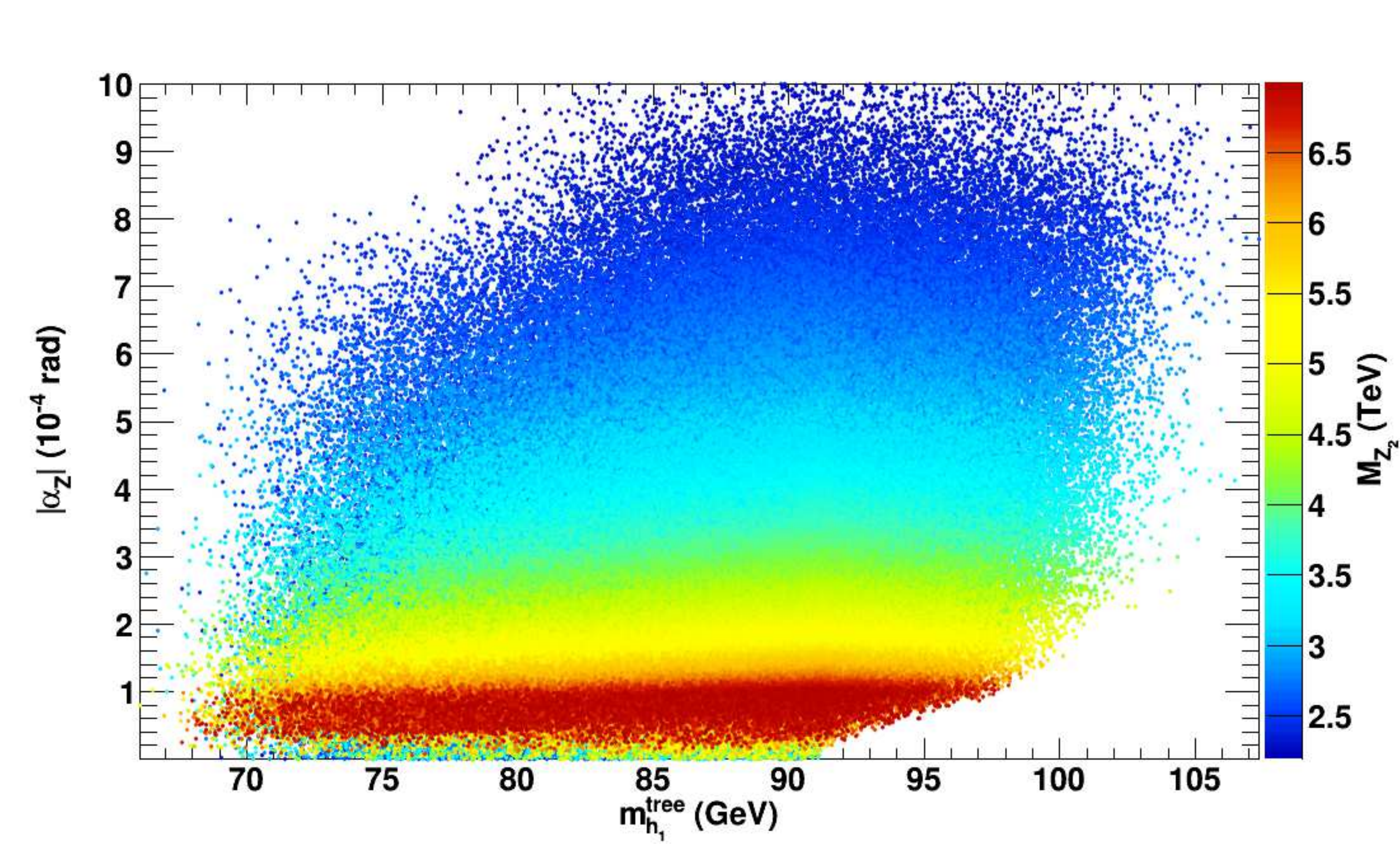}}
\subfloat[]{\label{fig:tree-level-h1-b}\includegraphics[width=8.cm,height=5.5cm]{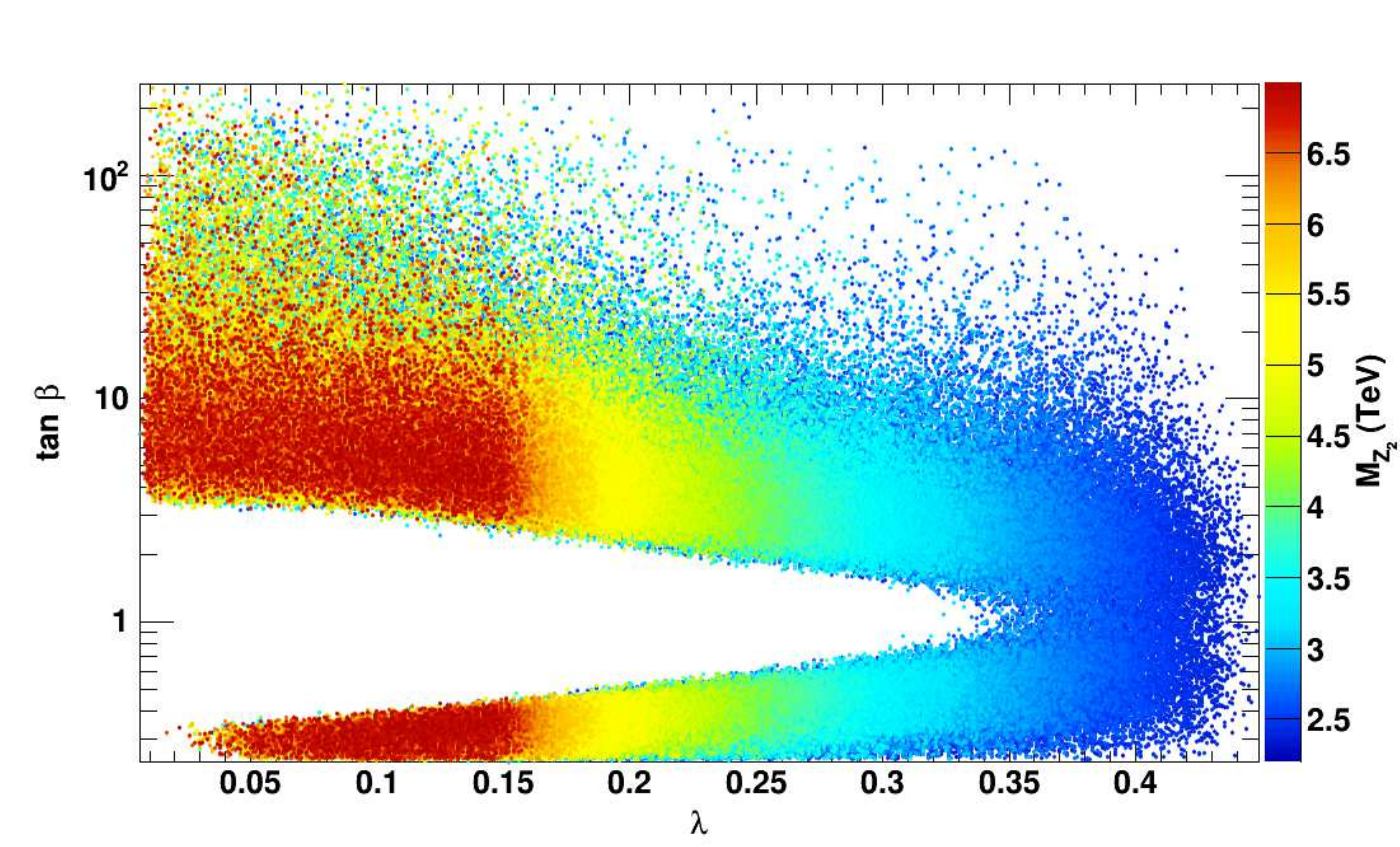}}
\caption{(a) $\alpha_Z$ as a function of the tree-level component of $m_{h_1}$ and (b) $\tb$ as a function of $\l$. For both plots $M_{Z_2}$ is taken as colour code.}
\label{fig:tree-level-h1}
\end{center}
\end{figure}

For all points that satisfy these sets of constraints in both scans, around $4 \times 10^5$, we 
 found that the maximum tree-level mass for the Higgs reached only $m_{h_1} \approx 107$~GeV and was above the $Z_1$ mass only for mixing angles $\alpha_Z > 2 \times 10^{-5}$, see figure~\ref{fig:tree-level-h1-a}.
 Thus a  contribution from the radiative corrections in the stop/top sector is still required to reach a Higgs mass of 125 GeV.
 Nevertheless the full range of values of $\tb$ is allowed.  Small values of  $\tb>1$ require a 
 large value of $\l$\  to compensate the small MSSM-like tree-level contribution to the light Higgs mass, see  figure~\ref{fig:tree-level-h1-b}. This also means that $v_s$, hence $M_{Z_2}$, cannot be too large given the range assumed for the $\mu$ parameter, see eq.~\eqref{eq:mueff}.
  Radiative corrections from the top/stop sector are expected to be large for $\tb<1$ since  the top Yukawa coupling increases as $1/\sin \beta$, which explains why a larger range  for $\l$ is allowed when $\tb<1$.

\begin{figure}[!htb]
\begin{center}
\centering
\subfloat[]{\label{fig:squarks-a}\includegraphics[width=8.cm,height=5.5cm]{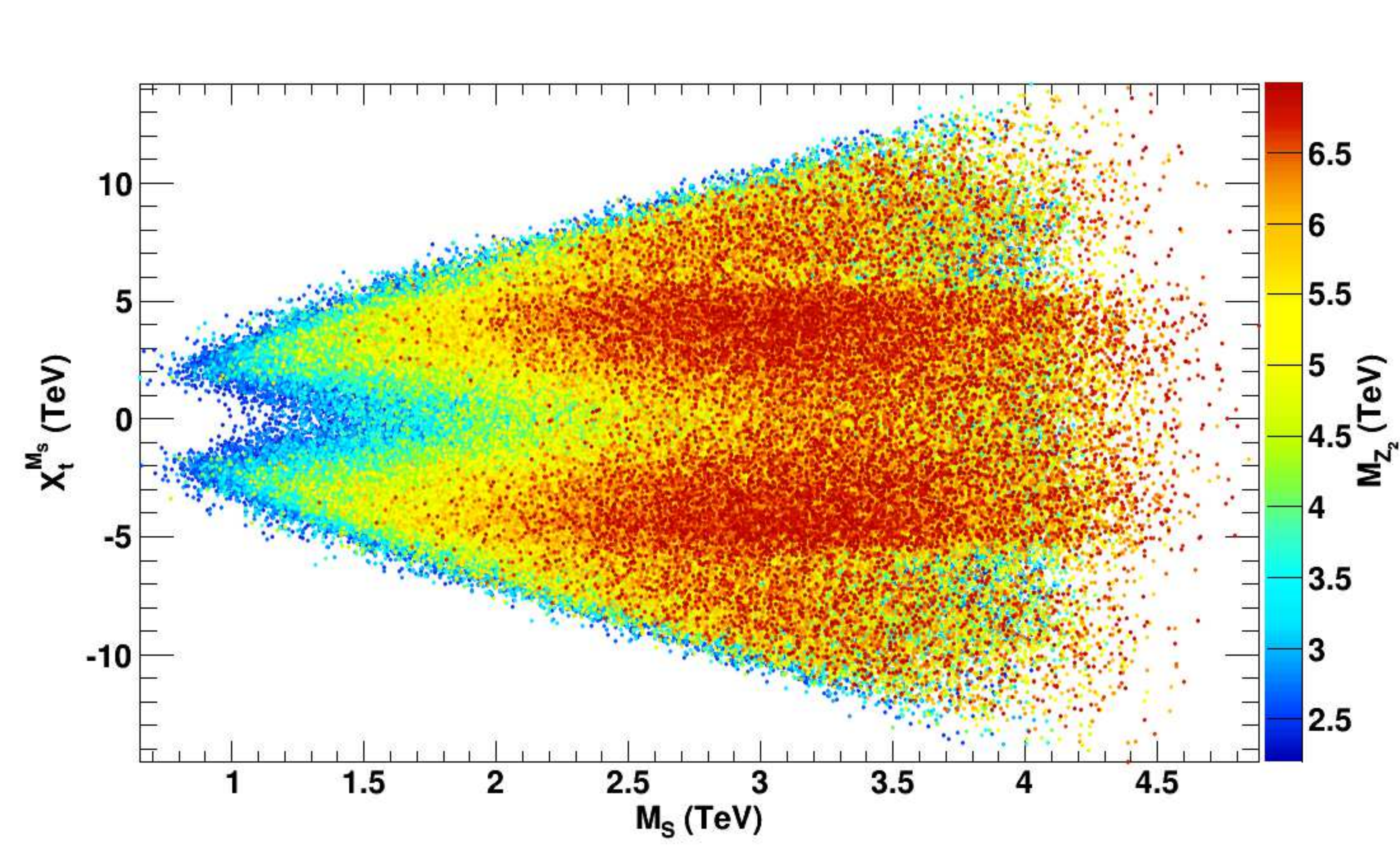}}
\subfloat[]{\label{fig:squarks-b}\includegraphics[width=8.cm,height=5.5cm]{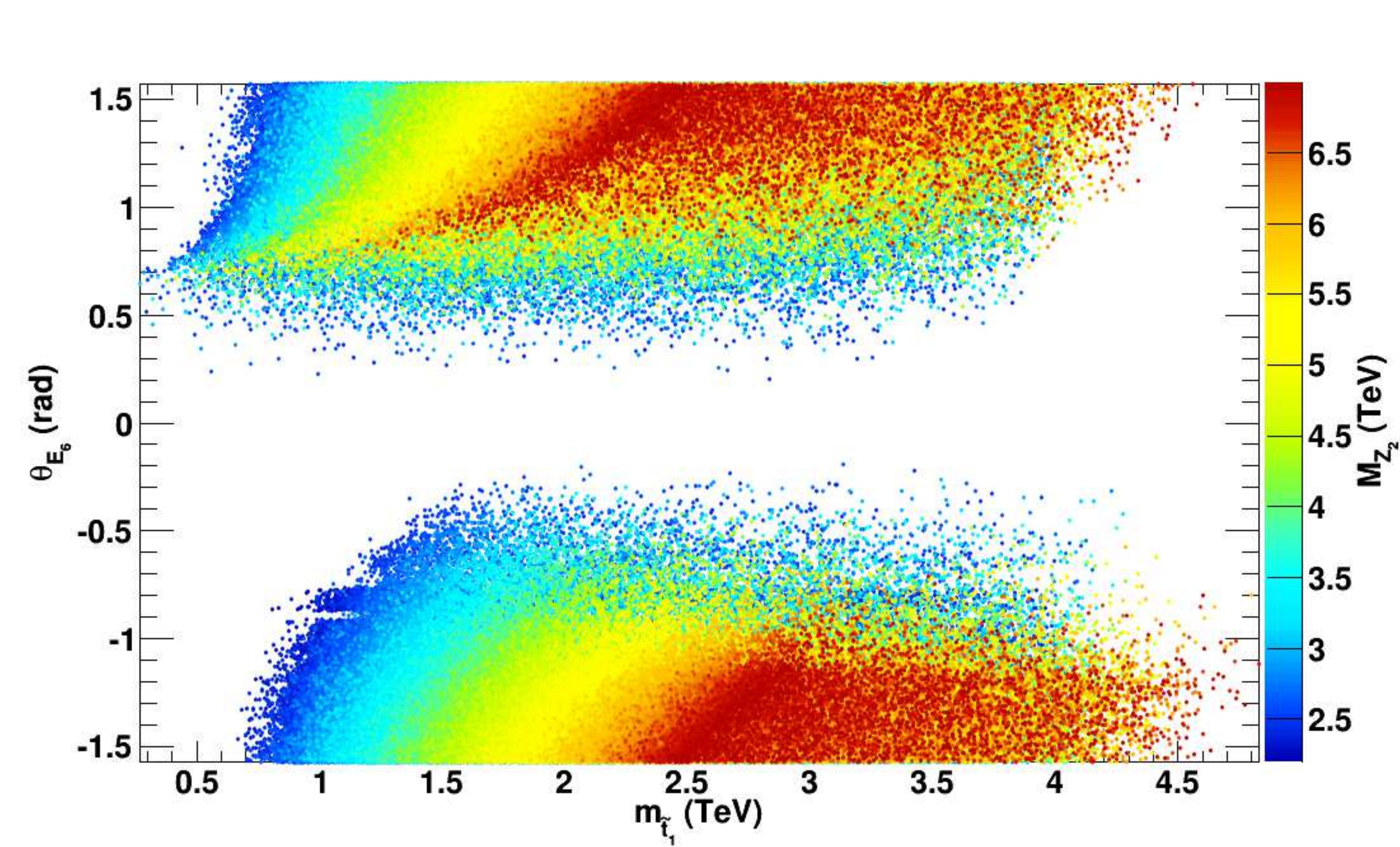}} \\
\subfloat[]{\label{fig:squarks-c}\includegraphics[width=8.cm,height=5.5cm]{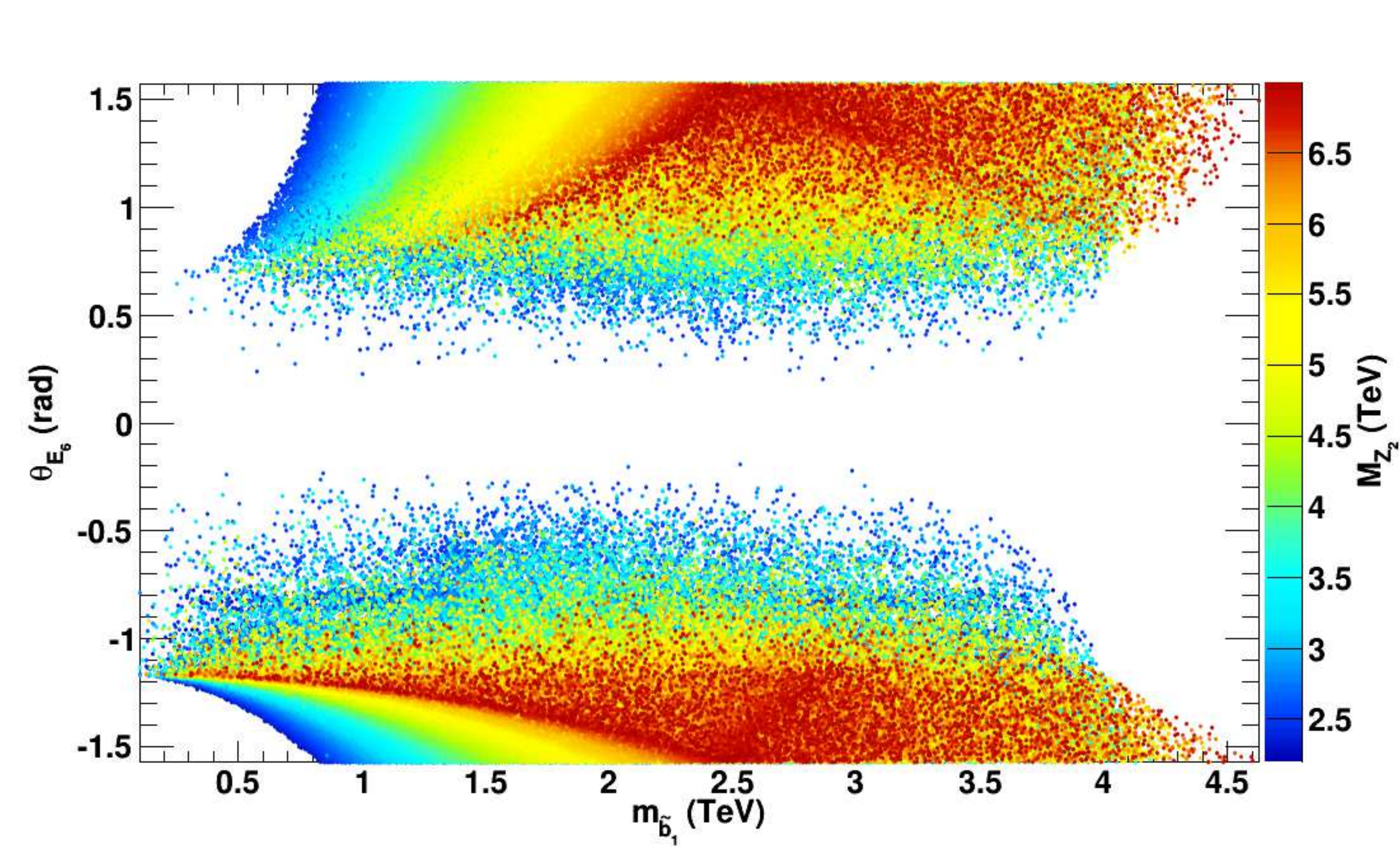}} 
\subfloat[]{\label{fig:squarks-d}\includegraphics[width=8.cm,height=5.5cm]{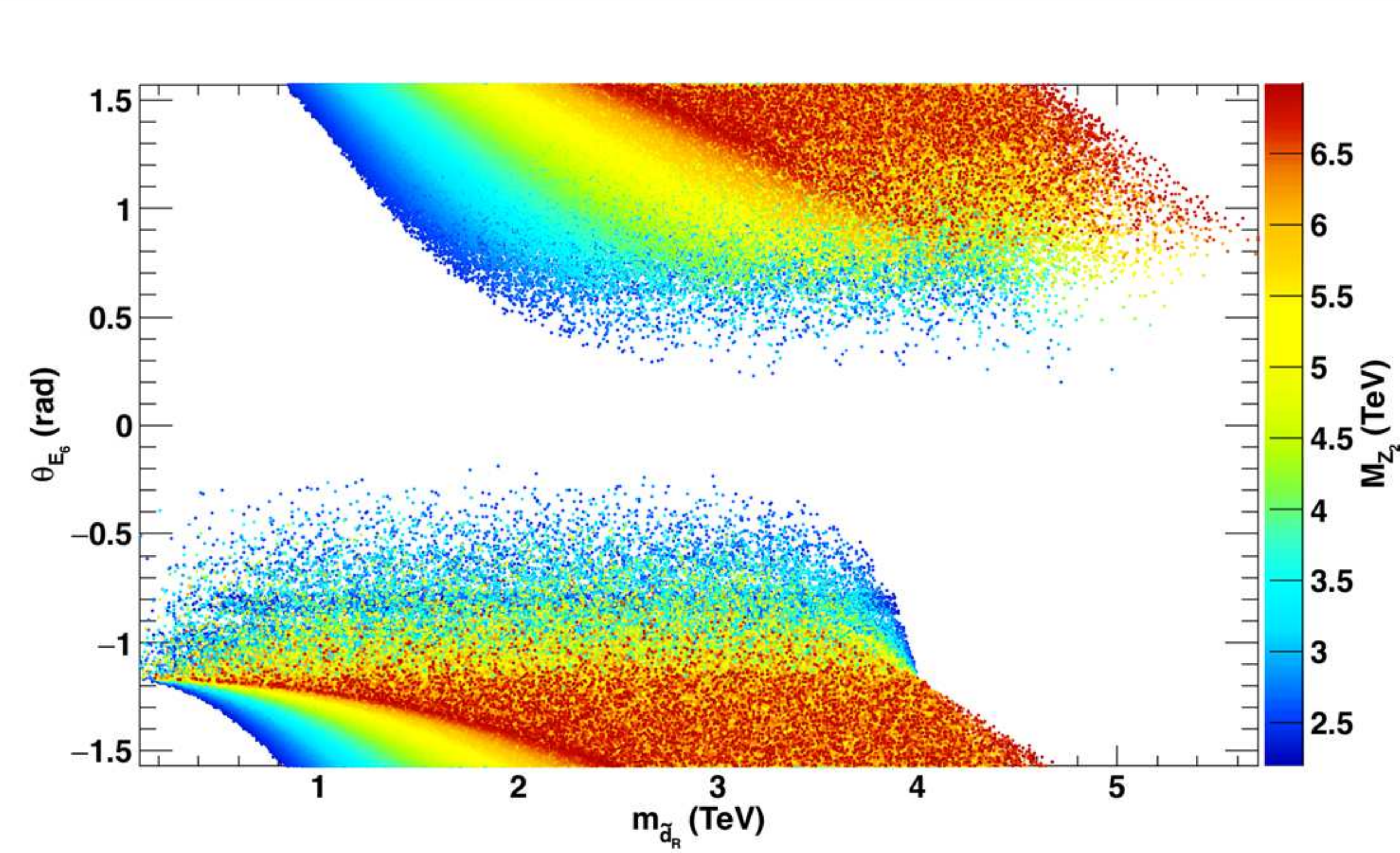}}
\caption{(a) $X_t$ as a function of $M_S$ and $\te6$ as a function of (b) $m_{\tilde{t}_1}$, (c) $m_{\tilde{b}_1}$ and (d) $m_{\tilde{d}_R}$. For all plots $M_{Z_2}$ is taken as colour code.}
\label{fig:squarks}
\end{center}
\end{figure}

It is well known that large one-loop corrections from the stop sector require  heavy stops and/or large mixing ~\cite{Hall:2011aa}.  The mixing parameter $X_t=A_t-\mu/\tb$ is indeed found to be large when $M_S=\sqrt{m_{\tilde{t}_1}m_{\tilde{t}_2}}<1$~TeV while heavy stops (associated with  large $M_S$) allow no mixing,  see figure~\ref{fig:squarks-a}. The heavier the $Z_2$ the larger the minimal value for the scale $M_S$ where zero mixing is allowed.  

The spectrum for supersymmetric particles differs significantly from the case of the MSSM and NMSSM, depending on the choice of $U(1)'$ charges.  
The lightest stop mass can be as light as 300 GeV for $\te6 \sim 0.66$ (figure~\ref{fig:squarks-b}), this value corresponds to the largest negative contribution to the stop mass from the $D$-term, see section~\ref{subsec:sfer}. When 
 $\te6<0$ the lightest stop is at least 670 GeV. Similar values are found for both LH and  RH up-type squarks, modulo mixing effects. 
Such light squark masses are well within the range of exclusion of LHC searches within the MSSM, hence the need to reinvestigate the impact of these searches within the UMSSM discussed in the next section. 
The $\tilde{d}_R$ mass receives a large negative $D$-term contribution for $\te6 = -\tan^{-1}(3\sqrt{3/5}) \approx -1.16$. For this value it can be as light as allowed by LEP (103 GeV), see figure~\ref{fig:squarks-d}. For $\te6> 0$ the RH d-squark is  above the TeV scale while the LH one can be light since $m_{\tilde d_L}= m_{\tilde u_L}$. This implies also that a light sbottom, say below 500 GeV, can be found for either value of $\te6$, see figure~\ref{fig:squarks-c}. In one case it is mostly LH and in the other RH. 
Note that an increase in the lower limit on the $Z'$ mass will lead to larger squark masses except for the specific values of $\te6$ where one gets a very large $D$-term contribution. 
Finally, the gluino mass is determined  by $M_3$, hence can also be well below the TeV scale.

\begin{figure}[!htb]
\begin{center}
\centering
{\includegraphics[width=0.65\textwidth]{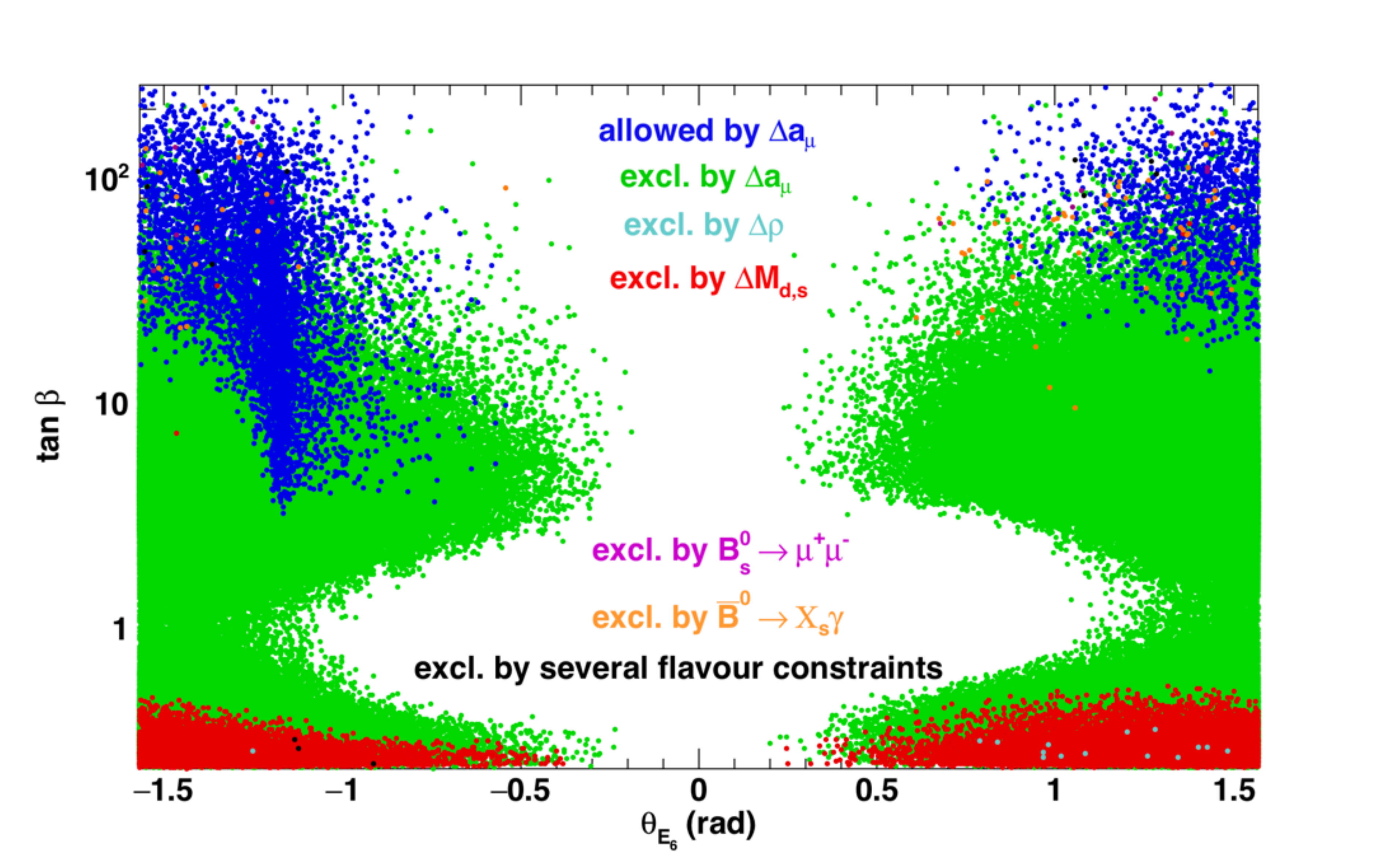}}
\caption{Points  of the scan in the $\tb$ - $\te6$  plane where the colour code shows the flavour process that provides the main exclusion.
The region that is compatible with $\Delta a_\mu$ is also displayed.
 The flavour observables are computed with the \NTools~routine adapted to the UMSSM.}
\label{fig:tbte6}
\end{center}
\end{figure}

The impact of the flavour constraints is best displayed in the $\tb-\te6$ plane, see figure~\ref{fig:tbte6}. As expected $\Delta M_s$ and $\Delta M_d$ are the most important constraints in our scans and exclude a large part of the parameter space when $\tb<1$, through the charged Higgs contribution~\cite{DaSilva:2013jga}. The contribution from Double Penguin diagrams to these observables enable exclusion of a few scenarios at large $\tb$. $\bsmu$ and $\bsg$ are important for scenarios at very large $\tb$ but they mostly fail to exclude points, especially for cases where the mass of heavy neutral and charged MSSM-like Higgs bosons is above several TeVs. Finally the New Physics contribution to the deviation of the $\rho$-parameter from unity  exclude only few points, mostly from the sfermion contributions. Actually the pure UMSSM contribution shown in eq.~\eqref{eq:rhoZZ'} can barely reach $10^{-4}$ for the allowed values for $\azz$ and $M_{Z_2}$ and is then negligible. Note that, as we will see in the next section, specific regions of the parameter space give large enough contributions to the anomalous magnetic moment of the muon.

\subsection{$\amu$}

\begin{figure}[!htb]
\begin{center}
\centering
\subfloat[]{\label{fig:amu-a}\includegraphics[width=8.cm,height=5.5cm]{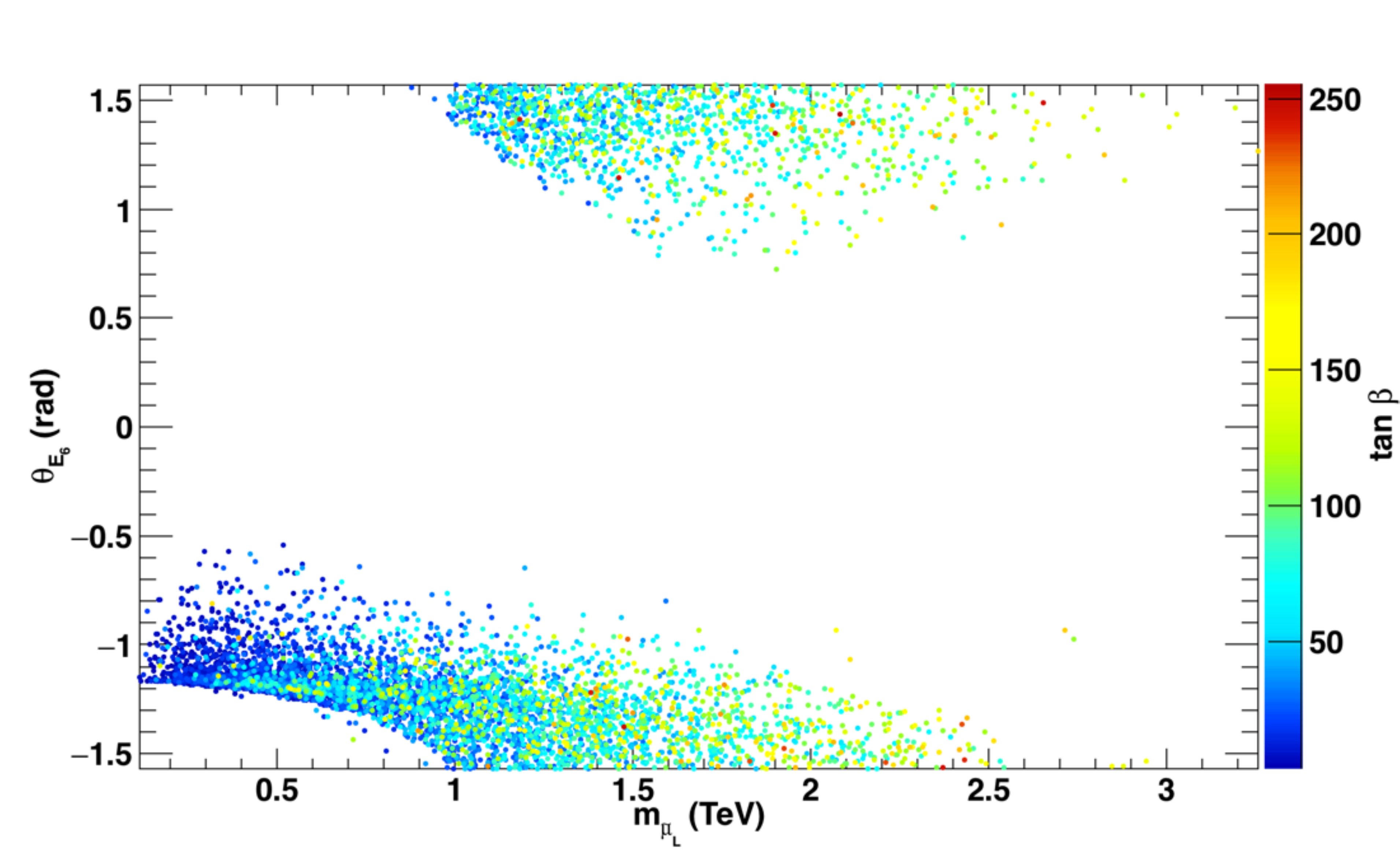}} 
\subfloat[]{\label{fig:amu-b}\includegraphics[width=8.cm,height=5.5cm]{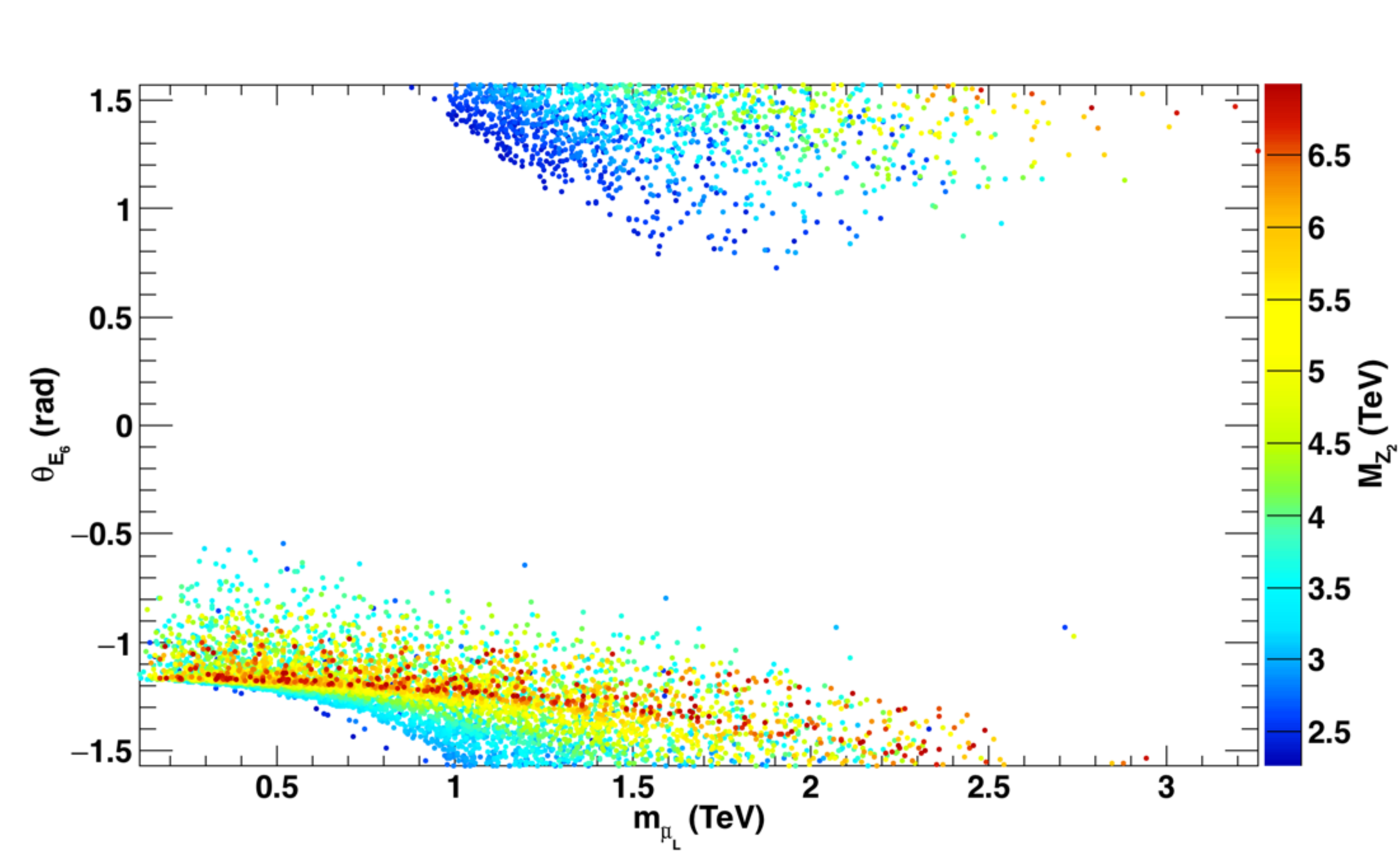}}
\caption{  Points allowed by $\amu$ in the  $\te6 - m_{\tilde{\mu}_L}$ plane, the colour code corresponds to (a) different values of $\tb$ and (b) $M_{Z_2}$.}
\label{fig:amu}
\end{center}
\end{figure}

Special conditions are required to get agreement with the value of $\amu$.
Indeed the discrepancy between the theoretical and experimental value requires a large contribution from 
New Physics. In the UMSSM this comes in particular from  diagrams involving smuon (LH sneutrino) and neutralino (chargino) exchange. 
A large UMSSM contribution requires either a  light smuon/LH sneutrino or an enhanced Yukawa for the muon. The latter is found at very large values of  $\tan\beta$, see figure~\ref{fig:amu-a}. 
A  light LH smuon mass arises  for $\te6 = -\tan^{-1}(3\sqrt{3/5}) \approx -1.16$  corresponding to  a large negative $D$-term contribution as explained in section~\ref{subsec:sfer}.  Future collider limits on the $Z'$ mass, say above 5 TeV,  will severely constrain scenarios for positive values of $\te6$ that are in agreement with the latest value of $\amu$,  see figure~\ref{fig:amu-b}. Note that the distribution of points in the $\te6 - m_{\tilde{\mu}_L}$ plane is similar to the one found in the general scan where consistency with the muon anomalous magnetic moment is not required, except that heavier sleptons are allowed in that case.

\subsection{Dark matter relic abundance}
\label{subsec:results_DM}

\begin{figure}[!htb]
\begin{center}
\centering
{\includegraphics[width=0.65\textwidth]{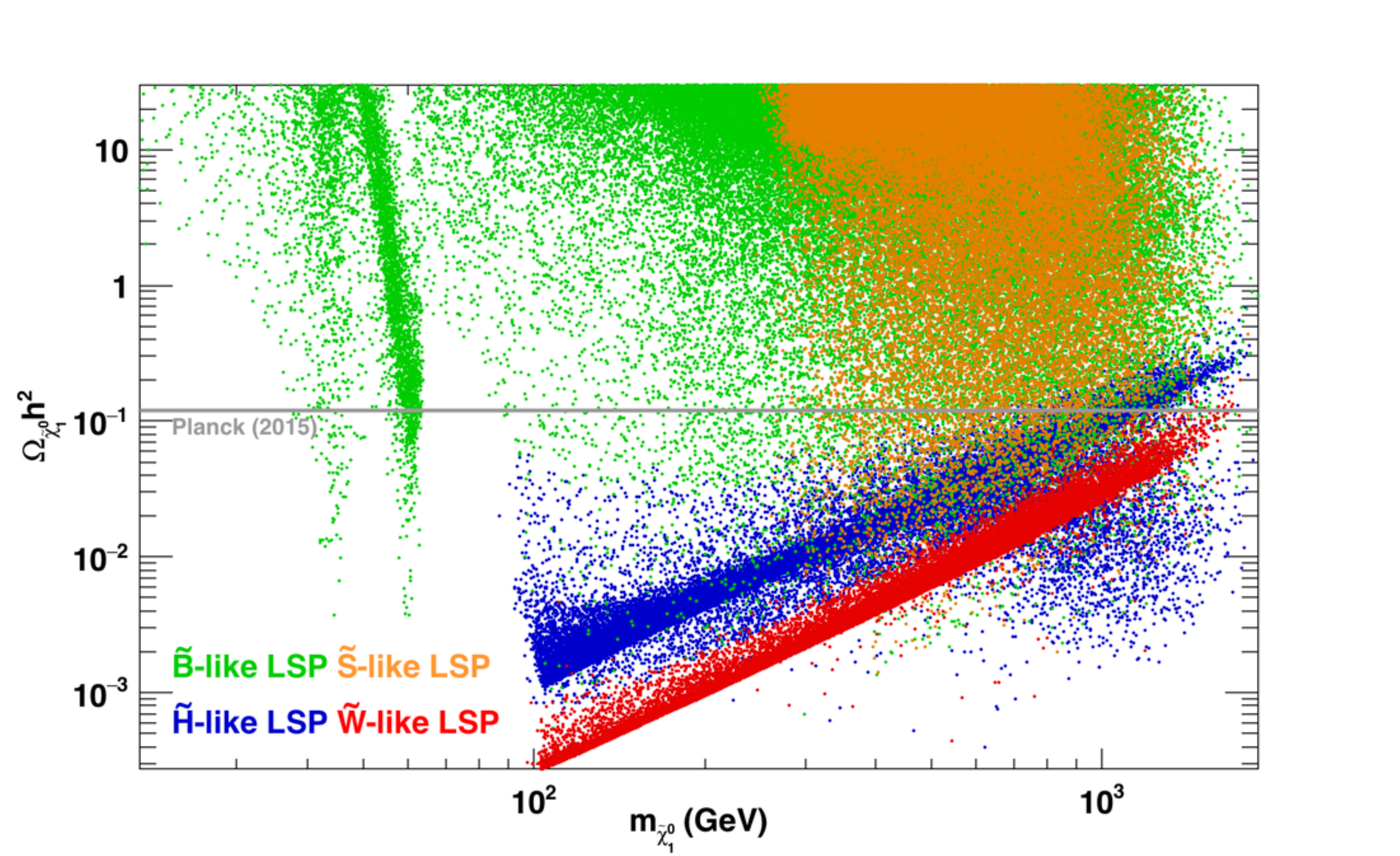}}
\caption{ Relic density of $\tilde{B}$ (green), $\tilde{W}$ (red), $\tilde{H}$ (blue) and $\tilde{S}$ (orange) LSP. The $2\sigma$ upper bound from \textit{Planck} is shown in grey.}
\label{fig:RDchi}
\end{center}
\end{figure}

\begin{figure}[!htb]
\begin{center}
\centering
\includegraphics[width=0.65\textwidth]{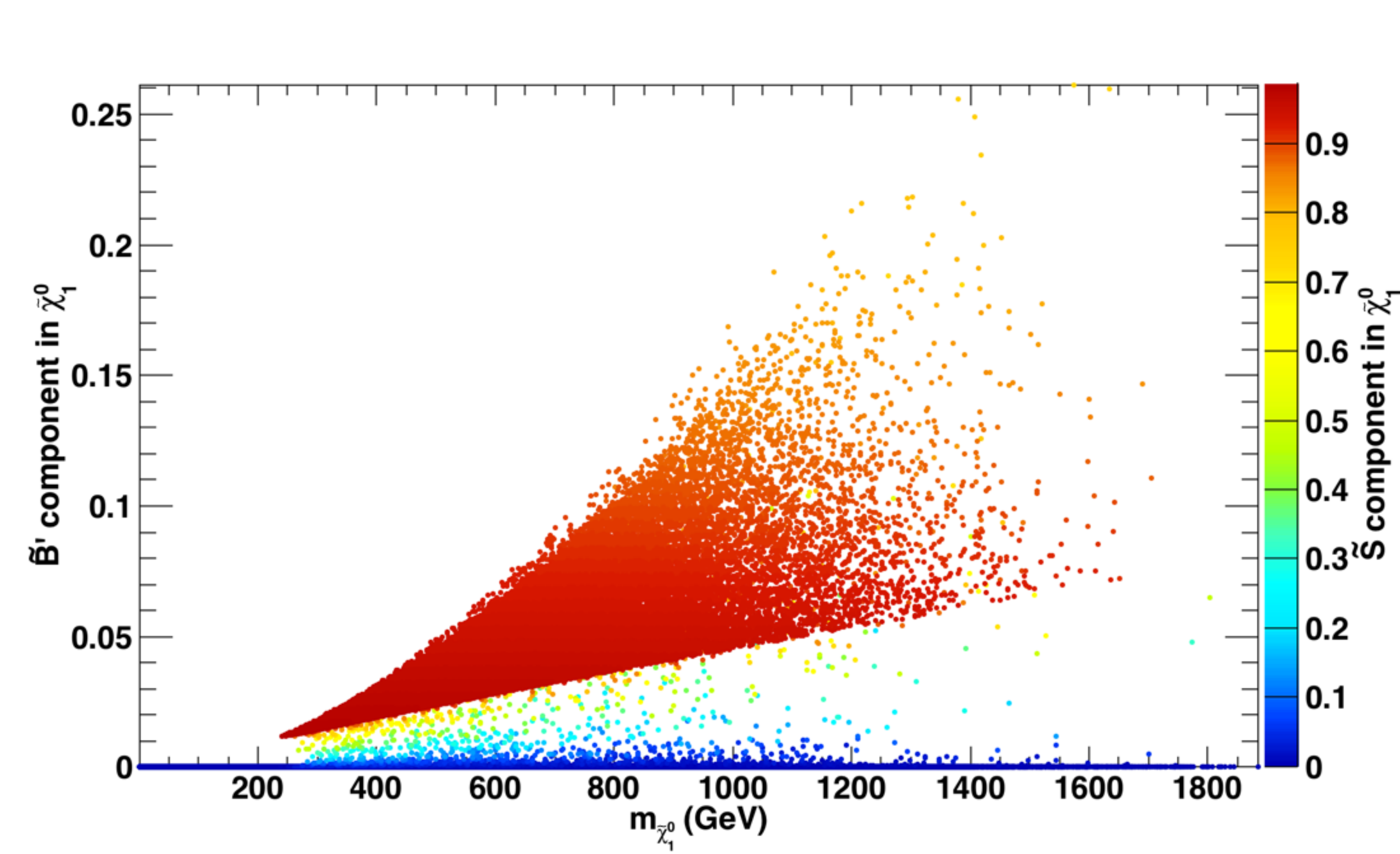}
\caption{$\tilde{B'}$ component in the neutralino LSP as a function of its mass with the $\tilde{S}$ component in the neutralino LSP as colour code.}
\label{fig:singlino_bino'}
\end{center}
\end{figure}

In this model the LSP can either be a neutralino or a RH sneutrino. 
The annihilation properties of the neutralino LSP are determined by its composition (figure~\ref{fig:RDchi}).  As in the NMSSM, the pure bino or singlino LSP is typically overabundant unless it can benefit from a resonance enhancement. Note that in this model the Higgs singlet is very heavy so that resonant annihilation of a  singlino  through the  Higgs singlet works only for heavy singlinos\footnote{For an analysis of a scenario with a light singlino DM see~\cite{Frank:2014bma}.}. The dominantly singlino LSP is found only for masses above 250~GeV. Some admixture of a higgsino/wino component or coannihilation processes can however reduce the relic density to $\Omega h^2\approx 0.1$  for any mass. Coannihilation can occur with gluinos or other gauginos as well as with sfermions. As in the MSSM the dominantly higgsino or wino LSP annihilates very efficienty into gauge boson pairs and therefore leads to an under-abundance of dark matter unless the higgsino (wino) LSP mass is roughly above 1 (1.5) TeV. 
Note that the $\tilde{B'}$ component of the LSP is never dominant, because the vev of the singlet, which mostly drives the mass of the $\tilde{S}$ and the $\tilde{B'}$,  eq.~\eqref{eq:neutralino}, is always above 6 TeV. For $|M'_1| \ll |v_s|$, $\tilde{S}$ and $\tilde{B'}$ are both shifted towards large masses whereas for $|M'_1| \gg |v_s|$ the singlino benefit from a seesaw-type mechanism which allows a singlino LSP down to 250 GeV. This close relation between $\tilde{B'}$ and $\tilde{S}$ is illustrated in figure~\ref{fig:singlino_bino'}.

We note that the fraction of points that satisfy the $2\sigma$ \textit{Planck} upper bound is much higher in the scan where we impose the constraint on $\amu$ than in the general scan. The main reason is that it is easier to satisfy the relic density upper bound with a bino LSP when the sleptons are light.

\begin{figure}[!htb]
\begin{center}
\centering
\includegraphics[width=0.65\textwidth]{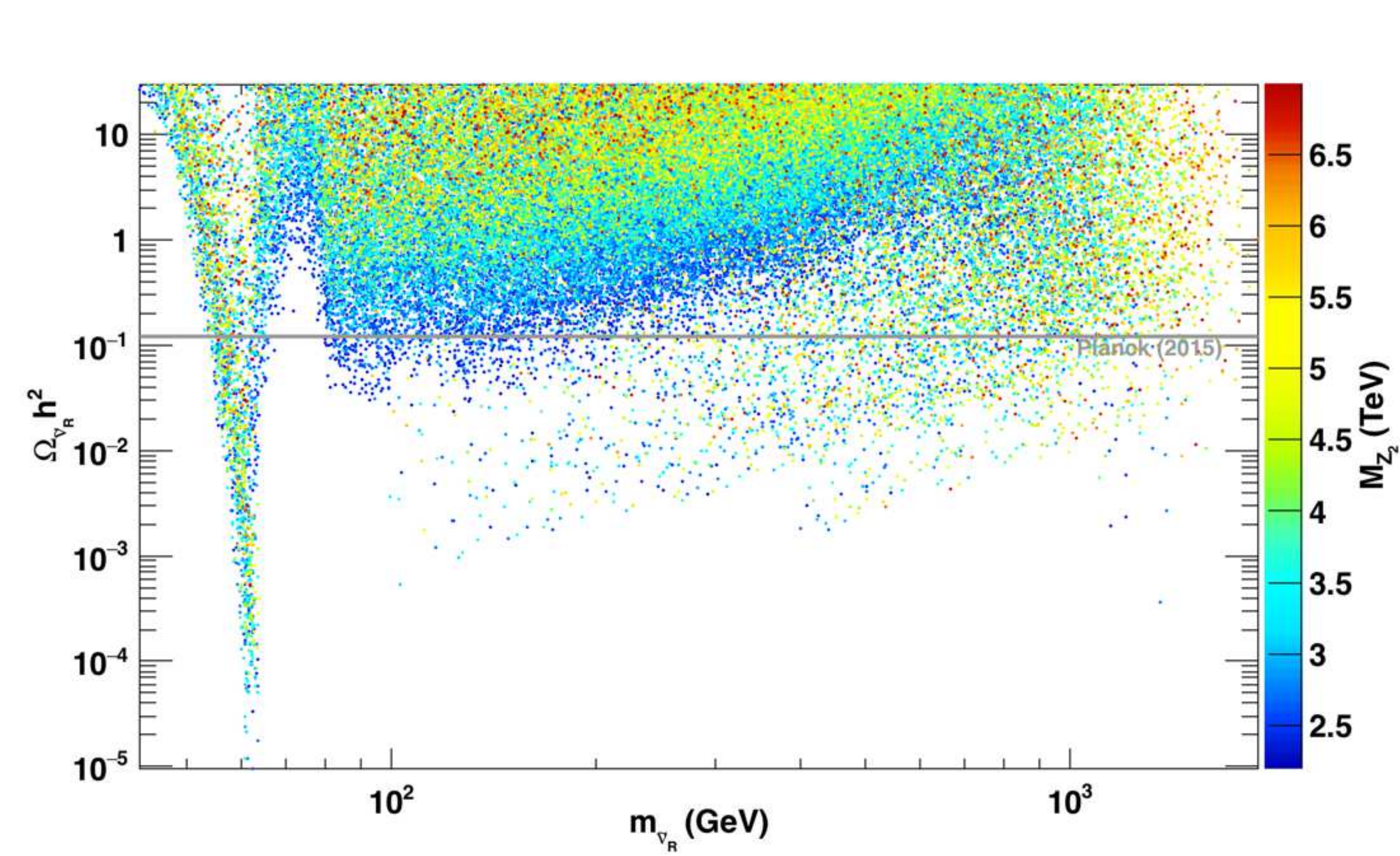}
\caption{Relic density for $\tilde{\nu}_{\tau R}$ LSP with $M_{Z_2}$ as colour code. The $2\sigma$ upper bound from \textit{Planck} is shown in grey.}
\label{fig:RDsneu}
\end{center}
\end{figure}

Sneutrino dark matter is typically overabundant as sneutrino annihilation channels are not very efficient. Agreement with the upper bound set by \textit{Planck} requires either $\msnu\approx m_{h_{1}}/2$ or $M_{Z_2}/2$ as found in~\cite{Belanger:2011rs}. The latter case requires $\msnu$ above the TeV scale when considering  current limits on the $Z'$ mass, here we consider DM below 2 TeV.  Annihilation into $W$ or $Z_1$ pairs through Higgs boson exchange was also found to be efficient enough for $\msnu \simgeq 100$~GeV~\cite{Belanger:2011rs}. However this process, which depends mostly on the singlet nature of the Higgs boson exchanged, will not give a large enough contribution if the lower limit on  $M_{Z_2}$ increases as shown in figure~\ref{fig:RDsneu}. Sneutrino LSP masses in the range $100-1000$~GeV are also allowed if some coannihilation mechanism, involving \textit{e.g.} the lightest neutralino or other sfermions, helps reduce the relic abundance. The low density of points in this region (see figure~\ref{fig:RDsneu}) reflects the fact that the importance of such coannihilation processes require the adjustment of uncorrelated parameters in the model.

\section{Impact of LHC searches for SUSY particles}
\label{sec:LHC_SMS_ll}

After having imposed the basic constraints, flavour constraints and an upper bound on the relic density $\Omega h^2< 0.1208$ (corresponding to the $2\sigma$ upper limit of eq.~\eqref{eq:omega}),
we next consider the impact of LHC searches for SUSY particles based on SMS results and using \smodelsnn. 
To analyse the impact of the SMS results we group the points into four categories.
Points excluded by \smodelsnn\ are labeled as excluded, points where the SMS results
apply but the cross section is below the experimental upper limit are labeled as not excluded.
Points where no SMS result applies, as explained in section~\ref{subsec:susy-constraints},
are labeled as not tested.
Finally points with long-lived particles cannot yet be tested in \smodelsnn.
Points that are not excluded are then examined in more details to determine the signatures that could best be used to further probe them with upcoming data.
We divide the study in three steps. First,  we consider scenarios with a neutralino LSP and find that the most stringent constraints on supersymmetric particles are obtained for light gluino or light squarks~\cite{Aad:2014wea,CMS:2013gea,CMS:2014ksa}.  Second, we concentrate on  points compatible with the measurements of the muon anomalous magnetic moment and that still have a neutralino LSP.  This dedicated scan provides a significant number of points with light sfermions and allows us to ascertain the impact of slepton searches. 
Finally we investigate scenarios with a RH sneutrino LSP, among these we do not characterize the ones  that are compatible with the muon $(g-2)$ because of the small number of points involved.
The possibilities to probe all points with long-lived charginos  are considered separately in section~\ref{sec:stable} regardless of the dark matter candidate.
Our results for the constraints on the SUSY spectra are presented in section~\ref{sec:afterLHC} where we combine all sets.

\subsection{Neutralino LSP}

In most points with a neutralino LSP, the LSP is actually either dominantly wino or higgsino, see figure~\ref{fig:RDchi}.
Points with a wino LSP are however mostly not considered in the \smodels\ analysis because they lead to long-lived charginos.
Therefore the most common configuration for the supersymmetric spectra relevant for SMS results is one with three  dominantly  higgsino particles with similar masses : the  LSP, the  second neutralino and the lightest chargino.
Moreover since the jets/leptons produced in the decay of the chargino (second neutralino) to the LSP are too soft to be detected the chargino (second neutralino) will often lead to a missing $E_T$ (MET) signature.
We will see that this has important consequences when using the SMS results.
In particular hardly any points can be excluded from  electroweakinos searches as only few can exploit the decay channel into real gauge/Higgs boson.
Furthermore we do not find constraints from decays into leptons via sleptons since sleptons are rarely light.

\subsubsection{Gluino constraints}
\label{sec:smodels:gluino}
\begin{figure}[!htb]
\begin{center}
\centering
\subfloat[]{\label{fig:gluinoExclusions-a}\includegraphics[width=0.52\textwidth]{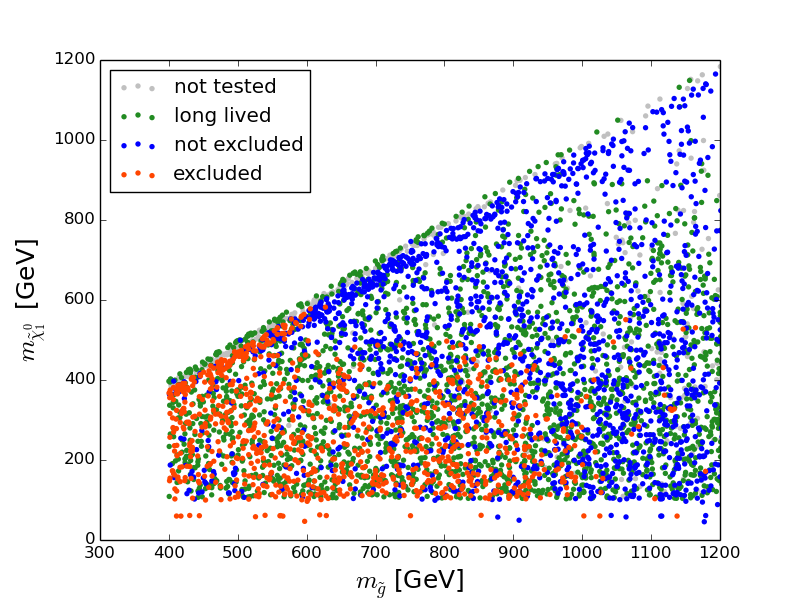}} 
\subfloat[]{\includegraphics[width=0.52\textwidth]{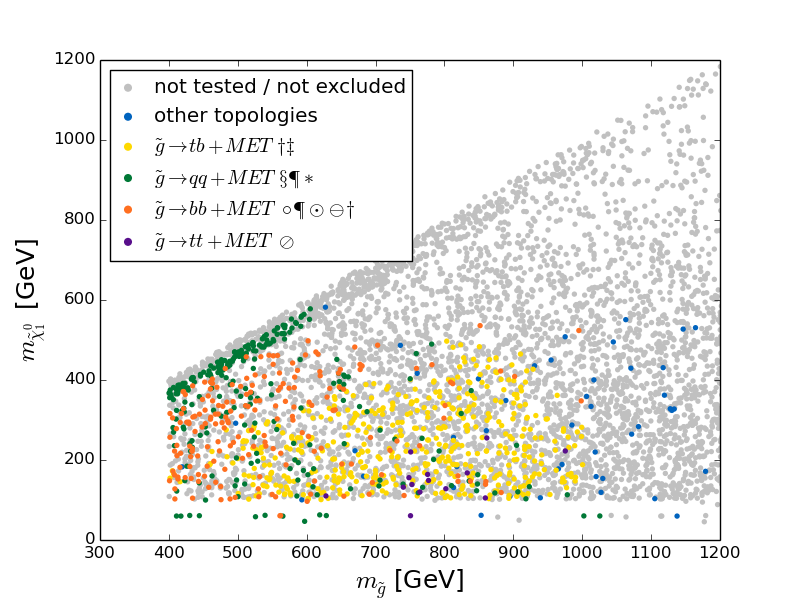}}
\caption{Exclusion with \smodels\ in the LSP - gluino mass plane. (a) shows whether a point can be tested, and excluded, as well as points which cannot be tested because of long-lived sparticles or other reasons. (b) shows  the most constraining topology for all excluded points. For the most frequently found topologies we specify the associated experimental searches : $\ddag$ =  \cite{ATLAS:2013tma}, $\dag$ = \cite{TheATLAScollaboration:2013tha}, $\ominus$ = \cite{CMS:zxa}, $\S$ = \cite{Aad:2014wea}, $\oslash$ = \cite{Chatrchyan:2013iqa}, $\ast$ = \cite{CMS:2013gea}, $\circ$ = \cite{CMS:2014wsa}, $\P$ = \cite{CMS:2014ksa} and $\odot$ = \cite{Chatrchyan:2013wxa}.}
\label{fig:gluinoExclusions}
\end{center}
\end{figure}

In figure~\ref{fig:gluinoExclusions} we show points with a neutralino LSP in the LSP and gluino mass plane for gluino masses up to $1200$~GeV.
On the left we show excluded points in red and allowed points in blue, moreover we indicate points
with long-lived sparticles that cannot be tested in \smodels~in green and points not tested for the other reasons mentioned before in grey.
The right panel indicates the topology giving the strongest constraint for each excluded point.

We find that gluino topologies, basically from gluino decaying into a pair of quarks and the LSP through  virtual squark exchange,  can exclude gluino masses up to $1100$~GeV
~\cite{TheATLAScollaboration:2013tha,Aad:2014wea,CMS:2014ksa}. The exclusions differ from those of a simplified model, since in general there
are many possible decay channels.
The decay branching ratios of the gluino depend strongly on the nature of the LSP. 
For a higgsino LSP,  the decay of the gluino via virtual stop is dominant because of the stronger
coupling which depends on the top mass, the final state is  $t\bar{t}\tilde\chi^0$ (when there is enough phase space)  and/or $t\bar{b}\tilde\chi^-$
where the chargino   is treated as an effective LSP.
The strongest constraints are found when phase space allows only the decay into the chargino final state, as there is one dominant
decay channel.
In other scenarios (non-higgsino LSP) there is no such strong preference for
one decay channel, and the signal cross section will be split up on several
simplified model topologies. Moreover mixed decays, where each gluino decays into different quark pairs and the LSP occur frequently and are not constrained by SMS. 
Hence the exclusion will be considerably weaker than for the pure simplified model exclusion.
For many configurations gluinos can decay to heavier gauginos
yielding topologies with long cascades not yet included in \smodelsnn.
Moreover each different topology resulting from such processes is typically suppressed because of multiple branching fractions.
Similarly points with gluino decaying via an on-shell sbottoms are not yet included in \smodelsnn~while  those decaying via an on-shell stops can be tested by SMS.  However we found that the cross sections are too small by two orders of magnitude for these points to be excluded.

Points with very light gluinos (below $500$~GeV) may remain allowed even for
light LSP (less than $200$~GeV) if the branching ratio $\tilde{g}\rightarrow t\bar{b}\tilde\chi^-$ 
is dominant. This is because constraints in the region where $m_{\tilde g} \le 500$~GeV 
available  from ATLAS~\cite{ATLAS_T1tbtb}  (where the chargino is considered degenerate with the LSP) are very weak. This search was also considered in CMS~\cite{Khachatryan:2015pwa}
but the results are not incorporated in \smodels~as digitized data are not yet available. Furthermore results for this topology  when the chargino is not degenerate with the LSP 
 are only available for one specific mass ratio and therefore cannot be used.

We also found that most points with a light gluino and a dominantly singlino LSP feature a very compressed spectrum. This follows from the relic density constraint that favours coannihilation as mentioned in section~\ref{subsec:results_DM}, thus these points will be hard to constrain from SUSY searches for gluinos.

\subsubsection{Squark constraints}
\label{sec:smodels:squark}
\begin{figure}[!htb]
\begin{center}
\centering
\subfloat[]{\includegraphics[width=0.52\textwidth]{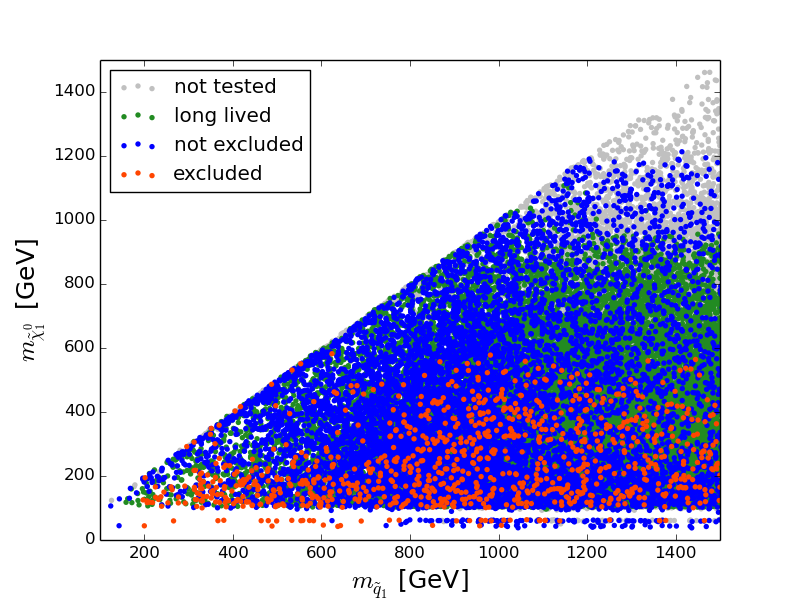}}
\subfloat[]{\includegraphics[width=0.52\textwidth]{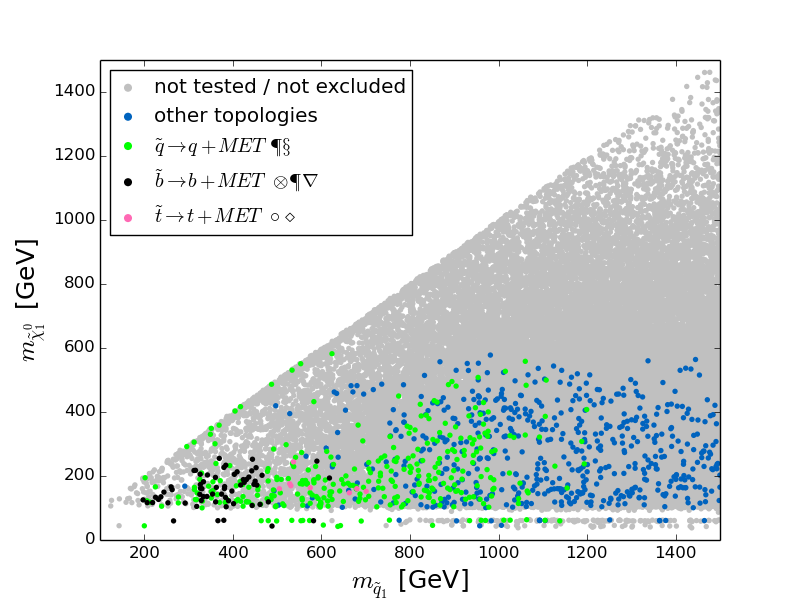}}
\caption{Exclusion with \smodels\ in the LSP - squark mass plane below 1.5 TeV, where we select the mass of the lightest squark (including stop, sbottom). (a) has same colour code as in figure~\ref{fig:gluinoExclusions-a}. (b) shows for all excluded points the most constraining topology. For the most frequently found topologies we show the associated experimental searches : $\S$ = \cite{Aad:2014wea}, $\circ$ = \cite{CMS:2014wsa}, $\P$ = \cite{CMS:2014ksa}, $\otimes$ = \cite{CMS:2014nia}, $\diamond$ = \cite{Chatrchyan:2013xna} and $\nabla$ = \cite{Aad:2013ija}.}
\label{fig:squarkExclusions}
\end{center}
\end{figure}

The model can naturally give light squarks, as was shown in section~\ref{sec:results}.
However we observe that these are poorly constrained by the SMS limits. We show the excluded vs. allowed points as well as the most constraining
topology for each point, here in the plane of the LSP and the lightest squark mass (including stop and sbottom), see figure~\ref{fig:squarkExclusions}.

A first observation is that 1$^\textrm{st}$ and 2$^\textrm{nd}$ generation squark topologies can
exclude points up to rather high squark masses (about $1200$~GeV) in excess
of the simplified models exclusions.
This is expected since a light gluino will enhance the squark production cross sections. Note that it was verified in~\cite{Edelhauser:2014ena} that SMS results can still be safely applied in this case. 
Points along the kinematic edge can in general only be excluded by one heavier
squark in the point, as such a compressed spectrum cannot be tested by the SMS results.

\begin{figure}[h]\centering
\includegraphics[width=0.57\textwidth]{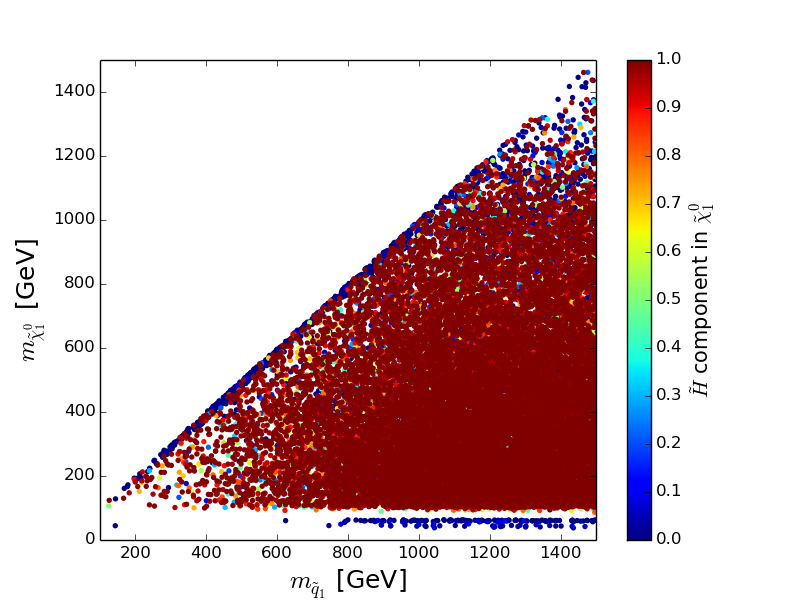}
\caption{Higgsino component of the LSP in the LSP - lightest squark mass plane below 1.5 TeV, for points that cannot be excluded by \smodels.}
\label{fig:squarkHiggsinos}
\end{figure}

We find however that many points with light squarks remain unconstrained,  even for large mass differences to the LSP.
The first reason for this is that the simplified model exclusions depend critically on the assumption that the 8 squarks of
the first and second generation are nearly degenerate.
This is not the case in the UMSSM, where because of the new $D$-term contributions the mass of the RH d-type squarks can differ significantly from the other squark masses.
Often their masses are not close enough to combine the production cross sections before comparing against an upper limit result \cite{Dreiner:2012gx}\footnote{{If several particles (such as squarks of different masses) contribute to the same topology, they will be combined if the corresponding masses are found to be compatible. The criterion is based on the difference in the experimental upper limits: if they differ less than 20\% (while the mass values may differ up to 100\%), the contributions are merged. For a more detailed explanation see~\cite{Kraml:2013mwa}. This is evaluated for each experimental result and may hence differ for different experimental analyses considering the same topology. Note that despite differences in the upper limits, the contributions of different mass configurations may still contribute to the same signal region. Using the appropriate efficiencies for each contribution might therefore improve the limits.}}.
We therefore find much weaker exclusions.
The second reason is again tied to  the nature of the LSP.
Recall that most points, and in particular the unexcluded ones,  feature a higgsino LSP,  as shown in
figure~\ref{fig:squarkHiggsinos},  and that important signatures of light squarks with a higgsino LSP  are not covered by existing SMS results. 

To identify  the main signatures for squarks that are not covered by SMS results, we discuss next the dominant  missing topologies, separately for first/second generation  and third generation squarks.
 A simplified model topology for which no matching experimental interpretation exists is labeled as ``missing topology".

\begin{figure}[!htb]
\begin{center}
\centering
\subfloat[]{\label{fig:diagrams-a}\includegraphics[width=0.25\textwidth]{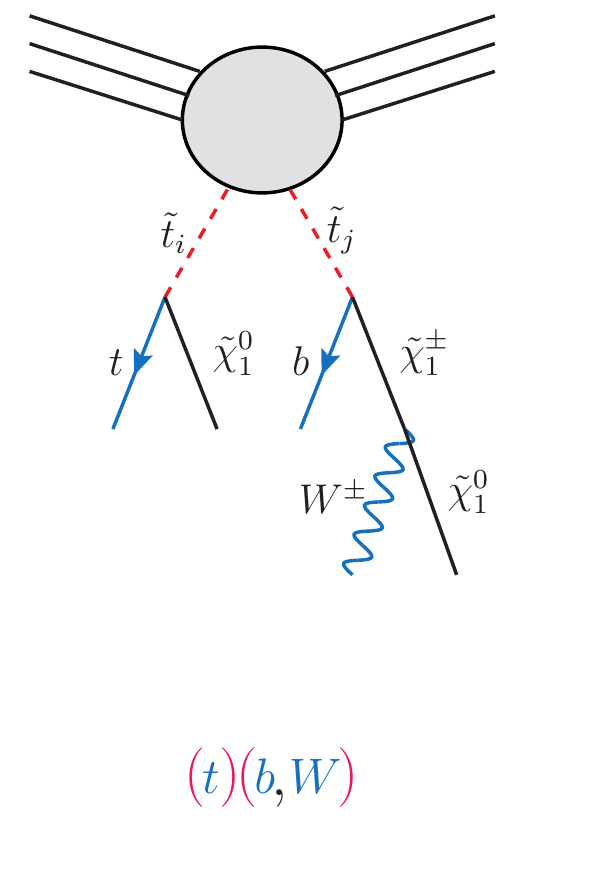}}
\subfloat[]{\label{fig:diagrams-b}\includegraphics[width=0.25\textwidth]{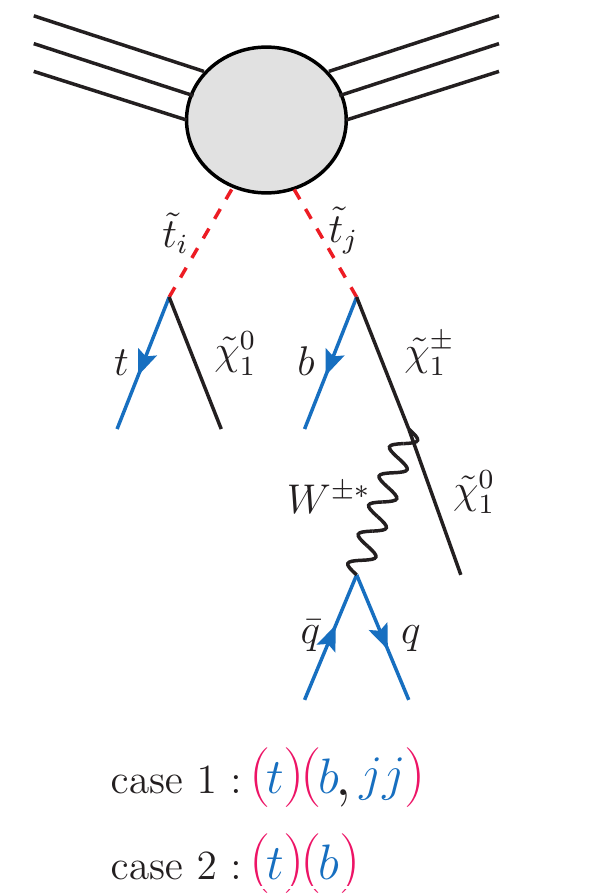}}
\caption{Sample diagrams illustrating missing topologies and their notation in the case of  stop pair production followed by an asymmetric decay to a neutralino LSP. Pair production, determining the branch structure is shown in red, R-even final state particles are indicated in blue.}
\label{fig:diagrams}
\end{center}
\end{figure}

The notation used for missing topologies keeps track of the  branch and vertex structure.  One branch is contained inside brackets, vertices are separated by a comma.  Only outgoing R-even particles in a given vertex are specified, light quarks and gluons appear as jets (denoted by ``j'') while third generation quarks are denoted by their name. MET from an outgoing DM candidate is always implied, and if no visible R-even particles appear in a branch it is denoted as ``(inv)''.
An example is stop pair production, with $\tilde{t}_i \rightarrow t \tilde{\chi}^0_1$ in one branch and $\tilde{t}_j \rightarrow b \tilde{\chi}^{\pm}_1$, $\tilde{\chi}^{\pm}_1\rightarrow W^{\pm(\ast)} \tilde{\chi}^0_1$ in the other ($i,j \in \{1,2\}$), illustrated in figure~\ref{fig:diagrams}.
This topology is denoted as ``$(t)(b, W)$" if the $W$ is on-shell (figure~\ref{fig:diagrams-a}).
In scenarios with an off-shell $W$ (as shown in figure~\ref{fig:diagrams-b}) only the decay products will be listed, e.g. ``$(t)(b, jj)$" for hadronic $W$ decay (case 1).
Finally if the mass gap between the chargino and the neutralino is smaller than the limit chosen for mass compression, the chargino decay is considered invisible, the topology is then listed as ``$(t)(b)$" (case 2).
\footnote{This notation directly translates to the \smodelsnn~nested bracket notation, where nested square brackets indicate the branches and vertices. The given example ``$(t)(b, W)$" is then written as \texttt{[[[t]],[[b],[W]]]}.}

\begin{figure}[!htb]
\begin{center}
\centering
\subfloat[]{\label{fig:squarkMissing-a}\includegraphics[width=0.52\textwidth]{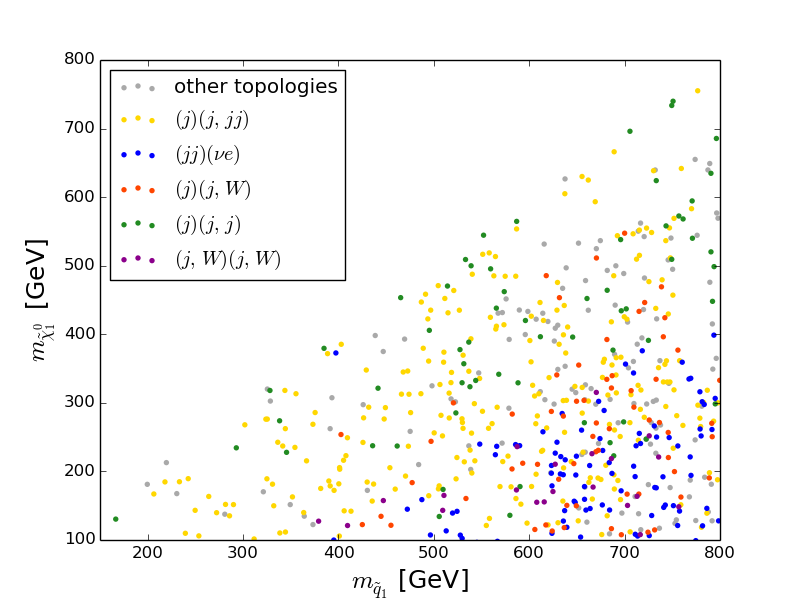}}
\subfloat[]{\label{fig:squarkMissing-b}\includegraphics[width=0.52\textwidth]{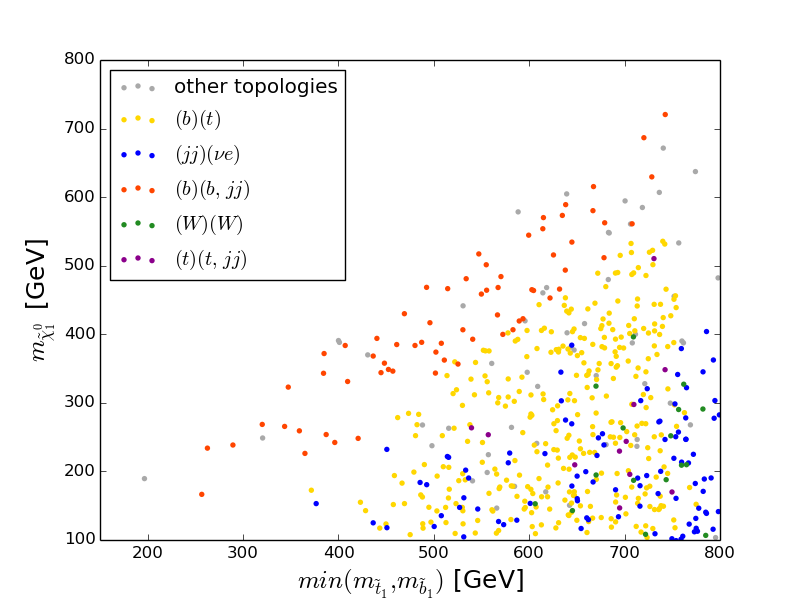}}
\caption{Missing topologies with highest cross section, for higgsino LSP, shown in (a) in the $m_{\tilde\chi_1^0} - m_{\tilde{q}_1}$ plane where $q_1$ is the lightest squark of the first and second generations.
In (b) the mass of the lightest third generation squark is used. For both plots only masses below 800 GeV are displayed. Here we have not considered direct higgsino production which cannot be tested by 8 TeV LHC results.}
\label{fig:squarkMissing}
\end{center}
\end{figure}

In figure~\ref{fig:squarkMissing-a} we show the missing topology  with the highest cross section for points labeled as not tested or not excluded and with light first/second generation squarks. Here we select only points where the higgsino fraction in the LSP is greater than 80\%. Moreover, to concentrate on topologies derived from squark production,  we have removed any topology where one branch is fully invisible, thus getting rid of direct higgsino production. Indeed  in  chargino-neutralino production, the neutralino LSP leads to an invisible branch, 
moreover a chargino can also  lead to an invisible branch when it is nearly degenerate with the neutralino since the soft jets that result from its decay cannot be detected. We further remove points in which the dominant missing topology has a weight
smaller than $1$ fb.
We find that a main missing topology consists of 4 jets + MET 
deriving from one squark decaying to $q\tilde\chi_1^0$ and the other to $q\tilde\chi_2^0$ with the neutralino further decaying 
 to the LSP via off-shell $Z_1$, giving the additional jets (mostly soft jets). 
A re-interpretation of the multijet analyses for this simplified model could be useful in constraining these points. Note that this topology is common in the compressed region where the squark - LSP mass difference is small.
We also find 3 jets + MET topologies, stemming from squark-gluino
production as described above.
These are found mainly when  both gluinos and squarks are light and the gluinos decay into a squark and a quark. Results for squark-gluino production within SMS exist only for almost mass degenerate gluinos and squarks. Note that for such points gluino pair production remains unconstrained as the gluino preferably decays to on-shell squarks, for which there are no  SMS results.
Similarly in scenarios where the gluino is lighter than the squark, squark pair
production remains unconstrained as they decay dominantly via gluinos. 

In case of larger squark - LSP mass splittings, we often find gauginos  with a mass between those of the squark and the higgsinos. In this configuration, the squark can decay either to the LSP or to a heavier gaugino, that then decays into the LSP and a gauge boson or a Higgs (real or virtual). In particular we find that an important missing topology is the one where each pair-produced squarks decays to a different channel, ``$(j)(j, W)$", but ``$(j, W)(j, W)$" is dominant in a few points. 

The limits on  the third generation squarks  are also much weaker than in the
simplified model. The reason is similar to the one invoked for gluino limits : with the higgsino LSP,  a stop may decay either to $t\tilde\chi^0$ or to $b\tilde\chi^+$. Therefore, only a fraction of the total cross section can be constrained by the
simplified model upper limit. Furthermore the ``mixed" decays, where one of the pair produced stop decays via top and the other
via bottom cannot be constrained as there are currently no SMS result for this channel.
This shows up as an important
missing topology, ``$(b)(t)$", in figure~\ref{fig:squarkMissing-b}.
The situation  will be improved when efficiency maps will be incorporated into \smodelsnn~\cite{smodels_efficiency}.
When the mass splitting between the stop and the LSP is below the top mass, the main missing topology is rather associated with sbottom  pair production  with one sbottom decaying to $b\tilde\chi_1^0$ and the other to  $b\tilde\chi_2^0$, followed by $\tilde\chi_2^0\rightarrow q\bar{q} \tilde\chi_1^0$ via an off shell $Z_1$ leading to ``$(b)(b, jj)$". 
Similarly ``$(t)(t, jj)$" appears at large mass splittings.
An important missing topology is the one  associated with chargino pair production with charginos decaying to the LSP and jets or leptons via a virtual $W$, ``$(jj)(\nu e)$".
We further find a few points where direct production of heavy charginos, decaying via $W$, gives the dominant missing topology ``$(W)(W)$".

Note that listing missing topologies with the largest cross section can sometimes be misleading as the background was not taken into consideration. 
 It is certainly possible that a signature with a smaller cross section gives a better signal to background ratio. Examples are leptonic vs. hadronic decays of the $W$, 
 or decays into b-quark  as compared to decay into  light jets.

\subsection{Neutralino LSP : $\amu$ and slepton constraints}
\label{subsec:slept_gmu}

\begin{figure}[!htb]
\begin{center}
\centering
\subfloat[]{\includegraphics[width=0.52\textwidth]{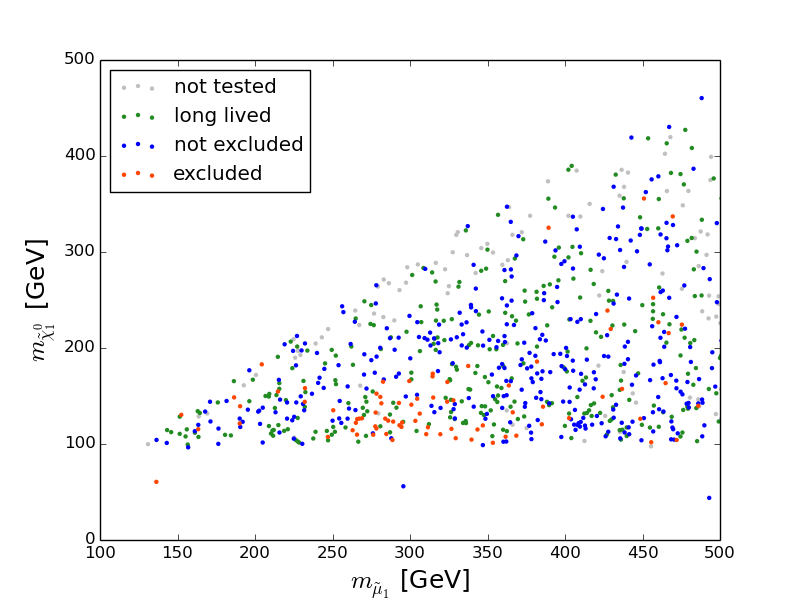}}
\subfloat[]{\includegraphics[width=0.52\textwidth]{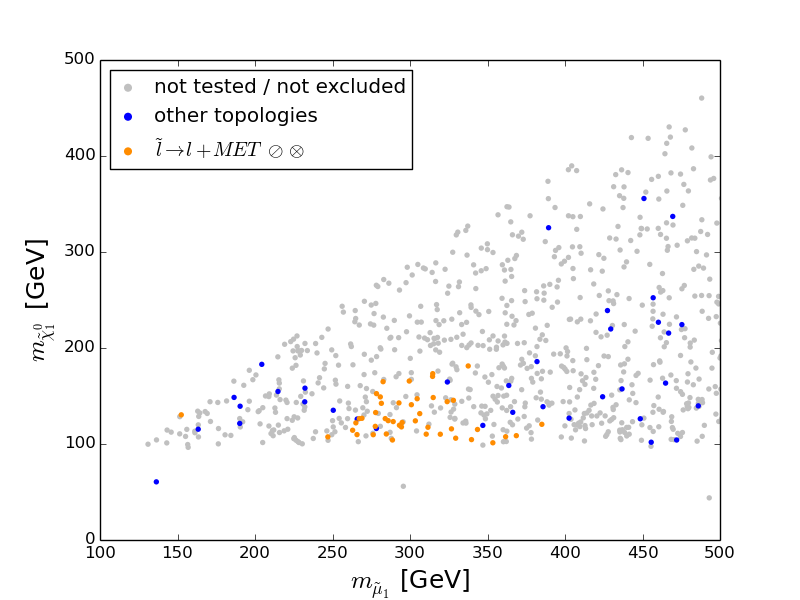}}
\caption{Exclusion with \smodels\ in the  $m_{\tilde\chi_1^0} - m_{\tilde\mu_1}$ plane below 500 GeV. (a) has same colour code as in figure~\ref{fig:gluinoExclusions-a}. (b) is showing for all excluded points the most constraining topology. For the most frequently found topology we show the associated experimental searches : $\oslash$ = \cite{Aad:2014vma} and $\otimes$ = \cite{Khachatryan:2014qwa}.}
\label{fig:smuonExclusions}
\end{center}
\end{figure}

We have separately studied points where the muon anomalous magnetic moment constraint is fulfilled.
Because of the light smuons (and selectrons) additional LHC constraints from slepton SMS topologies become relevant. These constraints played a marginal role in the general scan considering the small fraction of points with light sleptons.
We show the exclusions in figure~\ref{fig:smuonExclusions}.
Excluded points are found mainly in a small region in the mass plane, for light smuon masses $m_{\tilde\mu_1}$ between $250$ and $380$~GeV.
These exclusions are slightly weaker than the ones obtained in ATLAS and CMS \cite{Aad:2014vma,Khachatryan:2014qwa} which assume all sleptons decay into $l\tilde\chi_1^0$ while here sleptons can also decay into $\nu^l_L\tilde{\chi}_1^+$.
Moreover for weakly interacting particles we only compute the production cross section at leading-order while SMS include NLO cross sections.
A single point is excluded by the slepton SMS result despite a very small
mass difference between smuon and LSP.
However, in this case it is not actually the slepton production that is being
constrained, but pair produced charginos, each of them decaying to a left handed sneutrino which then decays invisibly to the neutralino.
The signature is identical to that of slepton pair production, giving a
2 lepton and missing energy final state and was discussed in \cite{Arina:2015uea}.
Other exclusions come from gluino and squark topologies, as described in
the previous section.

\subsection{RH sneutrino LSP}

In the case of a sneutrino LSP we have to bear in mind that since these sneutrinos are RH all decays of heavier sparticles must proceed via a neutralino.
When the neutralino is the NLSP it decays invisibly into $\nu_R \tilde \nu^*_R$ or $\bar{\nu}_R \tilde \nu_R$, therefore signatures resemble those associated with a neutralino LSP.
When decays through an on-shell neutralino are not allowed, we  effectively find additional
neutrinos in the decay vertex to sneutrino, 
for  example in the squark decay $\tilde{q}\rightarrow q\nu_R\tilde\nu_R$.
The signature is essentially the same as for a  squark decaying into a neutralino LSP since the neutrino will contribute only to the MET, but  the event kinematics can be changed due to the additional invisible particle in the vertex.  This issue remains to be  investigated  and these signatures are not treated in \smodels.

\begin{figure}[!htb]
\begin{center}
\centering
\subfloat[]{\includegraphics[width=0.52\textwidth]{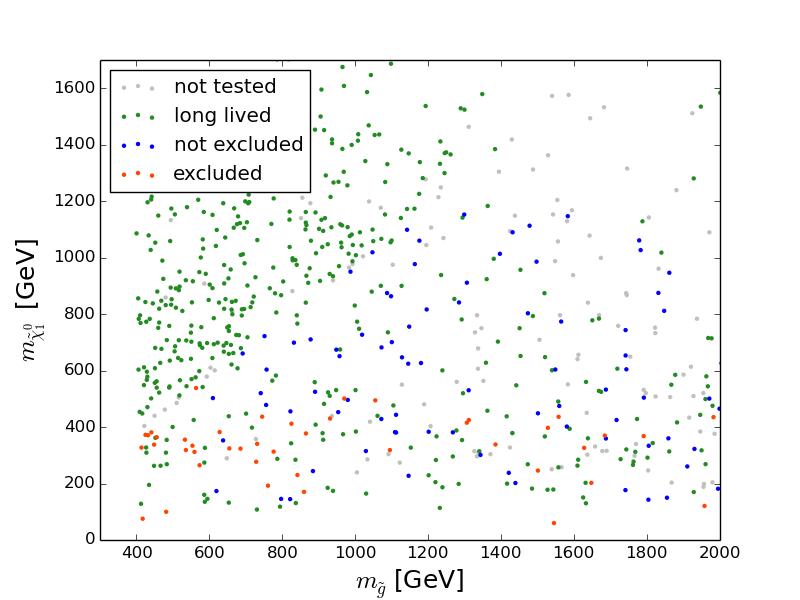}}
\subfloat[]{\includegraphics[width=0.52\textwidth]{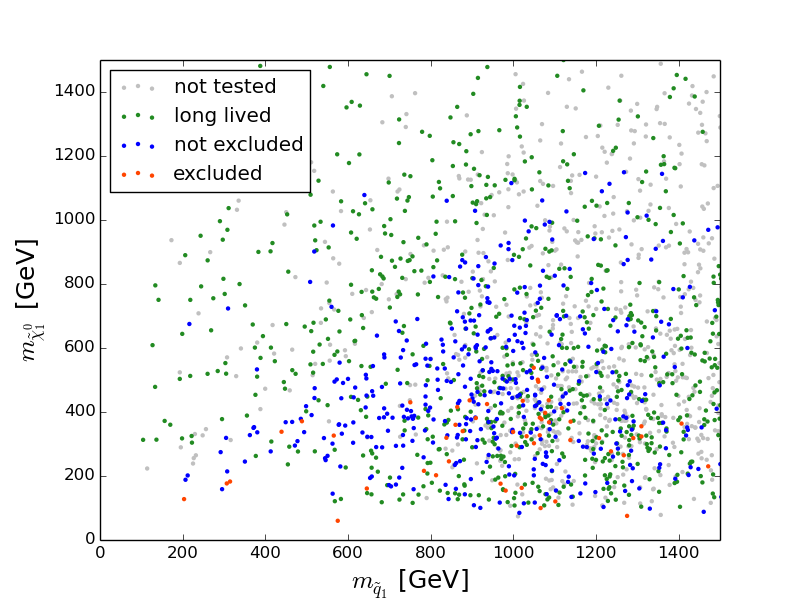}}
\caption{Exclusion with \smodels\ for points with RH sneutrino LSP (a) in the neutralino - gluino mass plane  and (b) in the neutralino - lightest squark mass plane. Points with very heavy neutralinos and squarks are not displayed.}
\label{fig:snuLSPstrong}
\end{center}
\end{figure}

\begin{figure}[h!]\centering
\includegraphics[width=0.25\textwidth]{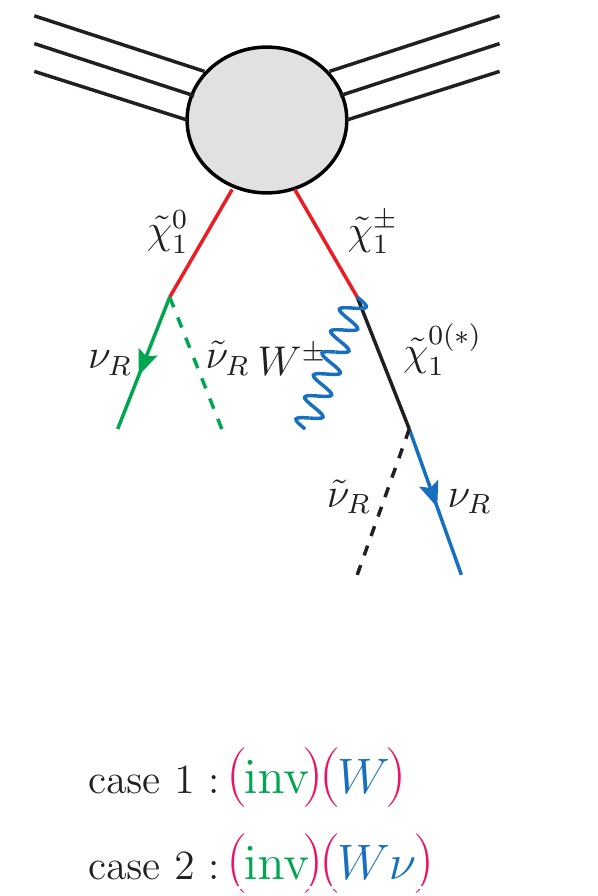}
\caption{Diagram illustrating missing topologies and their notation in the case of chargino-neutralino  production where the sneutrino is the LSP. Pair production, determining the branch structure is shown in red, R-even final state particles  are indicated in blue and the invisible branch is represented in green.
\label{fig:diagrams_sneutrino} }
\end{figure}

Figure~\ref{fig:snuLSPstrong} is showing points with a RH sneutrino LSP in the $\tilde \chi^0_1 - \tilde g$ and $\tilde \chi^0_1 - \tilde q_1$ mass planes.
One striking feature is that in this scenario 
there are many points with long-lived gluinos or squarks. Those are mainly found  when the lightest neutralino is heavier than the gluino or squark since decays into RH sneutrino LSP can only proceed via some virtual sparticle and are hence suppressed. Among the points that can be tested, only a small number  can actually be excluded.
The exclusion channels are similar to the ones  for the neutralino LSP discussed in sections~\ref{sec:smodels:gluino} and \ref{sec:smodels:squark} and involve a decay of a gluino or squark through a neutralino which further decays into the LSP. 
We find no exclusion from electroweak production. It is therefore instructive to consider the missing topologies. To clarify again the notation used for missing topologies, we show in 
 figure~\ref{fig:diagrams_sneutrino} the case of chargino-neutralino production for a sneutrino LSP. The neutralino decays to a neutrino and a sneutrino, making this branch entirely invisible, hence indicated by ``(inv)''.
The signature of the chargino decay will depend on whether the intermediate neutralino is on-shell or not, indicated by cases 1 and 2.
If the neutralino is on-shell its decay will  be invisible and it can be considered as an effective LSP, yielding the topology ``$(\mathrm{inv})(W)$".
If on the other hand the decay to on-shell neutralino is not possible, the chargino will effectively decay directly as $\tilde{\chi}^{\pm}_1\rightarrow W^\pm \nu_R \tilde{\nu}_R$, the topology is then described as ``$(\mathrm{inv})(W\nu)$". Recall that in the \smodelsnn~nested bracket notation,  this topology is denoted by  \texttt{[[],[[W,nu]]]}.

\begin{figure}[h]\centering
\includegraphics[width=0.79\textwidth]{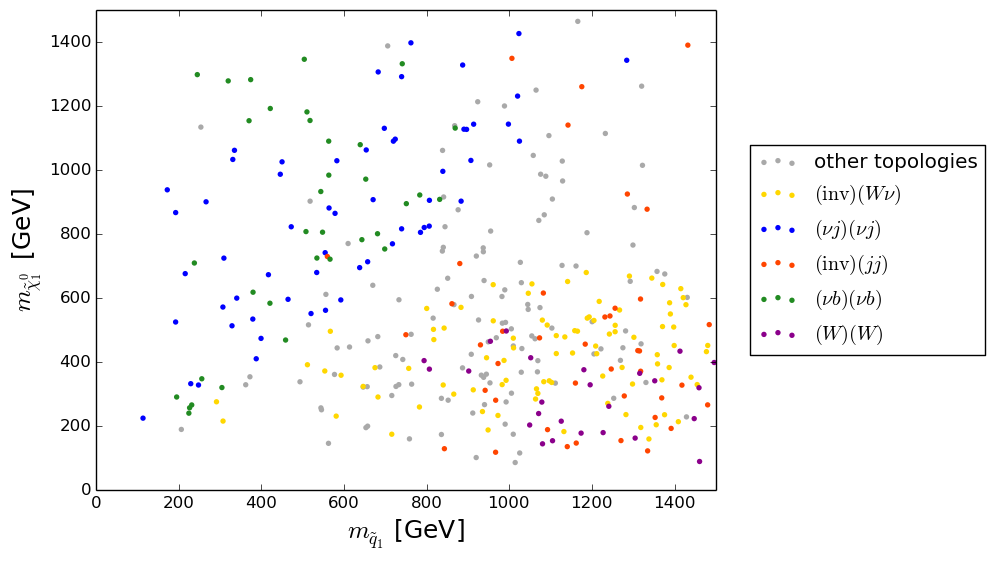}
\caption{Missing topologies with the highest cross section in the neutralino -  lightest squark mass plane, for points with a RH sneutrino LSP that are not excluded by \smodels~and that do not involve long-lived sparticles. }
\label{fig:snuLSPmissing}
\end{figure}

We show (for all not excluded or not tested points) the missing topology with the highest cross section, selecting only the five most frequent ones in each plane, see figure~\ref{fig:snuLSPmissing}.
At low neutralino masses topologies from neutralino-chargino production are often dominant, with
the charginos decaying  either directly to $W\tilde\nu_R\nu_R$, ``$(\mathrm{inv})(W\nu)$", or via $W^*\tilde\chi_1^0$, ``$(\mathrm{inv})(jj)$". In both scenarios the neutralino decay is invisible. Note that for the missing topologies we do not distinguish between LH or RH neutrinos. The SMS limits on chargino-neutralino production with $W^{(\ast)}$  final state cannot be applied for either topology. 
The reason is that SMS results assume that the process involves one of the heavier neutralinos which then decays via a gauge boson and the LSP, whereas here only the chargino decays into visible particles. Moreover, in the first case, there is an additional neutrino in the decay. 
In the second case the decay products of the $W^*$ are very soft because of the degeneracy between the lightest chargino and neutralino.
Pair produced charginos decaying  to $W\tilde\chi_1^0$ also provide an important topology, ``$(W)(W)$". Both the lightest and heaviest chargino can contribute to this topology. A similar topology with off-shell $W$'s also occurs although it is suppressed by the hadronic branching ratio. Note that current SMS results  for this topology only give weak constraints and are not yet included in \smodelsnn.

Finally topologies associated with squark pair production are also frequently dominant,
for light squarks and heavier neutralino we find squarks decaying directly to the right handed sneutrino, $\tilde{q}\rightarrow q\nu_R\tilde\nu_R$ with either a light quark or a b-quark, corresponding to the topologies
``$(\nu j)(\nu j)$" and ``$(\nu b)(\nu b)$" in figure~\ref{fig:snuLSPmissing}.
Missing topologies involving gluinos are similar to the ones in the neutralino LSP case, note however that when lighter than $\tilde\chi_1^0$  the gluino is likely to  be long-lived, or otherwise 
to decay via 4-body, $\tilde{g}\rightarrow jj\nu_R\tilde\nu_R$.

\subsection{Long-lived charged NLSP}
\label{sec:stable}

\begin{figure}[!htb]
\begin{center}
\centering
\subfloat[]{\label{fig:longlivedExclusions-a}\includegraphics[width=0.53\textwidth]{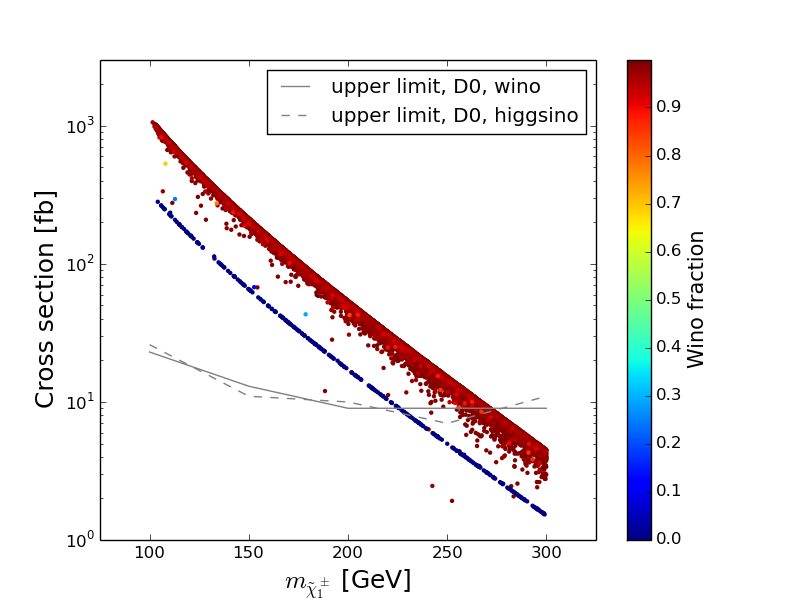}}
\subfloat[]{\label{fig:longlivedExclusions-b}\includegraphics[width=0.53\textwidth]{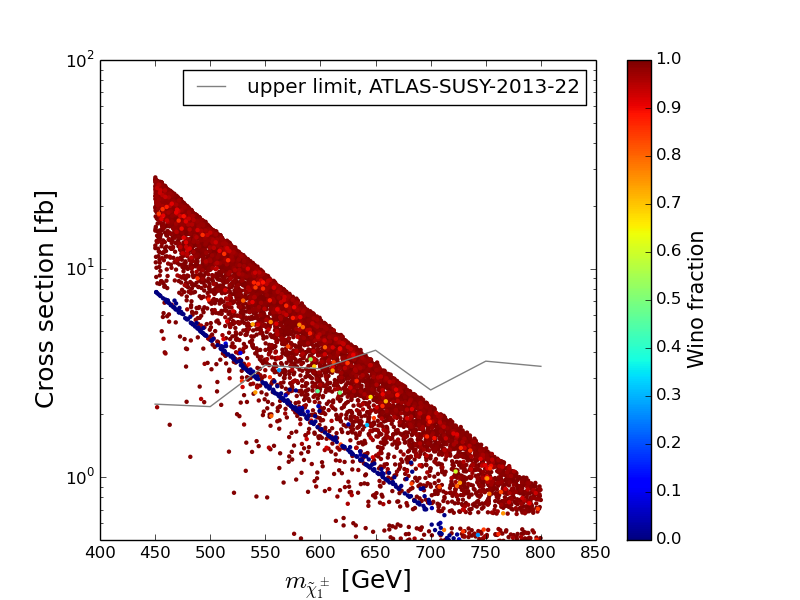}}
\caption{Points tested by searches for long-lived charginos:  (a) the chargino pair production cross section and the corresponding upper limits from D0 (b)  sum of chargino pair and chargino neutralino production cross sections and the corresponding upper limit from ATLAS. The colour code indicates the wino fraction in $\tilde\chi^\pm_1$.}
\label{fig:longlivedExclusions}
\end{center}
\end{figure}

The D0 collaboration has searched for pair produced long-lived charginos~\cite{Abazov:2012ina}, putting upper limits on the production cross section for
chargino masses between 100 and 300 GeV.
Since experimental limits are given separately for wino and higgsino-like chargino, we use the relevant result and in case of large mixing (\textit{i.e.} wino fraction in $\tilde\chi^\pm_1$
between 0.3 and 0.7) we apply the more conservative limit. Note that the limit is only marginally different in the two cases.  Results are shown in figure~\ref{fig:longlivedExclusions-a}.
We find that long-lived charginos lighter than about 230 GeV are excluded.

In addition, the ATLAS collaboration has searched for long-lived charginos from
either pair production of charginos or chargino-neutralino production~\cite{Chatrchyan:2013oca},
yielding upper limits on the combined cross section for chargino masses between 450 and 800 GeV.
We have checked that the less constrained chargino-neutralino
contribution is never larger than in the scenario considered by ATLAS, thus ensuring that the application of the upper limit is always conservative.
Note that in addition to chargino pair production we generally consider only $\tilde\chi_1^\pm \tilde\chi_1^0$ production, except when  this is essentially zero then we include also  $\tilde\chi_1^\pm \tilde\chi_2^0$ production. This may occur if the LSP is bino or singlino are degenerate in mass with the chargino\footnote{This degeneracy can follow from imposing the relic density upper limit 
which in this case will be satisfied because of the contribution from efficient  coannihilation channels involving the chargino and heavier neutralinos~\cite{Belanger:2005kh}.}. 
Results are shown in figure~\ref{fig:longlivedExclusions-b}.
We find that even at low masses some points cannot be excluded, because
interference between light squark exchange diagrams lead to small production cross sections.
However, a large number of points, with chargino masses up to about 650 GeV,
can be excluded. Note that in both cases we have used linear interpolation between the given data points.
We expect that smaller masses (below 450 GeV) should be excluded as well, but existing searches in that mass range  consider long-lived staus (ATLAS) or
long-lived leptons (neutral under $SU(2)_L$, see~\cite{Chatrchyan:2013oca}) and were not applicable here.

\begin{figure}[!htb]
\begin{center}
\centering
\subfloat[]{\includegraphics[width=0.53\textwidth]{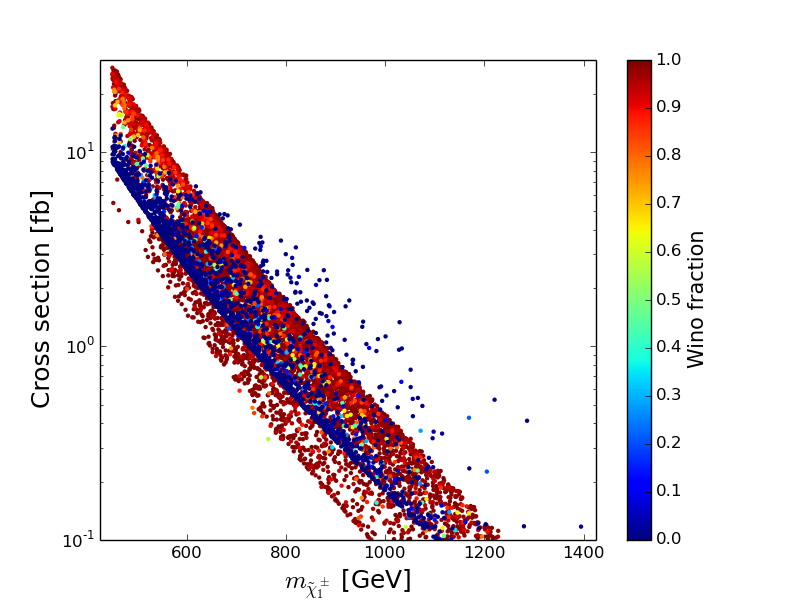}}
\subfloat[]{\includegraphics[width=0.53\textwidth]{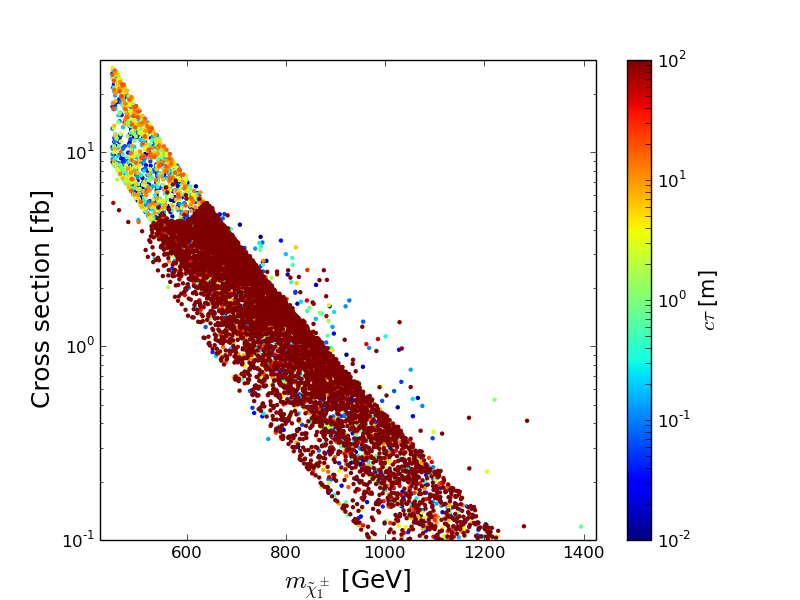}}
\caption{Chargino pair production cross section at 13 TeV, for unexcluded points with long-lived charginos where the colour code indicates either (a) the wino fraction in $\tilde\chi^\pm_1$ or (b) $c\tau<100$~m.}
\label{fig:longlived13TeV}
\end{center}
\end{figure}

Finally we point out the potential of such a search at 13 TeV.
In figure~\ref{fig:longlived13TeV},  the cross section for pair production of charginos with decay lengths  $c\tau > 10$~mm is displayed. Here all points that have not yet been excluded are shown.
We find that about one order of magnitude improvement over the current limit would allow to probe a large fraction of the points with a long-lived chargino below the TeV scale.
Note that in this figure we have included long-lived charginos decaying either inside or outside the detectors. Each category includes a significant number of points. Therefore both types of searches could be used  to test the model further.

\section{Summary after LHC constraints}
\label{sec:afterLHC}

\subsection{Exclusion potential of current LHC searches on the UMSSM}

\begin{figure}[!htb]
\begin{center}
\centering
\subfloat[]{\includegraphics[width=8.cm,height=5.5cm]{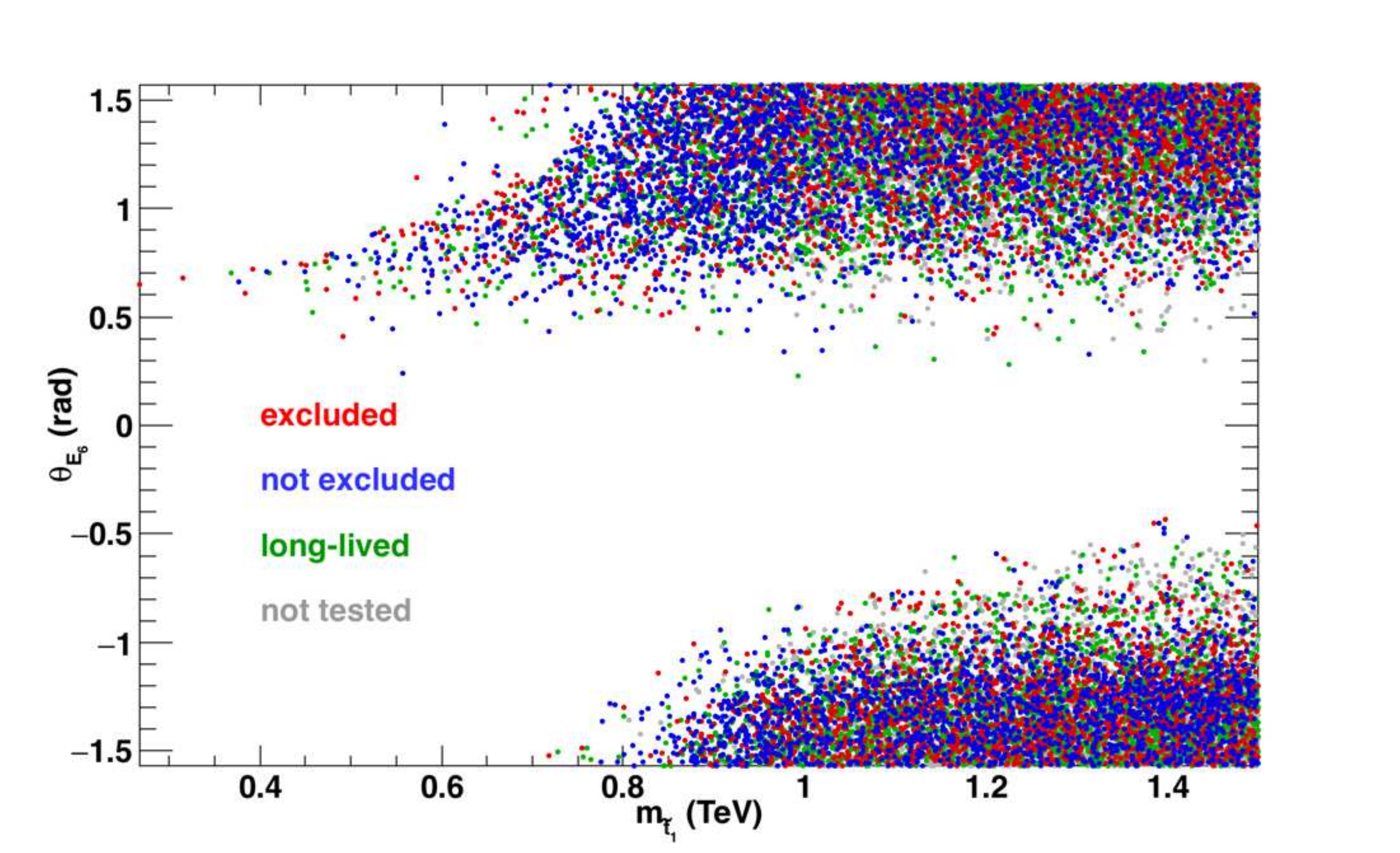}} 
\subfloat[]{\includegraphics[width=8.cm,height=5.5cm]{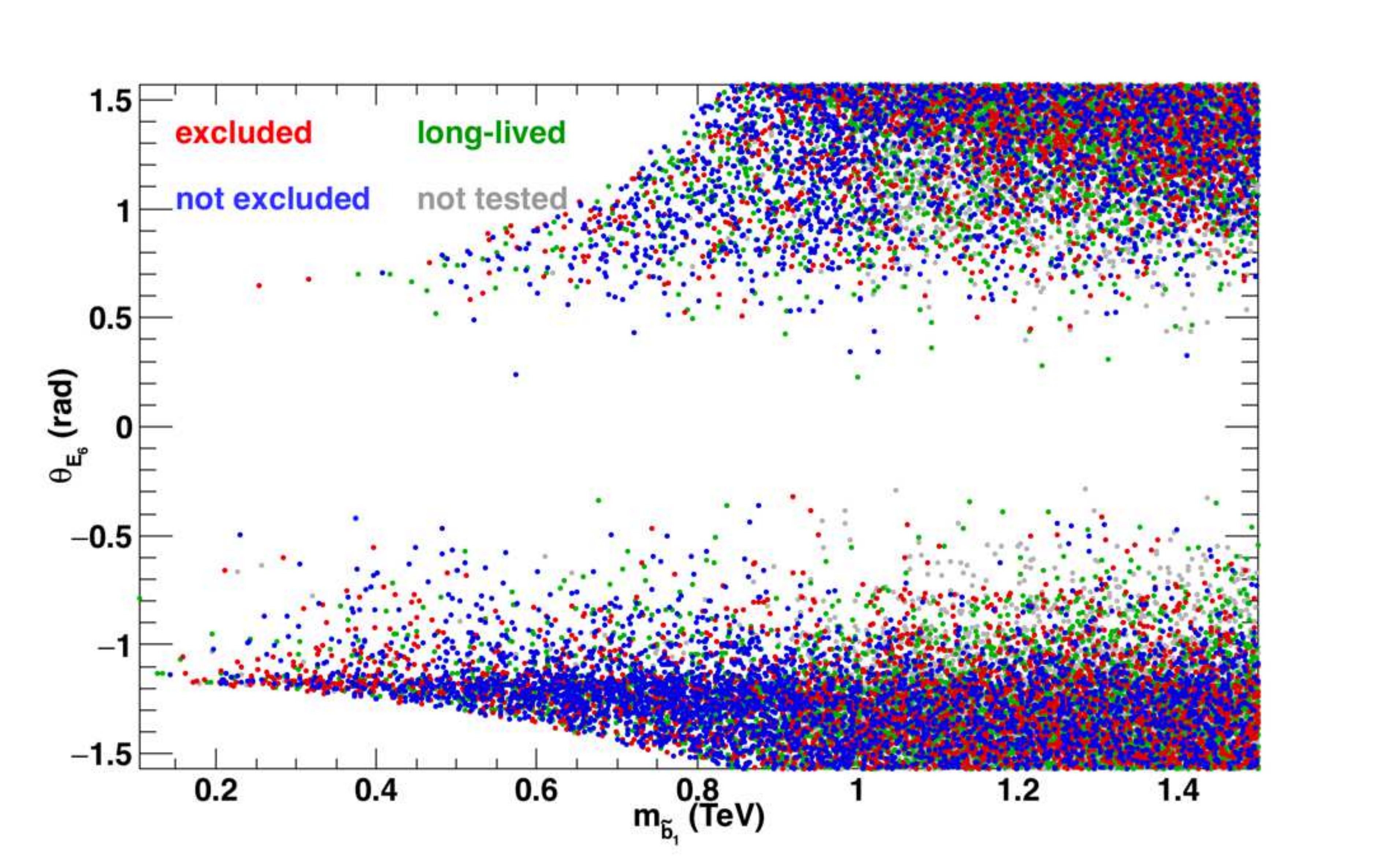}} \\
\subfloat[]{\includegraphics[width=8.cm,height=5.5cm]{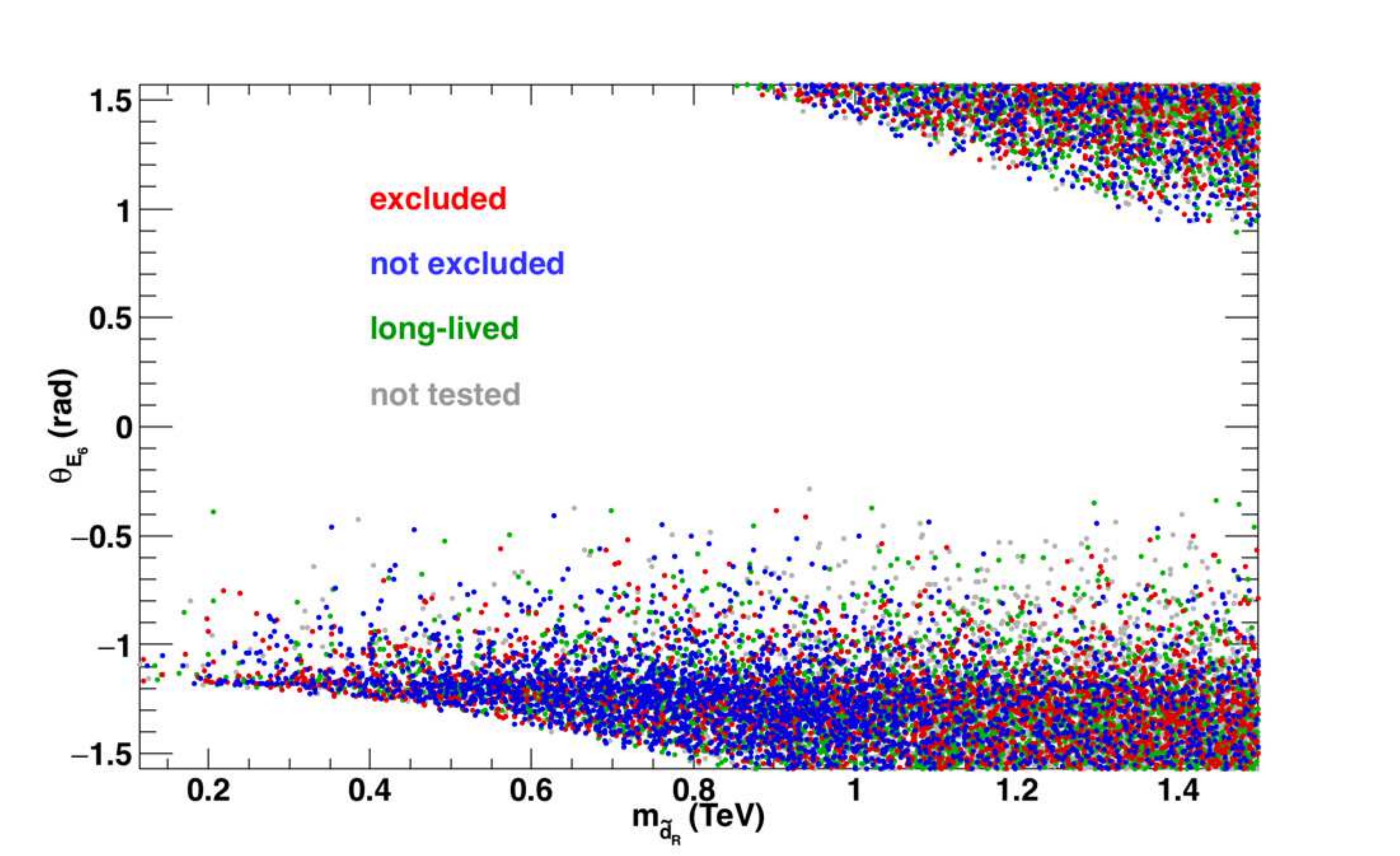}} 
\subfloat[]{\label{fig:sfermions_after_SModelS-d}\includegraphics[width=8.cm,height=5.5cm]{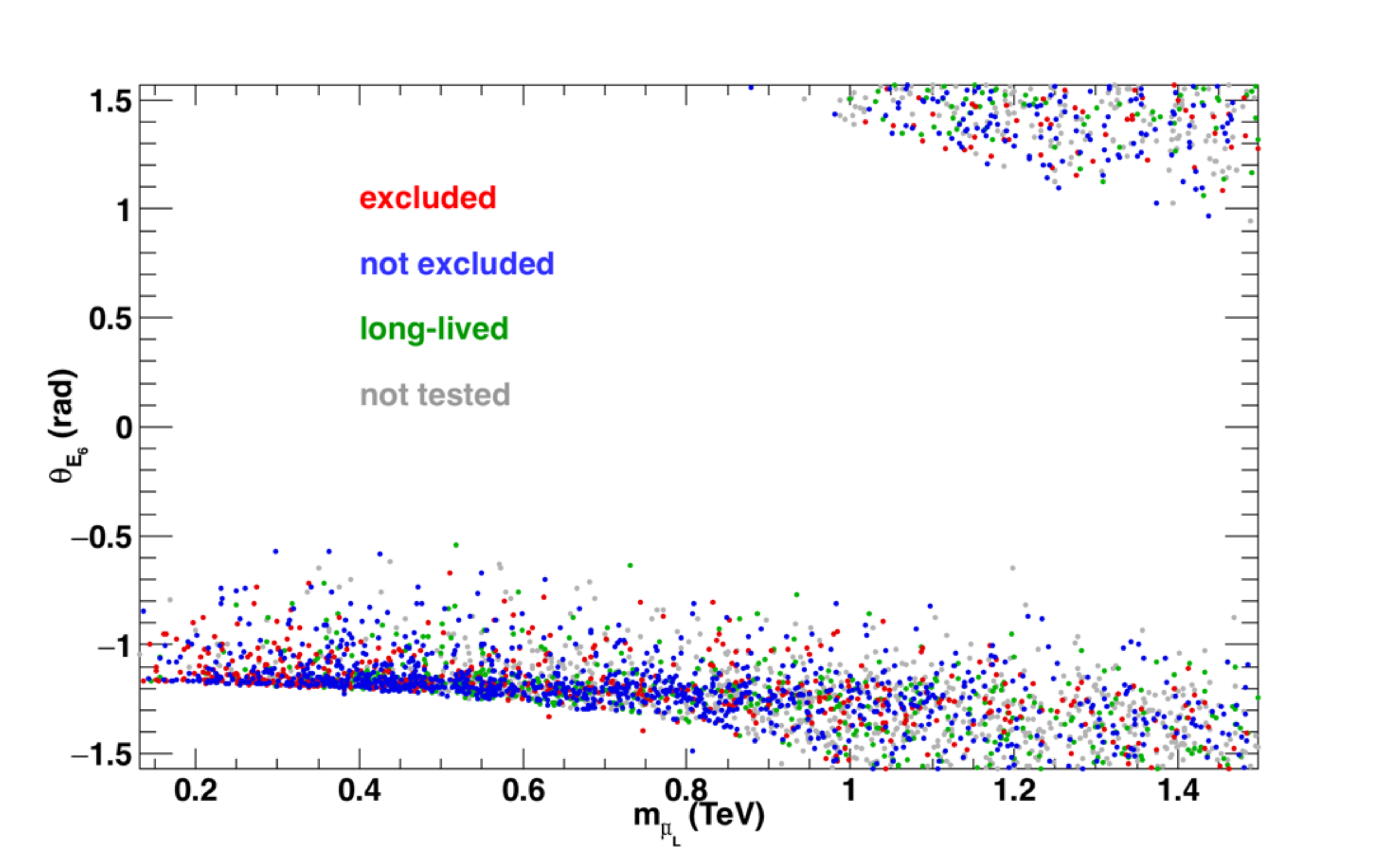}} 
\caption{Points in the $\te6 - m_{\tilde{f}}$ plane for  (a) $\tilde{f} = \tilde{t}_1$, (b) $\tilde{f} = \tilde{b}_1$, (c) $\tilde{f} = \tilde{d}_R$ and (d) $\tilde{f} = \tilde{\mu}_L$. For all plots $m_{\tilde{f}} < 1.5$~TeV and the colours correspond to not tested (grey), long-lived (green), not excluded (blue) and excluded (red) configurations using \smodels~and the searches for long-lived charginos. For (d) only points satisfying $\amu$ are represented.}
\label{fig:sfermions_after_SModelS}
\end{center}
\end{figure}

To summarize the impact of the LHC constraints on the sfermion spectrum we display in figure~\ref{fig:sfermions_after_SModelS} the excluded/non-excluded points in the plane $\te6 -m_{\tilde f}$ for $\tilde f \in \{\tilde{t}_1, \tilde{b}_1, \tilde{d}_R\}$ as well as $\tilde f = \tilde{\mu}_L$ for the sample where the muon anomalous magnetic moment constraint is imposed.
Among the non excluded points those that satisfy all constraints have a different colour code than those that are associated with a long-lived NLSP or that are not tested by \smodels. 
In all cases the excluded points  are scattered and represent only a fraction of all points. 
It should be stressed again that many scenarios with squark masses well below 1 TeV are allowed. 
When the agreement with $\amu$ is not required we found that 45\% (41\%) of the points that were confronted with the LHC limits had a long-lived sparticle in the case of a neutralino (RH sneutrino) LSP, 16\% (17\%) were tested by \smodelsnn~of which 10\% (11\%) were excluded. The remainder of the points was not testable by \smodelsnn~either because of too low cross sections or lack of SMS result. 
We additionally found that 42\% (24\%) of the sample with long-lived NLSP were excluded by long-lived chargino searches. 
In the case where the muon anomalous magnetic moment constraint is required and for a neutralino LSP the amount of points tested by \smodelsnn~is larger (34\%, out of which 11\% are excluded), whereas the fraction of points with long-lived sparticles is smaller (30\%, out of which 44\% can be excluded).
Extending these searches for long-lived charginos to the full mass range would therefore clearly provide a powerful probe of the model. Moreover, when the RH sneutrino is the LSP a large fraction of the points involves long-lived gluinos and squark. These scenarios could test the model further, but require reliable limits on R-hadrons. Note that to facilitate the interpretation of limits on long-lived charginos, it would be useful if limits on the direct chargino pair and neutralino-chargino production were separately provided by the experimentalists.

Many of the points  that are in agreement with the measured value of the  muon anomalous magnetic moment, even those associated with a very light smuon cannot be completely excluded, see figure~\ref{fig:sfermions_after_SModelS-d}. The LHC13TeV with higher luminosity will allow to extend the reach for smuons in the conventional lepton + MET channel. Unfortunately the light charginos that are present in this case cannot be probed easily as once again  they are often dominantly higgsino hence almost degenerate with the LSP.

\subsection{Suggestions for future LHC searches}

Conventional searches for a new $Z'$ provide the most distinctive signature of the UMSSM. In addition we have pointed out in previous sections many additional signatures  that are still unconstrained by current SMS results. Here we summarize the main missing topologies found for each scenario.

As expected, the most distinctive SUSY signatures in the UMSSM are found in the case of a RH sneutrino LSP. We have already stressed that long-lived gluinos or squarks are fairly common in such scenarios and could therefore provide further constraints on the model. We have also found that the following  topologies could be used to probe the model either by reinterpreting current LHC data or by exploiting Run II data. 

\begin{itemize}

\item mono-$W$, ``$(\textrm{inv})(W\nu)$", from chargino neutralino production with $\tilde\chi^0_1 \to \nu_R\tilde\nu_R$ and $\tilde{\chi}^{\pm}_1 \rightarrow W^\pm \nu_R \tilde\nu_R$.
The single $W$ can be  energetic enough to lead to visible decay products, leptons or jets as long as there is a large mass difference between the chargino and sneutrino LSP.  Such a topology occurs also for a neutralino LSP (as in the MSSM)  but only when there is a large  $\tilde{\chi}^{\pm}_1 -\tilde\chi_1^0$ mass splitting to allow $\tilde{\chi}^{\pm}_1 \rightarrow W^\pm \tilde\chi^0_1$, which is not the most common configuration after imposing DM constraints.

\item  dijets + MET, ``$(\nu j)(\nu j)$"  or $b\bar{b}$ + MET, ``$(\nu b)(\nu b)$",  from squark pair production with $\tilde q \to q\nu_R\tilde\nu_R$ where $q$ here stands for either light jets or b-jets. This occurs when the squark is lighter than all neutralinos and therefore has to decay directly to the sneutrino LSP. Such a configuration is clearly only possible in a model with a sneutrino LSP. 
The dijet + MET signature is of course common to squark pair production in the MSSM, however  it remains to be seen how the additional $\nu_R$ in the decay will affect the efficiencies, hence could lead to different exclusions than in the case of  the neutralino LSP.
\end{itemize}

Other important missing topologies include  dijets + MET, ``$(\textrm{inv})(j \; j)$",  from chargino neutralino production with the same decays as the mono-$W$ above  except that the $W$ is off-shell leading to soft final states as well as $W$ pairs + MET, ``$(W)(W)$", from chargino pair production with $\tilde{\chi}^{\pm}_1 \rightarrow W^\pm \tilde\chi^0_1\rightarrow W^\pm \nu_R \tilde\nu_R$.  These topologies also arise in the MSSM with a neutralino LSP and are poorly constrained from searches at Run I partly due to the small production cross section. The situation should however improve after accumulating more data  in Run II.

When the  neutralino is the LSP,   most SUSY signatures are the same as  found in the MSSM. However we stress that having imposed only the upper limit on the dark matter relic density,  most of our scenario have a wino/higgsino-like LSP. Thus, the SUSY signatures  can differ from the bino LSP assumed in several SMS results. In particular, the chargino decay can be invisible and this will have an impact on many SUSY searches. In addition this implies that a significant fraction of the scenarios have a long-lived chargino and/or neutralino, hence the importance of searches for stable charged particles at collider scale and for displaced vertices. 

Many of the topologies that could not be constrained by current SMS results   are associated with asymmetric decays, that is the pair produced particles have two different decay chains whereas most SMS results assume identical decays for both particles. We emphasize here the missing topologies for the case of the higgsino LSP since it is hard to probe.

\begin{itemize}

\item  3 jets + MET, ``$( j ) (j , j)$",  from gluino-squark production  with $\tilde{g}\to \tilde{q}q$ and $\tilde{q}\to q\tilde\chi_1^0$.
Current  SMS interpretations exist only for scenarios where  the gluino and squarks are almost mass degenerate, and both decay directly to jets and LSP.
Similarly the topology 4 jets + MET, ``$( j ,j) (j , j)$",  from gluino pair production  with $\tilde{g}\to \tilde{q}q$ and $\tilde{q}\to q\tilde\chi_1^0$ arises from a process
with a large production cross section that is not constrained by SMS results since  the  gluino decays via on-shell first or second generation  squarks.
Note that both these topologies are of special interest in the UMSSM where  the limits on light squarks from direct squark production  are much weaker because the squarks are not necessarily all degenerate.

\item $bt$ + MET,  ``$(b)(t)$" ,  from stop (sbottom) pair production with asymmetric decays, $\tilde{t}\to t\tilde\chi_1^0$ and  $\tilde{t}\to b\tilde\chi^+_1$  ($\tilde{b}\to b\tilde\chi_1^0$ and  $\tilde{b}\to t\tilde\chi^-_1$) when the chargino is nearly degenerate with the LSP. This signature is a generic feature of models with wino/higgsino LSP and  light third generation  squarks ~\cite{Belanger:2015vwa,CMS:2014wsa}.

\item 4 jets + MET, ``$(j) (j, j j)$",  from squark pair production with asymmetric decays. Here one squark decays directly to the LSP,   $\tilde q \rightarrow q \tilde{\chi}^0_1$ 
while the other decays  via heavier neutralino, $\tilde q \rightarrow q \tilde{\chi}^0_2$ and $\tilde{\chi}^0_2 \rightarrow Z_1^{\ast} \tilde{\chi}^0_1$.
A re-interpretation of the multi-jet analysis to study the effect of the soft jets from the virtual $Z_1$ on the efficiency would be useful.

\item 2$b$ + 2 jets + MET, ``$(b) (b, jj)$",  from sbottom  pair production with asymmetric decays, same as above. Note that this topology is found for small mass difference between the sbottom and the LSP.   Similarly the 2$t$ + 2 jets, ``$(t) (t, jj)$", from stop  pair production is also a missing topology.

\item 2 jets + $W$ + MET, ``$( j ) (j, W)$",  from squark pair production with asymmetric decays. One squark decays to LSP $\tilde{q}\to q\tilde\chi_1^0$ while the other decays 
$\tilde{q}_u\to q_d\tilde\chi_1^+$  with  $\tilde\chi_1^+\to W^+\tilde\chi_1^0$  or $\tilde{q}\to q\tilde\chi_2^0$  with  $\tilde\chi_2^0\to W^-\tilde\chi^+_1$ when the chargino decays invisibly.
We find this when the mass splitting between the squark and the  LSP is large. Note that typically there would be similar channels with $Z_1$ or $h_1$ in the final state instead of a $W$ reducing the cross section for each single channel. 

\item 2 jets + $WW$ + MET, ``$( j, W) (j, W)$", from squark pair production, here  both squarks can decay to chargino, followed by $\tilde\chi_1^\pm \to W^\pm\tilde\chi_1^0$ as above.
This channel has been considered at the LHC but is not yet included in the \smodelsnn~database, moreover results are available only for specific mass relations. It
would be preferable to provide SMS results that allow for interpolation over wide range of masses in different mass planes. 

\end{itemize}

In addition the signature $WW$ + MET  from chargino pair production  will be useful in constraining the model, although it currently gives only weak limits~\cite{Aad:2014vma}. 
Similarly the signature 2 jets + lepton + MET from charginos decaying via virtual $W$'s is often found, current data do not put useful constraints but these searches should be improved in the next Run.

\section{Couplings and signal strengths for the Higgses}
\label{sec:Higgs_coup_sig}

\begin{figure}[!htb]
\begin{center}
\centering
\subfloat[]{\label{fig:reduced-a}\includegraphics[width=8.cm,height=5.5cm]{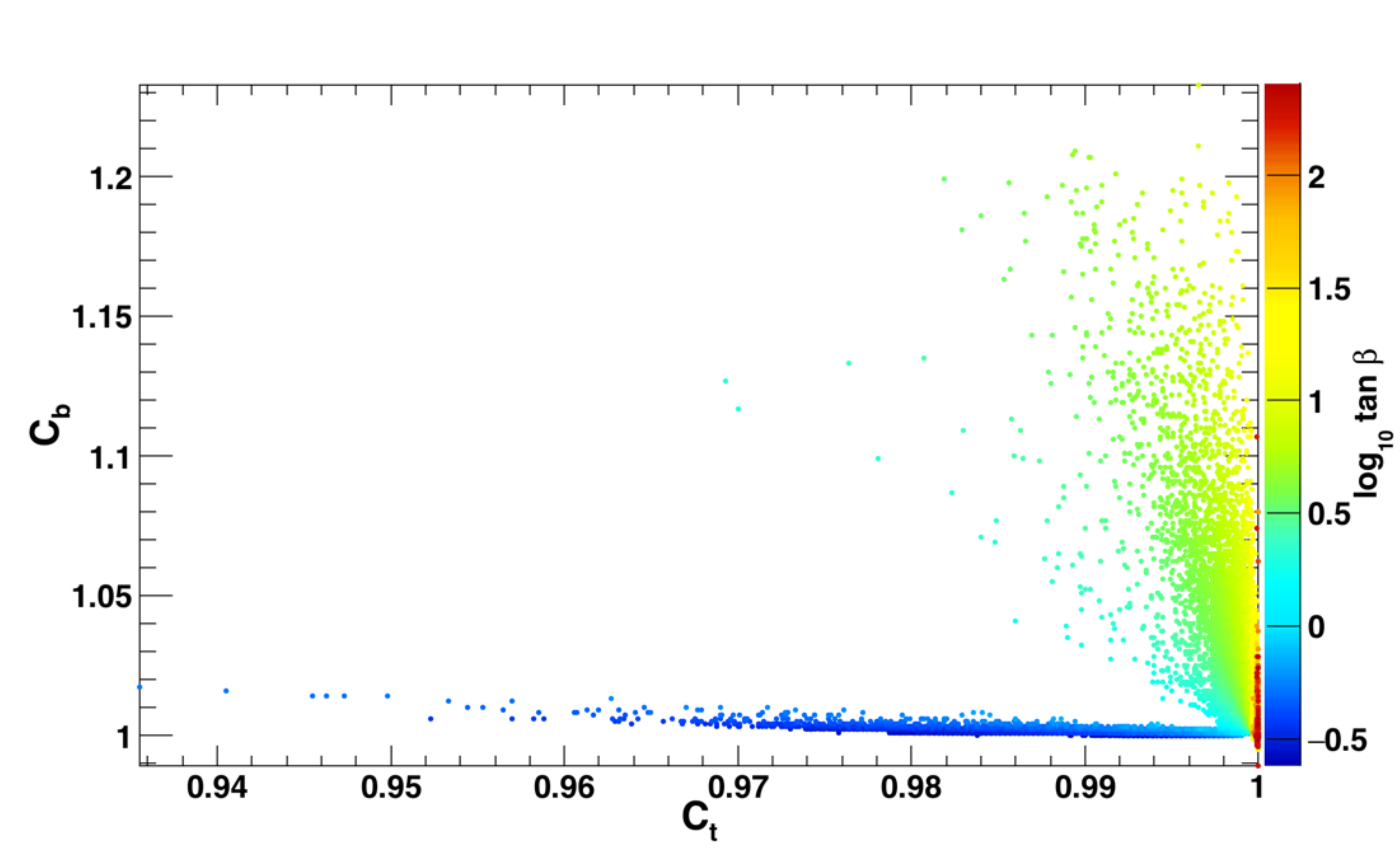}}
\subfloat[]{\label{fig:reduced-b}\includegraphics[width=8.cm,height=5.5cm]{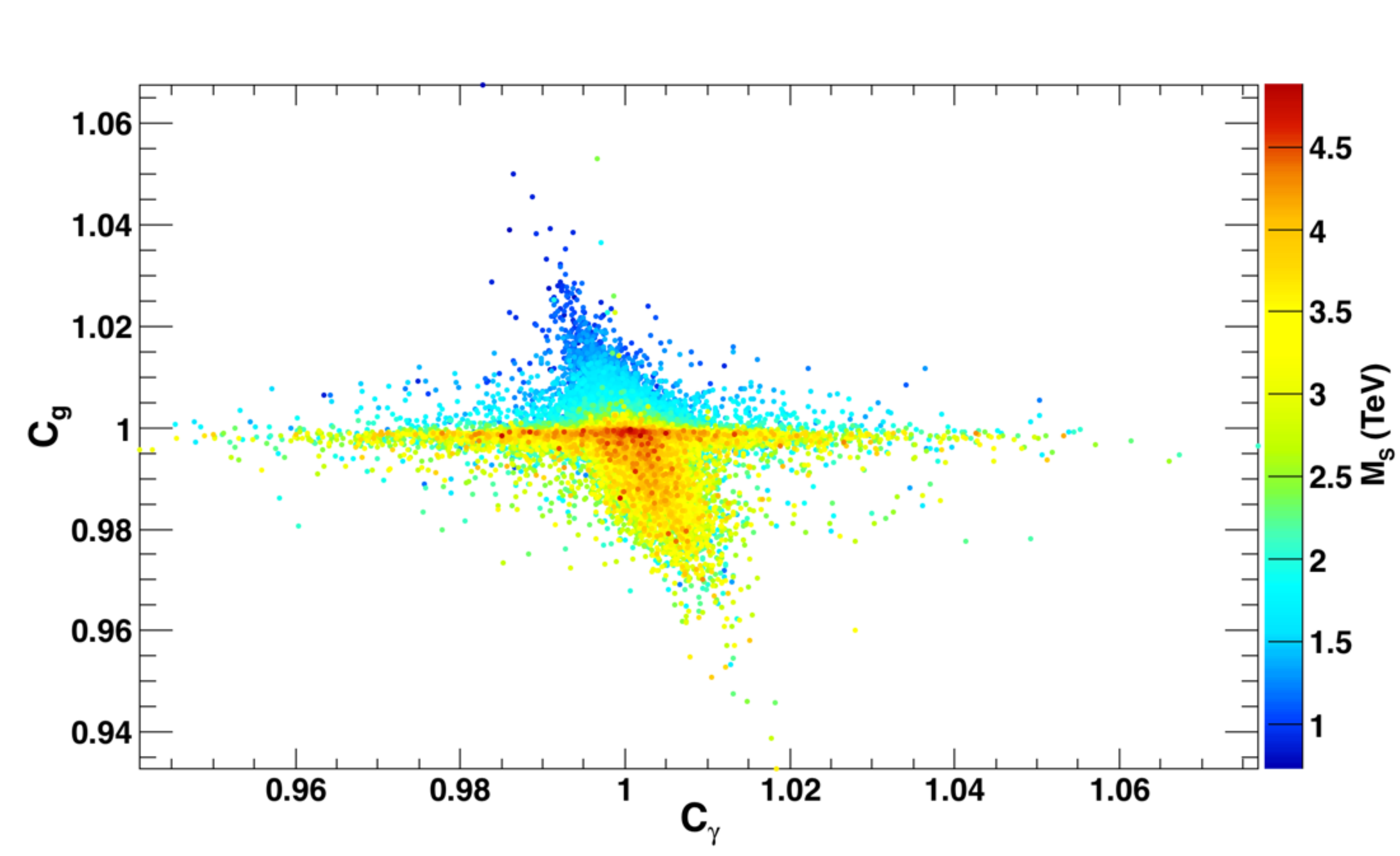}}
\caption{ Reduced couplings of the light Higgs for all points satisfying collider and DM relic density constraints : (a) $C_b$ vs. $C_t$  with $\tan\beta$ as colour code, (b) $C_g$ vs $C_\gamma$ with $M_S$ as colour code.}
\label{fig:reduced}
\end{center}
\end{figure}

 In the UMSSM a lightest Higgs scalar with a mass of 125 GeV is easily found. Typically this lightest scalar is doublet-like and behaves roughly as the SM Higgs. Measurements of the Higgs couplings at the LHC  Run II could therefore provide additional probes of the model. For all points of the UMSSM scan that successfully pass all collider constraints we have computed the  reduced couplings of the 125 GeV Higgs. The reduced couplings are defined as scaling factors of the couplings in the UMSSM relative to their SM counterparts. 
We find that  the $h_1W^+W^- (h_1Z_1Z_1)$ couplings deviate by at most 1\% from the SM couplings while there is more room for deviations in the quark couplings. 
The $h_1b\bar{b}$ reduced coupling ($C_b$) can be as large as 1.2 for large values of $\tb$ while the $h_1t\bar{t}$ coupling ($C_t$) can be suppressed by at most 5\% for low values of $\tb$, see figure~\ref{fig:reduced-a}. Note that the couplings are generation universal. Modifications of the quark couplings induce a correction to the loop-induced couplings of the Higgs to gluons ($C_g$) and photons ($C_\gamma$). In particular since the top quark gives the largest contribution to $C_g$ in the SM, we expect a reduction in $C_g$. Furthermore, this should be correlated with a mild increase in $C_\gamma$ as observed in figure~\ref{fig:reduced-b}. Supersymmetric particles can also contribute to the loop-induced coupling, for example light squarks can lead to $C_g>1$ although the effect is again below 5\%. Light sleptons and charginos will only contribute to $C_\gamma$. Again the effect typically does not exceed 5\%. 

\begin{figure}[!htb]
\begin{center}
\centering
\subfloat[]{\label{fig:signal_strength-a}\includegraphics[width=8.cm,height=5.5cm]{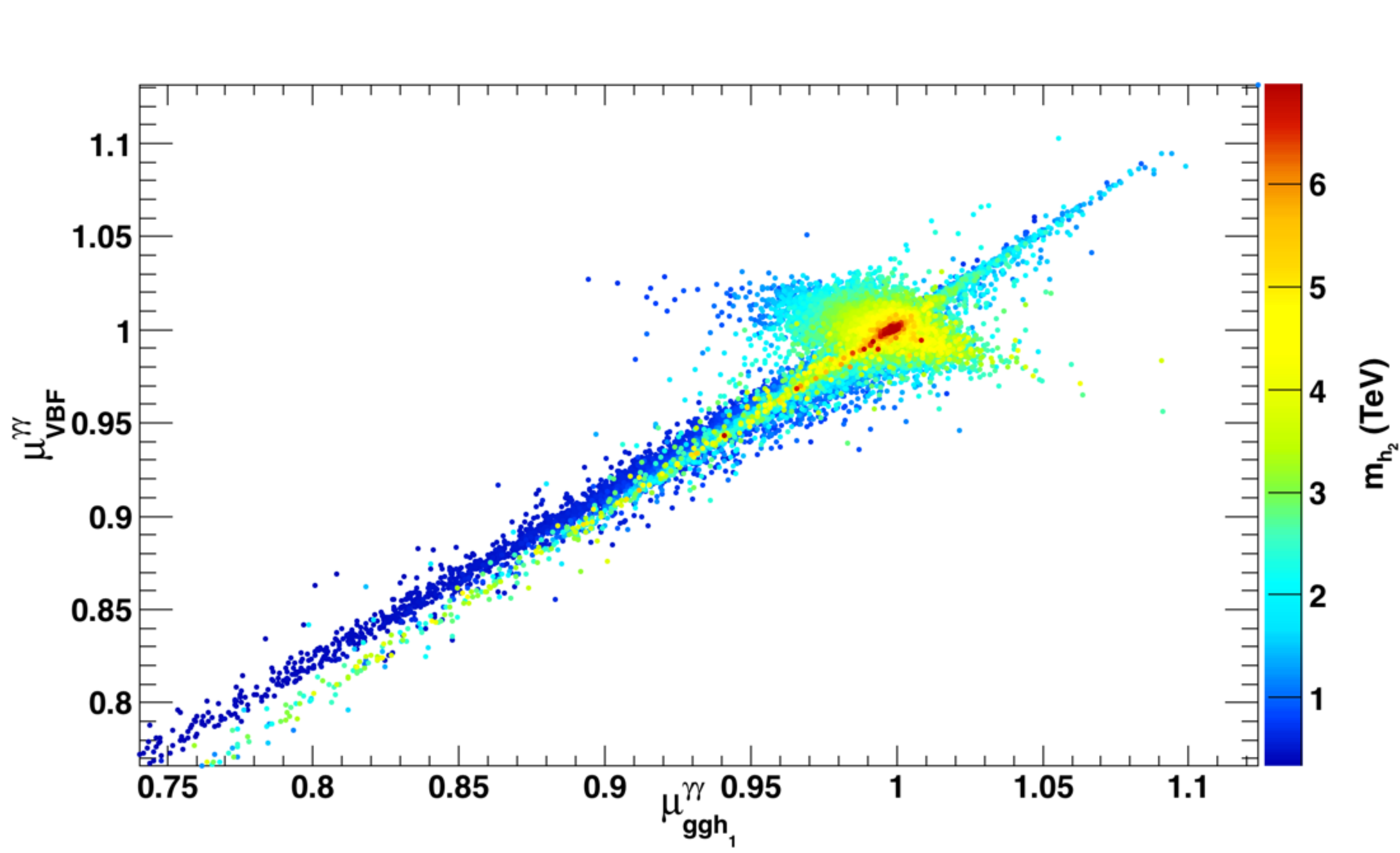}}
\subfloat[]{\label{fig:signal_strength-b}\includegraphics[width=8.cm,height=5.5cm]{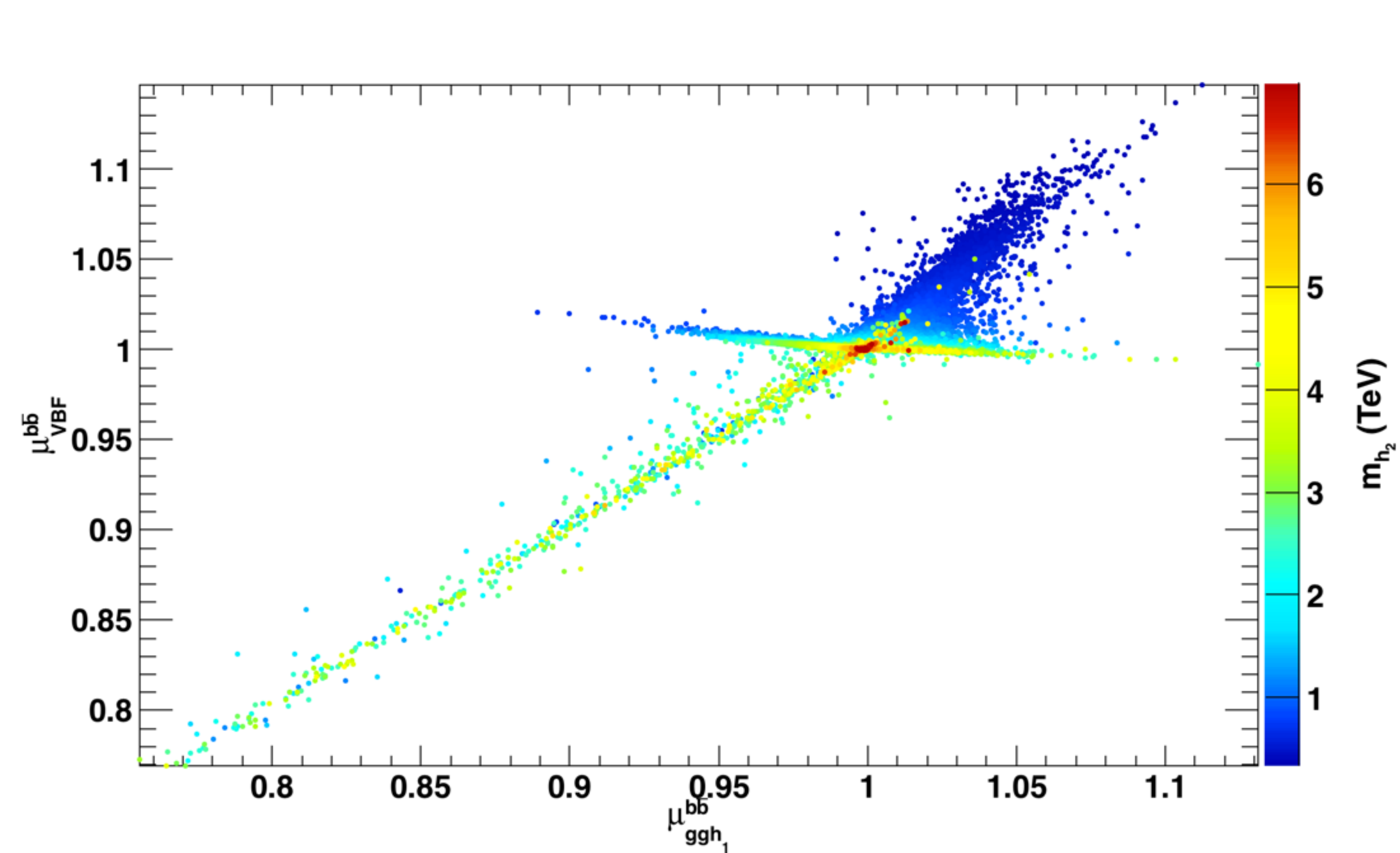}}
\caption{Signal strengths of the light Higgs into (a) $\gamma\gamma$  and (b) $b\bar{b}$  in the Vector boson fusion (VBF) vs. gluon fusion for all points satisfying collider and DM relic density constraints. For both plots $m_{h_2}$ is taken as colour code.}
\label{fig:signal_strength}
\end{center}
\end{figure}

The effect on the Higgs signal strength can be much larger than on the reduced coupling. The signal strength in one channel is defined as the production cross section times branching ratio in the UMSSM relative to the SM expectation for a Higgs of the same mass.
An increase in the total width, through an increase of the dominant $b\bar{b}$ partial width, will lead to a reduced branching ratio, hence to a reduced signal strength, in all other decay channels.
Furthermore, when new decay modes are possible (here it means invisible decays into the LSP) the total width of the Higgs increases, thus reducing the signal strengths in all channels. 
For example the signal strength for the two-photon mode in gluon fusion $\mu_{ggh_1}^{\gamma\gamma}$ can be reduced by 25\% as compared to the SM expectation, see figure~\ref{fig:signal_strength-a}. Because this large reduction comes from the total width we expect it to be completely correlated with the signal strength in the $W$ fusion mode. A comparison with the signal strengths for the $b\bar{b}$ mode, figure~\ref{fig:signal_strength-b}, clearly shows that this reduction can be correlated with the one in the $b\bar{b}$ channel (when the invisible width is large) or with an increase in the signal strength in the $b\bar{b}$ channel when $C_b>1$.  The invisible width of $h_1$ is found to be below 25\%. Recall that
 current  limits from direct searches are 58\% in CMS~\cite{Chatrchyan:2014tja} while preliminary results from ATLAS in the vector boson fusion mode set the limit at 29\%~\cite{ATLAS-CONF-2015-004}.   A stronger limit of 12\% is obtained from  global fits to the Higgs~\cite{Bernon:2014vta}, however the latter applies only when all Higgs couplings are SM-like. In future runs, it is expected that the LHC could
probe directly an invisible width of 17\%~\cite{Ghosh:2012ep}.

Other probes of the Higgs sector can be performed by searching for  the heavy Higgses at the LHC. After applying all constraints described in section~\ref{sec:constraints}, which in particular include heavy Higgs searches at LHC8TeV, we find that  
the lowest allowed  value for the mass of $h_2$ is around 340 GeV  and   can reach several TeV's.
Below the TeV scale,  the pseudoscalar is typically nearly degenerate with the doublet-like $h_2$ and the value of $\tan\beta$ ranges from 2-40 with a large fraction of the points with $\tan\beta <10$ because of flavour and  direct search limits.   

\begin{figure}[!htb]
\begin{center}
\centering
\subfloat[]{\label{fig:Hhhmh2-a}\includegraphics[width=8.cm,height=5.5cm]{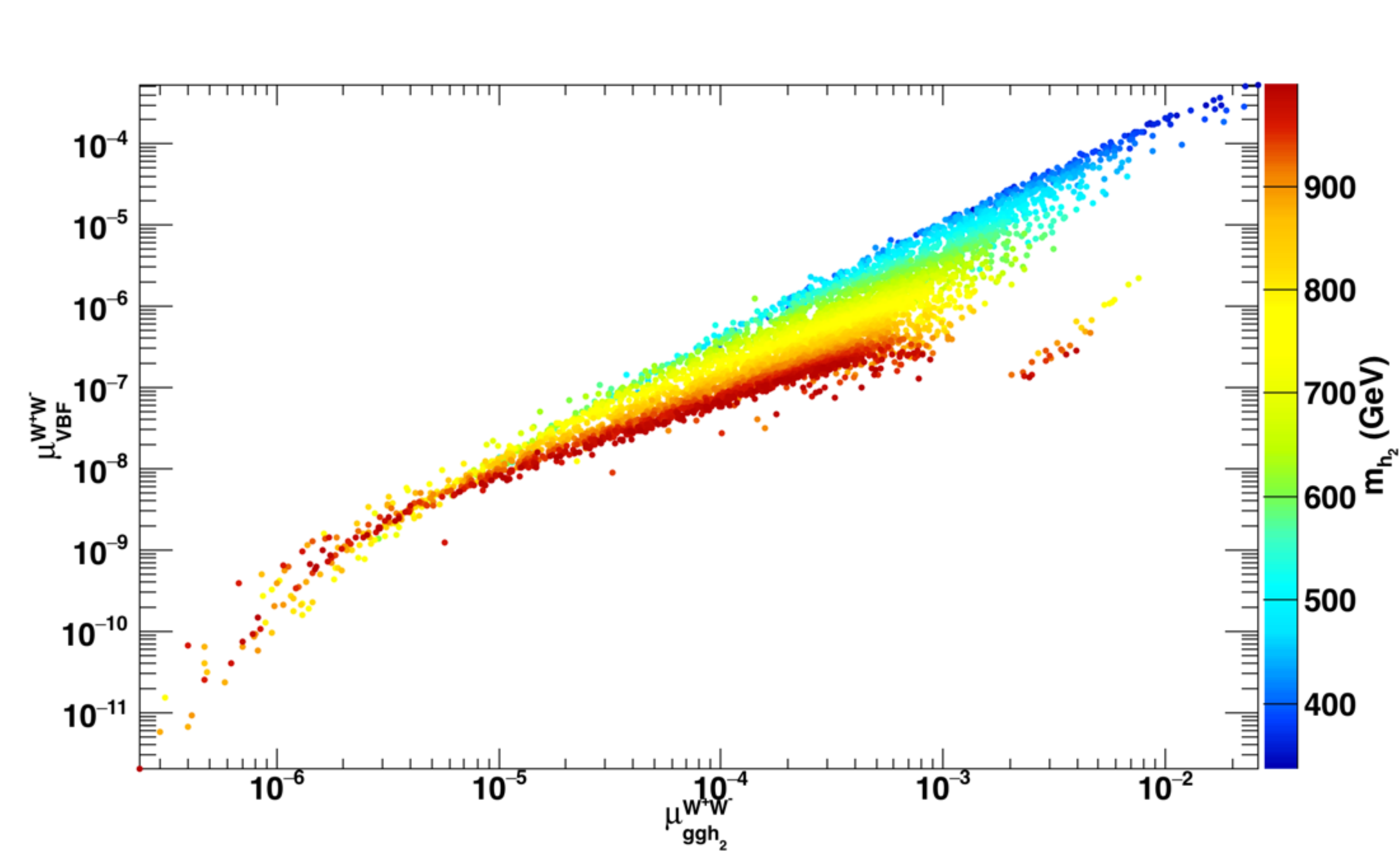}}
\subfloat[]{\label{fig:Hhhmh2-b}\includegraphics[width=8.cm,height=5.5cm]{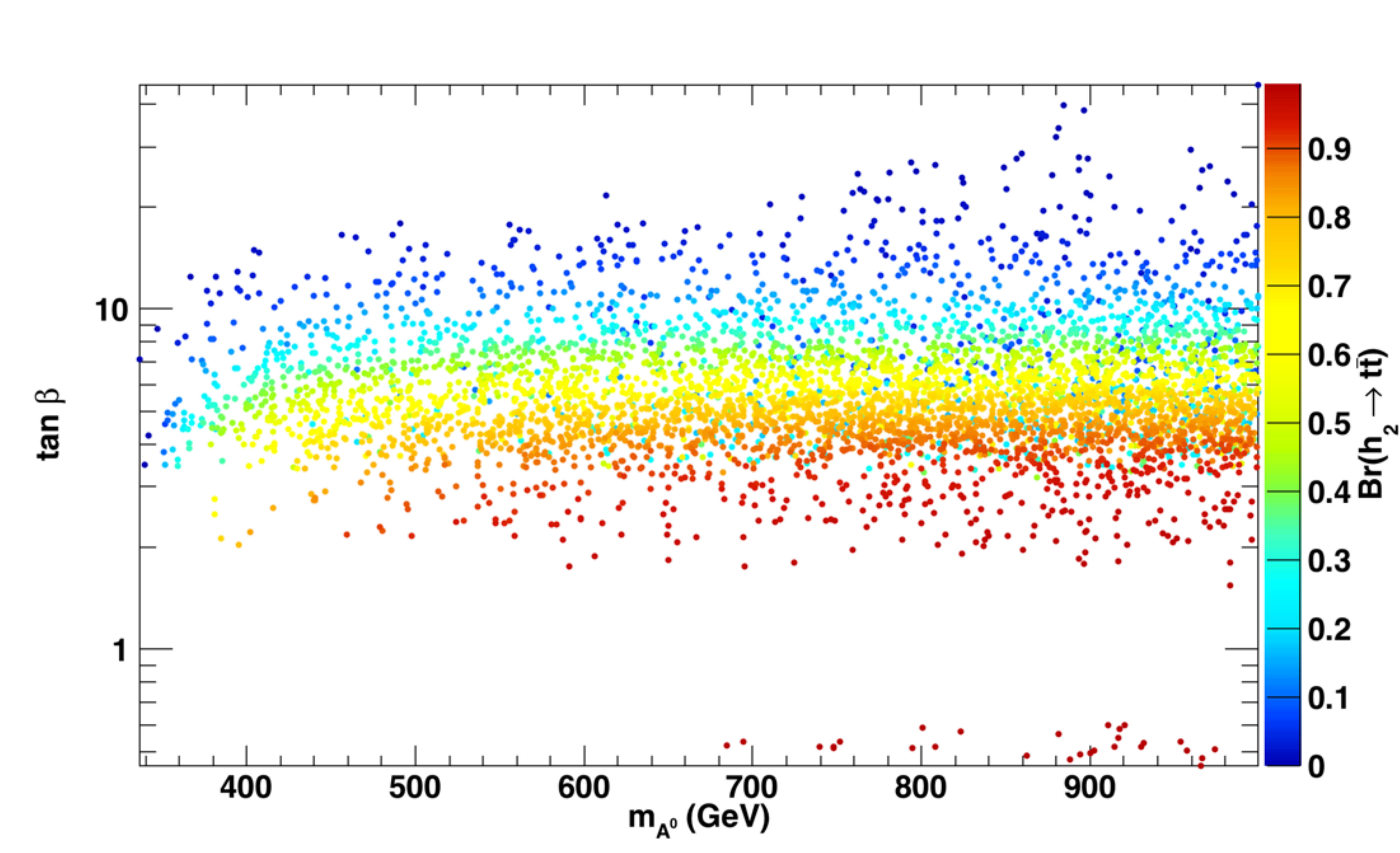}}
\caption{For all points satisfying collider and DM relic density constraints are displayed (a) the signal strengths for $h_2$ in gluon fusion and VBF modes in the $W^+W^-$ final state, with $m_{h_2} < 1$~TeV as colour code.
(b) $\mathscr{B}(h_2 \to t\bar{t})$ in the $\tan\beta - m_{A^0}$ plane with $m_{A^0} < 1$~TeV. 
\label{fig:Hhhmh2}}
\end{center}
\end{figure}

To compare with the recently released limits on searches for heavy Higgs in the $W^+W^-$ channel we have computed the signal strengths for $h_2\rightarrow W^+W^-$ in both the VBF and gluon fusion mode. 
We expect this signal strength to be quite low (as we have argued above the coupling $h_1W^+W^-$ is SM-like). This means that in the decoupling limit the $h_2W^+W^-$ coupling is suppressed, $\cos(\alpha-\beta) \approx 0$ in the MSSM notation. Indeed we find that the signal strength is suppressed in the gluon fusion channel,
$\mu^{\rm WW}_{\rm gg}(h_2)<0.03$, due to the small branching into gauge boson final state and obviously even more so in the VBF production mode where the signal strength is well below 
$10^{-3}$, see figure~\ref{fig:Hhhmh2-a}. Thus $h_2$ easily escapes current limits. Note furthermore that the largest signal strengths are found for low values of $\tan\beta$ and for $h_2$ much below the TeV scale, a region where potentially the $t\bar{t}$ channel offers a better probe, as discussed below.  

In the sub-TeV region, preferred decay channels of $h_2$  are usually in the $b\bar{b}$ ($\tau^+\tau^-$) final states for moderate to large values of $\tan\beta$, as in the MSSM. However, for low values of $\tan\beta$,  $h_2$ can decay exclusively into $t\bar{t}$, see figure~\ref{fig:Hhhmh2-b}. Moreover decays into the lightest Higgs can also be large (as much as 50\% when $m_{h_2} < 360$~GeV) but drop rapidly reaching at most 10\% when $m_{h_2} > 460$~GeV. 
In the MSSM it was shown that searches for heavy Higgs in the  $t\bar{t}$ ($hh$) channel offer good discovery potential at LHC13TeV for small values of $\tan\beta$, when $m_{h_2}< 1(0.5)~{\rm TeV}$~\cite{Djouadi:2015jea}.  Hence, such searches should also probe of the UMSSM model further. However, decays of  $h_2$ into supersymmetric particles can affect the main SM particle signatures. 
In particular decays into electroweakinos can reach 84\% (86\%) for the neutralino (RH sneutrino) LSP scenarios, while the invisible decay of $h_2$  into the neutralino LSP reaches at most 15\%. 
Large branching fractions into electroweakinos are expected when the kinematically accessible states have a large higgsino/gaugino component, hence when $\mu,M_2$ are small.
These decay modes could therefore provide additional search channels for a second Higgs, see \textit{e.g.}~\cite{Bisset:2007mi}. 
 For a Higgs below the TeV scale, the decays into sfermions are generally kinematically forbidden. When they are allowed the branchings never reach the percent level and are therefore negligible.

\section{Dark matter  probes}
\label{sec:DM_ID-DD}

\begin{figure}[!htb]
\begin{center}
\centering
\subfloat[]{\label{fig:RD_after_SModelS-a}\includegraphics[width=8.cm,height=5.5cm]{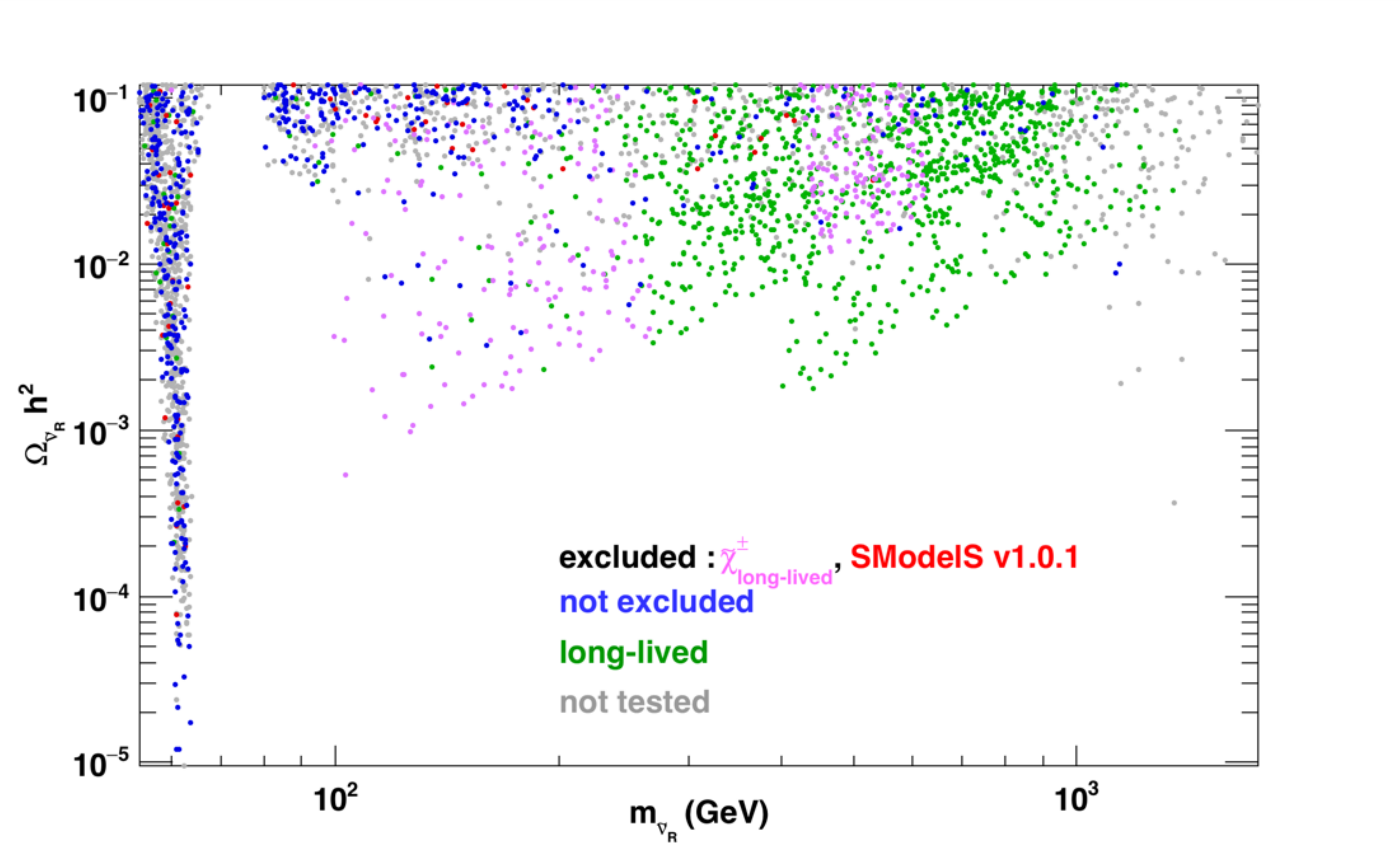}}
\subfloat[]{\label{fig:RD_after_SModelS-b}\includegraphics[width=8.cm,height=5.5cm]{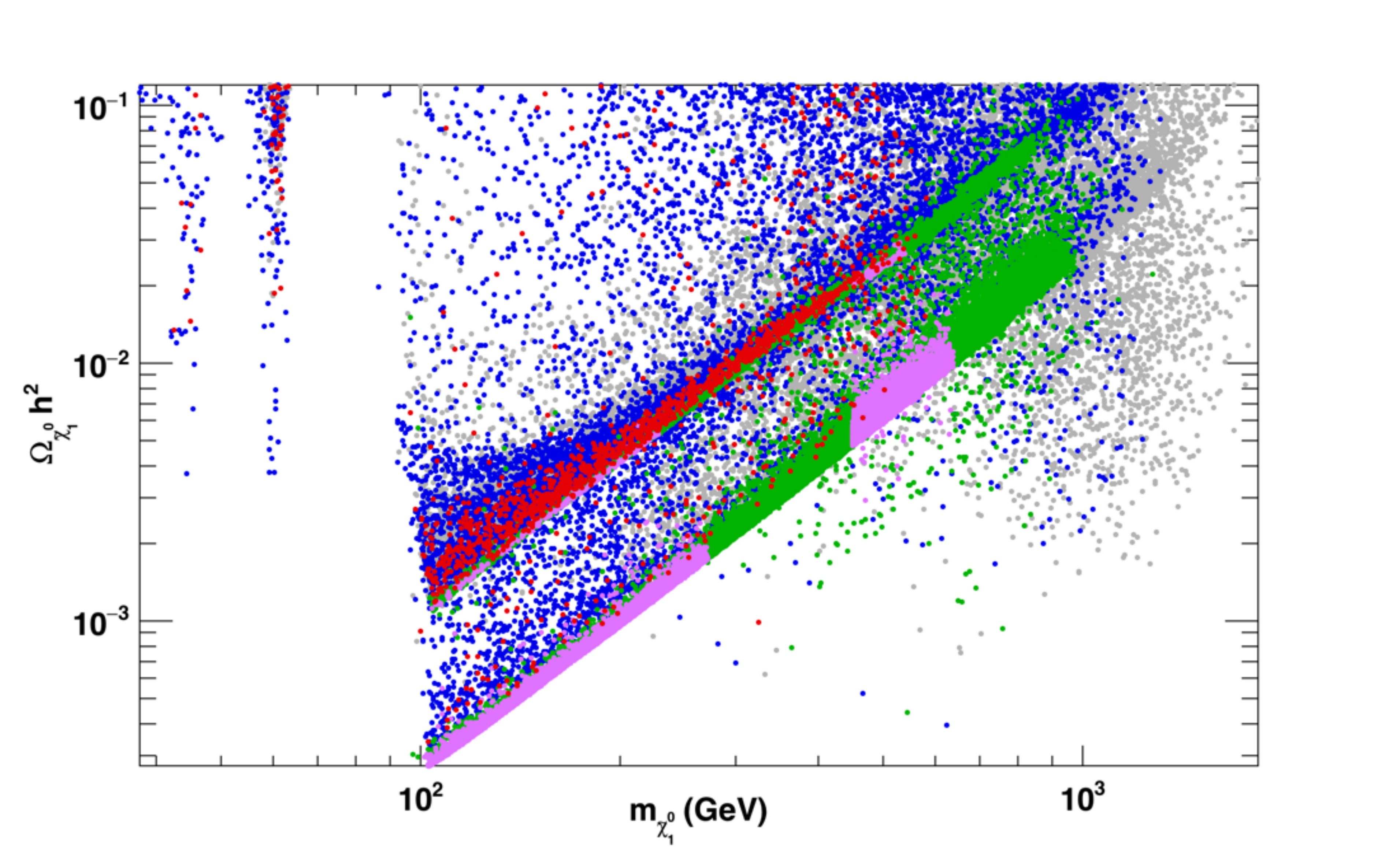}}
\caption{Relic density for (a) $\tilde{\nu}_{\tau R}$ or (b) $\chi^0_1$ LSP with the same colour code as in figure~\ref{fig:sfermions_after_SModelS} except for the configurations excuded by searches for long-lived charginos which are highlighted in pink.}
\label{fig:RD_after_SModelS}
\end{center}
\end{figure}

The correlation between LHC constraints and DM relic abundance on both the neutralino and RH sneutrino LSP scenarios is summarized in figure~\ref{fig:RD_after_SModelS}.
Clearly there is a strong preference for a RH sneutrino around 60 GeV, see figure~\ref{fig:RD_after_SModelS-a}.
 Moreover for the neutralino LSP case, the wino scenario (the lower branch in figure~\ref{fig:RD_after_SModelS-b}) is strongly constrained by searches for long-lived charginos. We now consider the predictions for DM observables.

Direct searches for dark matter through their scattering on nuclei in a large detector provide a complementary method to probe supersymmetric dark matter. 
When examining the predictions for dark matter searches  we use a rescaling factor to take into account cases where the LSP constitutes only a small fraction of the dark matter. 
The $2\sigma$ deviation from the central value measured by \textit{Planck} is $\Omega h^2 = 0.1168$, hence we define the rescaling factor $\xi$ as

\begin{eqnarray}
\xi = 
\begin{cases} 
\frac{\Omega_{\textrm{LSP}} h^2}{0.1168} \quad \textrm{for} \ \Omega_{\textrm{LSP}} h^2 <0.1168, \\   
1  \qquad \quad \, \textrm{for} \ \Omega_{\textrm{LSP}} h^2 \in [0.1168, 0.1208].
\end{cases}
\end{eqnarray}

\begin{figure}[!htb]
\begin{center}
\centering
\subfloat[]{\label{fig:DD_after_SModelS-a}\includegraphics[width=8.cm,height=5.5cm]{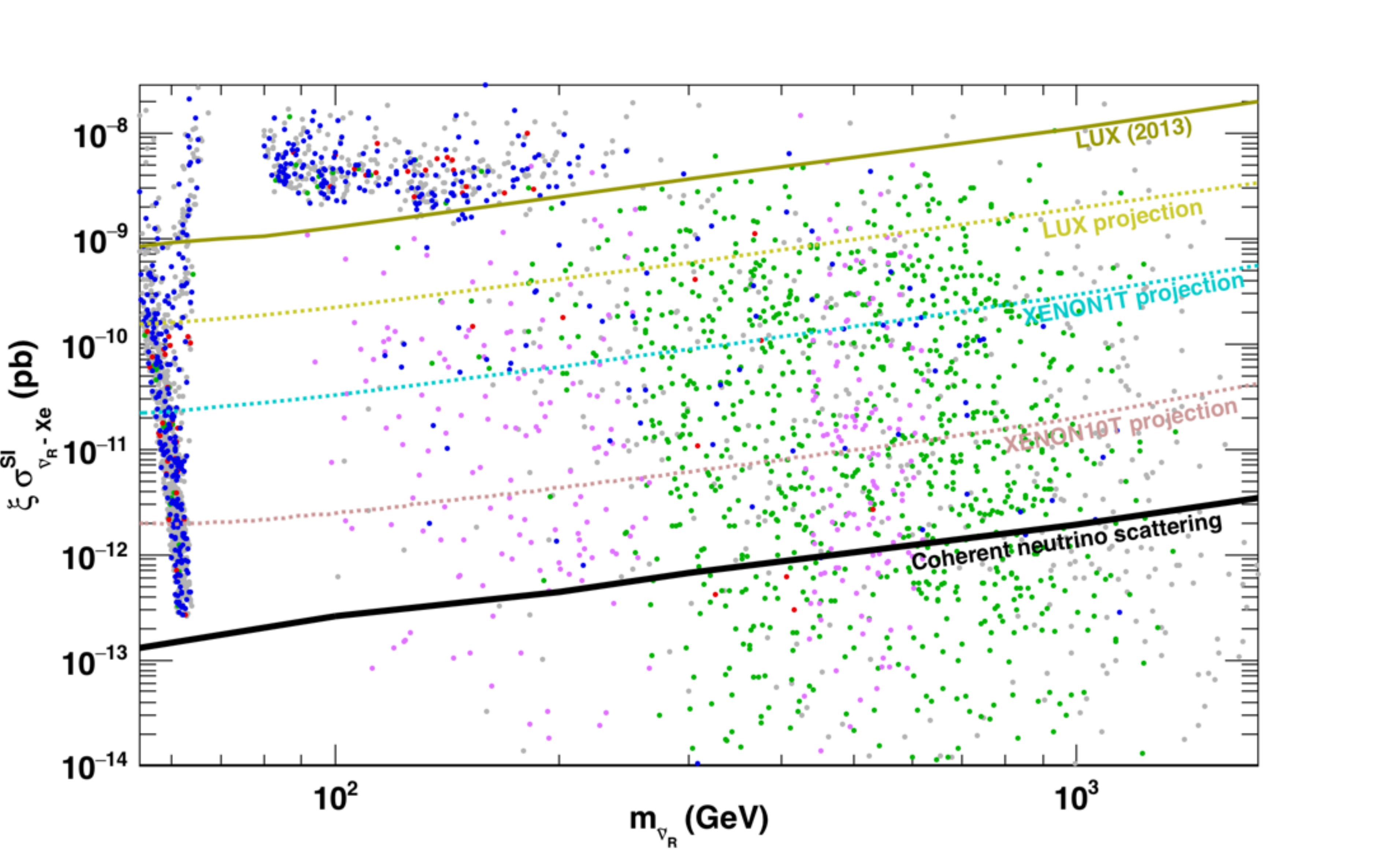}}
\subfloat[]{\label{fig:DD_after_SModelS-b}\includegraphics[width=8.cm,height=5.5cm]{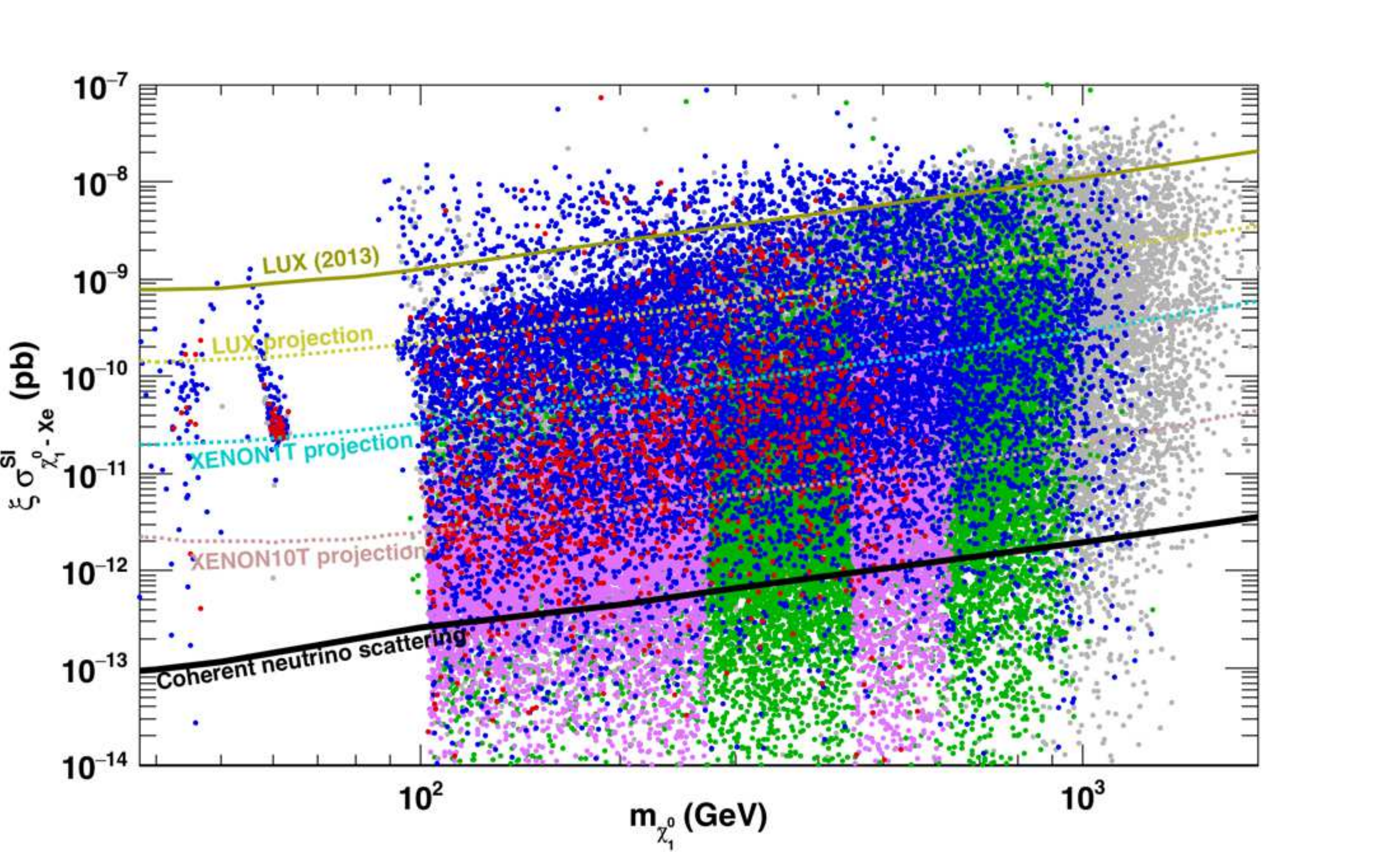}}
\caption{Rescaled direct detection cross section for (a) $\tilde{\nu}_{\tau R}$ or (b) $\tilde\chi^0_1$ LSP with the same colour code as in figure~\ref{fig:RD_after_SModelS}. LUX exclusion (dark beige), projections from future large detectors as well as the neutrino background are also displayed.}
\label{fig:DD_after_SModelS}
\end{center}
\end{figure}

\begin{figure}[!htb]
\begin{center}
\centering
\subfloat[]{\label{fig:DDllctau}\includegraphics[width=8.cm,height=5.5cm]{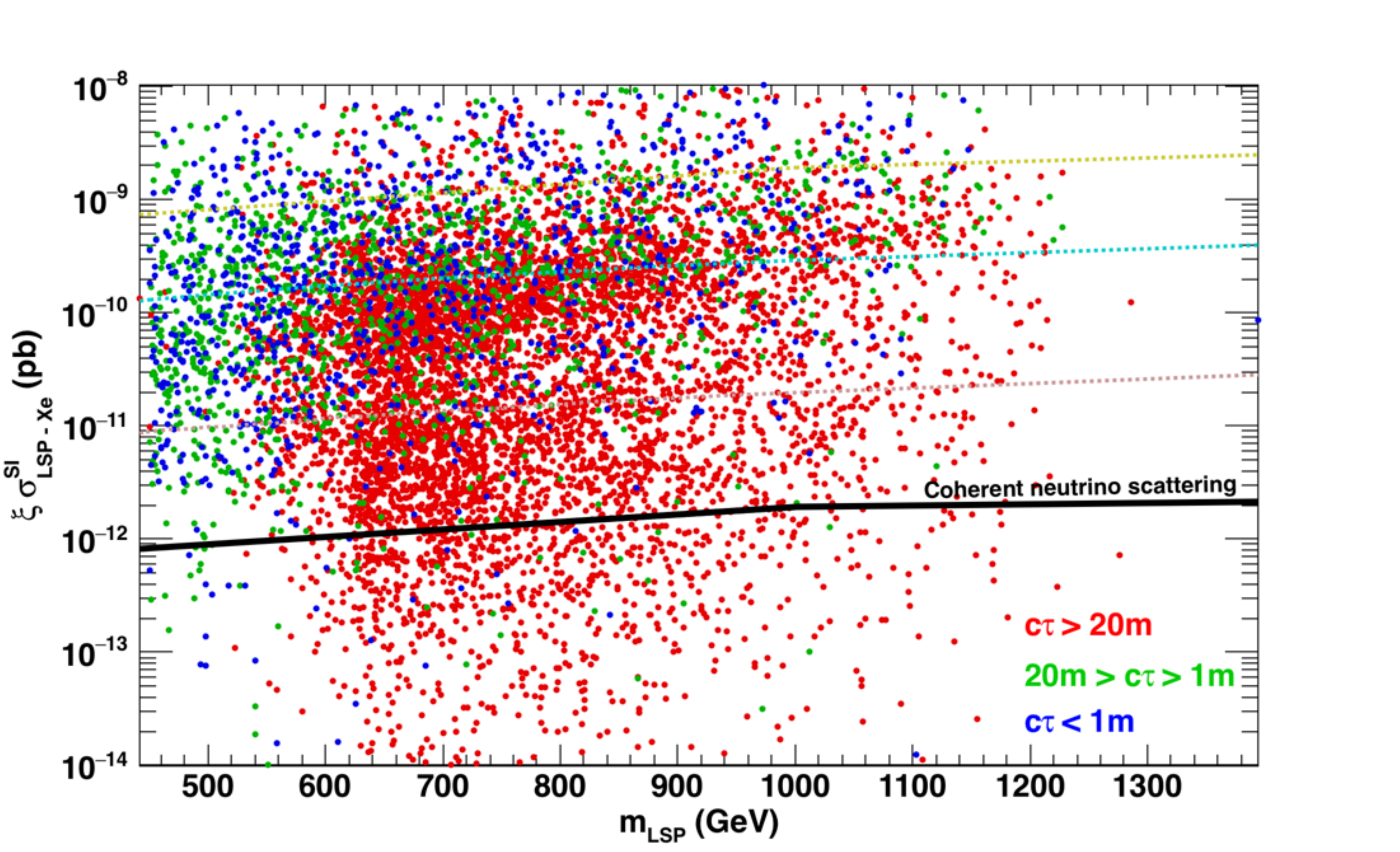}}
\subfloat[]{\label{fig:DDchi1}\includegraphics[width=8.cm,height=5.5cm]{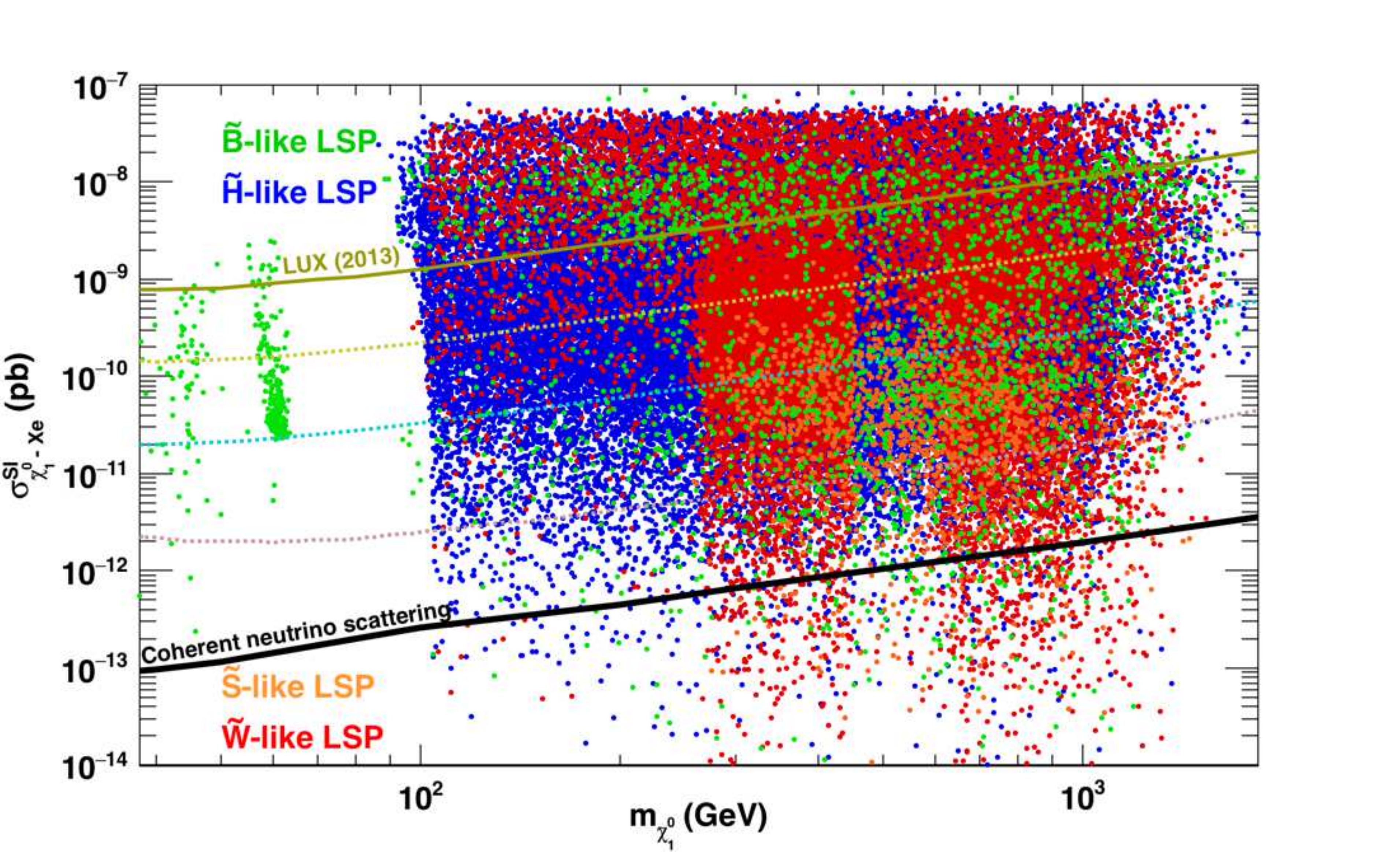}}
\caption{(a) Rescaled direct detection cross section with a $\tilde{\nu}_{\tau R}$ or a $\tilde\chi^0_1$ LSP for the same set of points as in figure~\ref{fig:longlived13TeV} and after applying the LUX bound. Chargino decay lengths $c\tau > 20$~m (red), between 20 m and 1 m (green) and below 1 m (blue) are represented. (b) Direct detection cross section without rescaling for $\tilde{B}$ (green), $\tilde{W}$ (red), $\tilde{H}$ (blue) and $\tilde{S}$ (orange) LSP, for all points satisfying collider and DM relic density constraints.}
\end{center}
\end{figure}

Figure~\ref{fig:DD_after_SModelS-a} shows that most of the RH sneutrino LSP points with a mass around 100 GeV are excluded by LUX. Those with a mass near 60 GeV escape the LUX upper limit  but are generally within the reach of the future Xenon1T detector. Other RH sneutrinos, which as we have argued before benefit from coannihilation and are therefore associated with a compressed spectrum,  are safely below current exclusions. In some cases the predicted cross section is even below that of the coherent neutrino background and can therefore never be probed by direct detection. 
The scenarios with a neutralino LSP are hardly probed by LUX, see figure~\ref{fig:DD_after_SModelS-b}, only some of the mixed bino/higgsino points are excluded. The Xenon1T will be able to probe many more points, although  a large fraction is beyond the reach of even a 10 ton detector, if not below the coherent neutrino background. These points are dominantly wino (hence labelled as long-lived)  or singlino LSP. 
It is  interesting to note that many of the points that are out of reach of direct detection detectors are associated with long-lived sparticles. To illustrate the complementarity with collider searches, we show in  figure~\ref{fig:DDllctau} the points with a long-lived chargino which could be probed at LHC Run II, that is  the points  in figure~\ref{fig:longlived13TeV} for which the cross section for chargino pair production is above $0.1$~fb. Clearly, many of the points with charginos stable at the collider scale have a direct detection cross-section below the reach of Xenon1T and even below the neutrino background. Note that the lowest value for the direct detection is about four orders of magnitude below the neutrino background (not shown in the figure). It should also be emphasized that many points with a chargino lifetime that leads to displaced vertices (in blue and green in figure~\ref{fig:DDllctau}) are also beyond the reach of ton-scale detectors, hence we stress again the importance of probing  these signatures at colliders.
A quite different conclusion would be  reached if we set the rescaling factor $\xi=1$, that is assuming some regeneration mechanism for the neutralino LSP. As shown in figure~\ref{fig:DDchi1}, some of the mixed wino points are not allowed by LUX and a large fraction are within the reach of Xenon1T. However even with these optimistic assumptions we find a few scenarios  with a cross section below that of the coherent neutrino background that will never be tested.

\begin{figure}[!htb]
\begin{center}
\centering
\subfloat[]{\label{fig:ID-a}\includegraphics[width=8.cm,height=5.5cm]{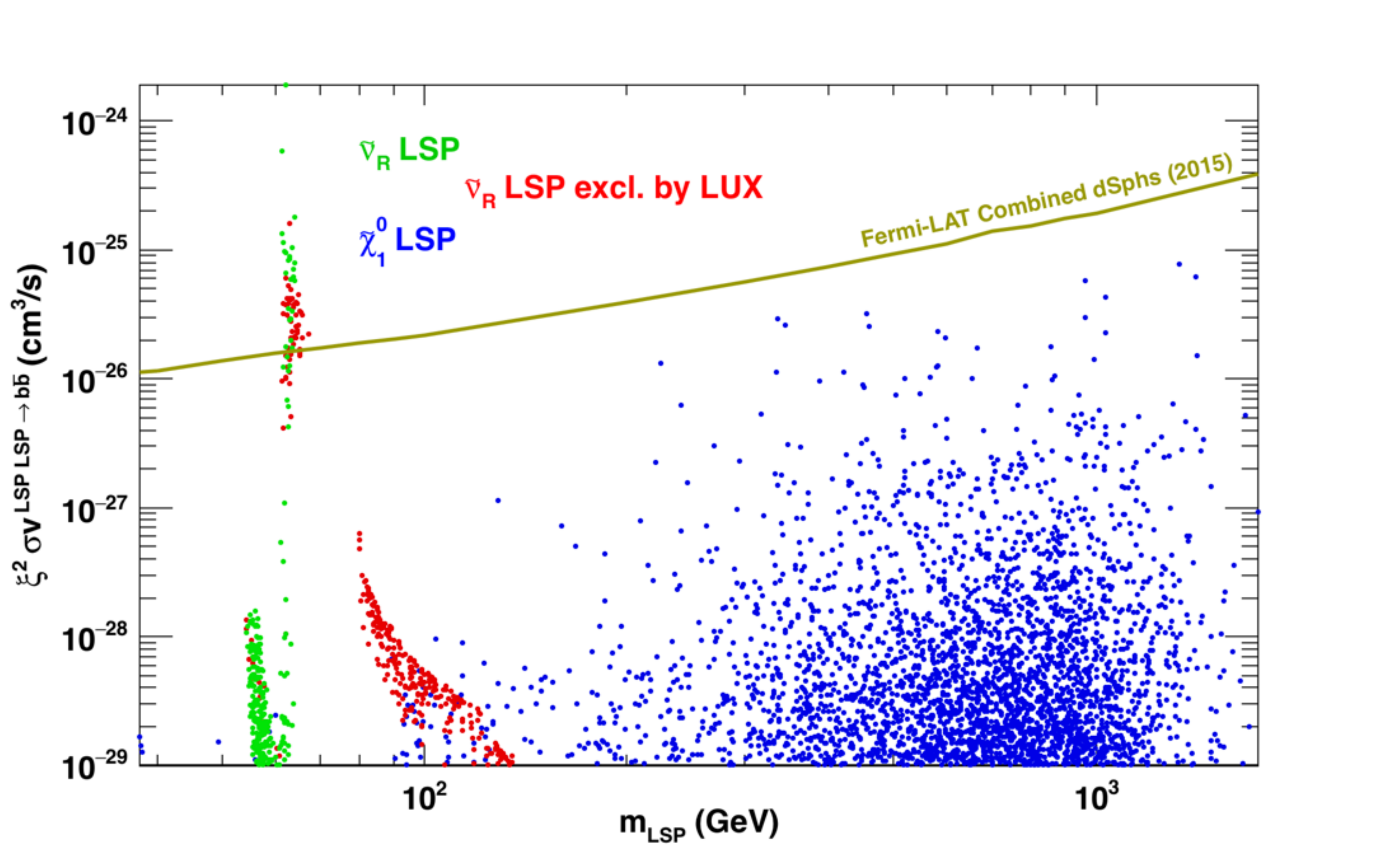}}
\subfloat[]{\label{fig:ID-b}\includegraphics[width=8.cm,height=5.5cm]{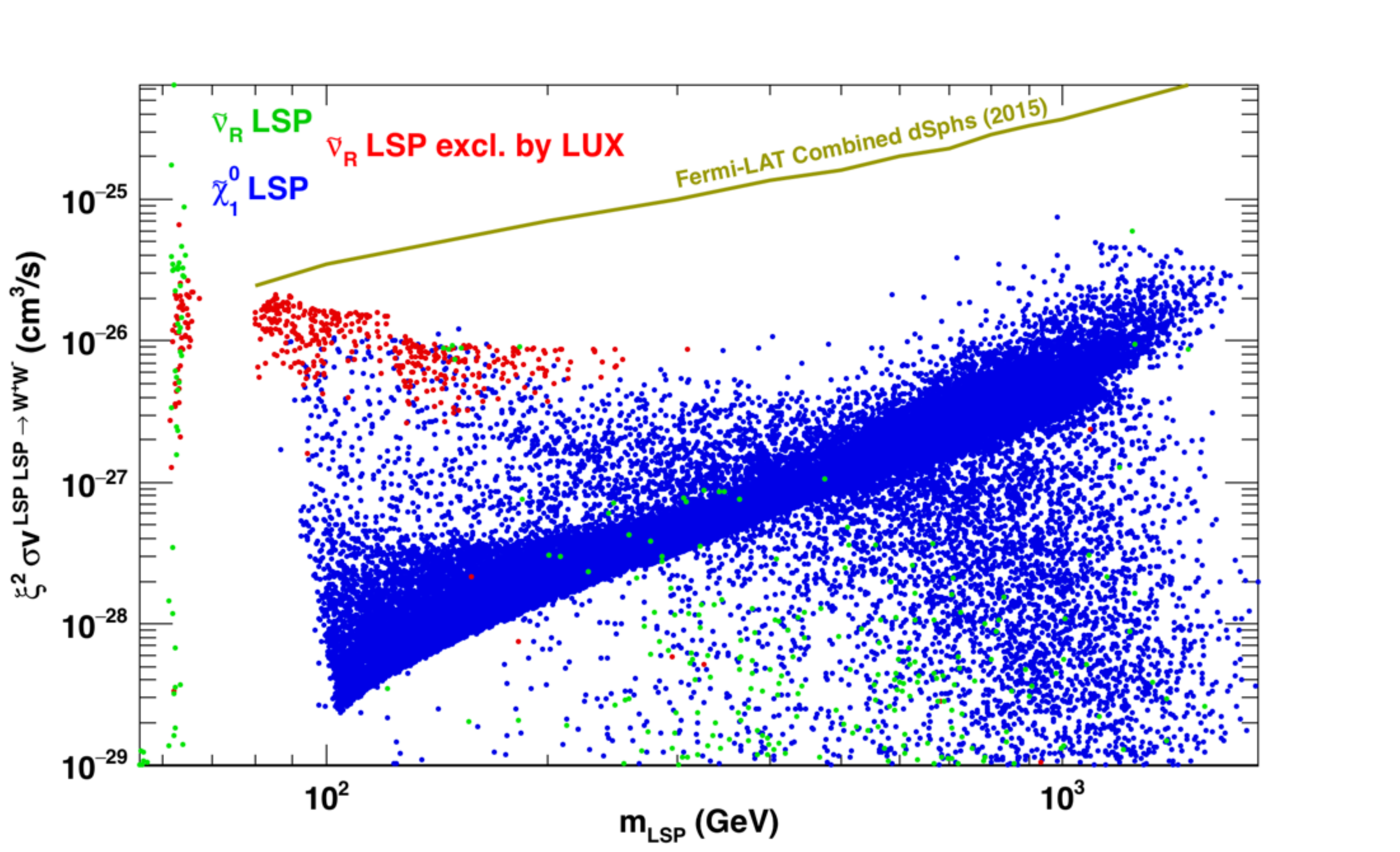}}
\caption{Rescaled annihilation cross section into (a) $b\bar{b}$ (b) $W^+W^-$, for $\tilde{\nu}_{\tau R}$ (green, red when excluded by LUX) or $\chi^0_1$ (blue) LSP,  for all points satisfying collider and DM relic density constraints. In both cases the FermiLAT limits are displayed.}
\label{fig:ID}
\end{center}
\end{figure}

\begin{figure}[!htb]
\begin{center}
\centering
\includegraphics[width=0.65\textwidth]{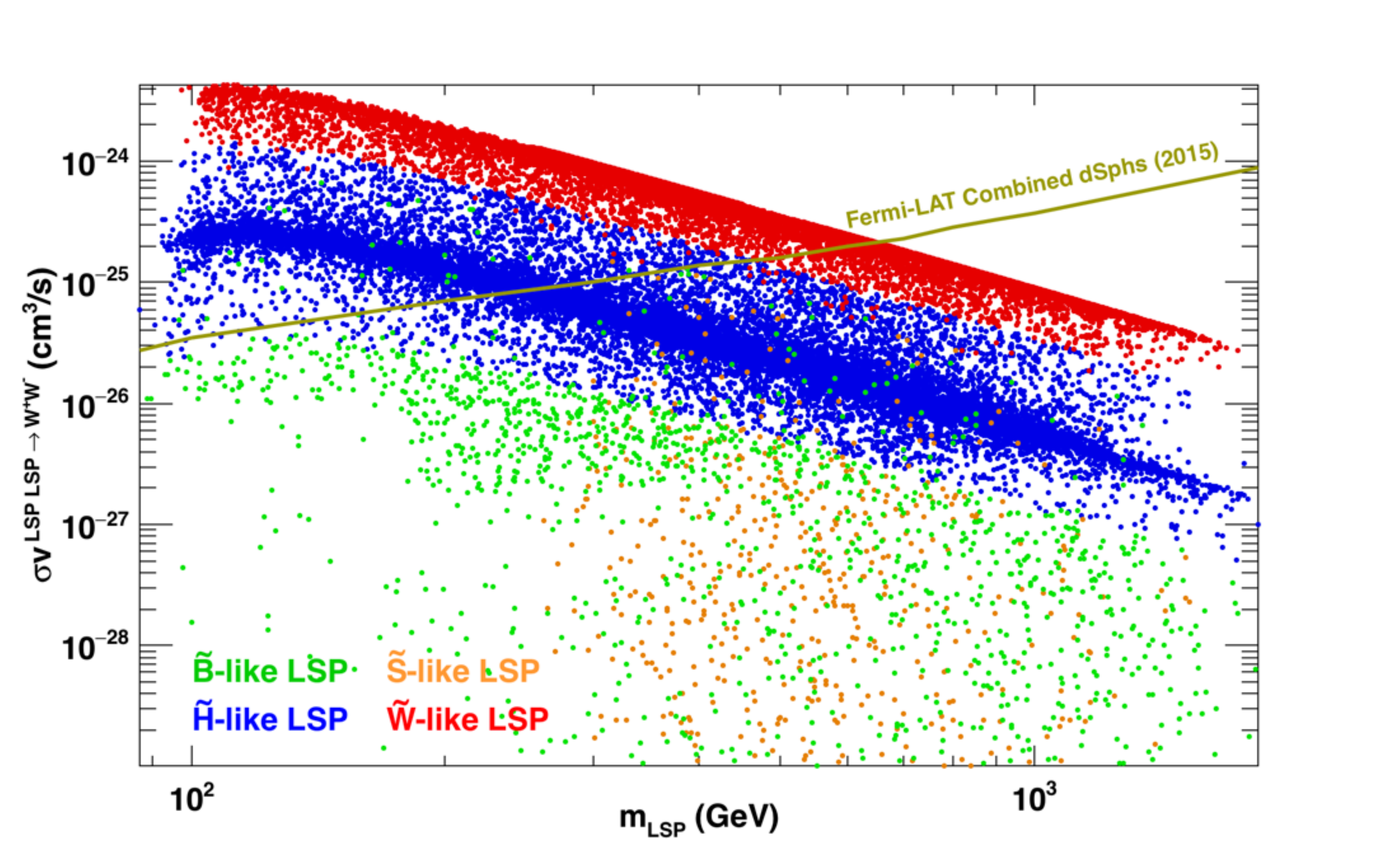}
\caption{Annihilation cross section into $W^+W^-$ without rescaling for $\tilde{B}$ (green), $\tilde{W}$ (red), $\tilde{H}$ (blue) and $\tilde{S}$ (orange) LSP,  for all points satisfying collider and DM relic density constraints.}
\label{fig:IDNoResc}
\end{center}
\end{figure}

We also explore the possibility to probe DM with indirect detection, in particular using the limits obtained from observations of photons from dwarf spheroidal galaxies in the Milky Way by FermiLAT.
For this we again rescale the cross section for points where the RH sneutrino or neutralino LSP cannot explain all the cosmologically measured dark matter. This means introducing a suppression by a factor $\xi^2$. 
We find that only a few points with a RH sneutrino LSP with a mass near 60 GeV are excluded by the FermiLAT limit from the $b\bar{b}$ channel, see figure~\ref{fig:ID-a}. Some of these points were also excluded by LUX,  however the predicted cross sections for most of the points are suppressed by at least two orders of magnitude as compared to current limits. For heavier LSP's the preferred annihilation channel is into $W$ pairs. After applying the rescaling factor no exclusion can be obtained. Again, the predicted cross section is generally two orders of magnitude below the current limit, see figure~\ref{fig:ID-b}. A quite different conclusion is reached if one does not apply the rescaling factor, then all winos with a mass below 500 GeV are excluded by FermiLAT as well as  most of the higgsino LSP's with a mass below 200 GeV, see figure~\ref{fig:IDNoResc}. This is not a specific feature of the UMSSM and was already observed in the MSSM both for photons~\cite{Williams:2012pz,Cohen:2013ama} and also antiprotons~\cite{Belanger:2012ta}. Note that the singlino LSP scenarios cannot be probed in this channel even without the rescaling.

\section{Conclusion}
\label{sec:conclusion}

We have reexamined the viability of the UMSSM model to describe physics beyond the standard model and dark matter after the results of the LHC Run I on the Higgs, on flavour observables and on new particle searches.
We found compatibility with all latest experimental results for  large regions of the parameter space for both a neutralino or RH sneutrino DM and explored potential future probes of the model at Run II of the LHC and in direct detection.
Imposing only the upper bound on the relic density favors either sneutrino DM near 60 GeV or  neutralino DM with a large wino and/or higgsino component - hence associated with either a long-lived chargino or a chargino which decays primarily into virtual $W$ and the LSP. The latter feature is common also to the MSSM  and  has important consequences both for squark and gluino searches as well as for electroweakino searches at the LHC, since the chargino will either have a missing energy signature or yield soft decay products.  
 
The most crucial test of the model in Run II of the LHC is the search for a  $Z'$ gauge boson, the most prominent characteristic of the model.
Another landmark of the model is the possibility of a RH sneutrino LSP.
The collider signatures often resemble those of the MSSM  but it remains to be seen whether the additional neutrino that appears in the decays affects the kinematics of the process, for example  this is the case for  the dijet + MET signature from squark production when the squark can only decay to the sneutrino LSP.  
We have also found a $W$+invisible channel  that could lead to a single energetic lepton from chargino/neutralino production. Due to the presence of the sneutrino LSP, the $W$ can be energetic enough to lead to visible decays even when the chargino and lightest neutralino are nearly degenerate, thus this is quite a distinct signature from the MSSM. Another striking feature of the sneutrino LSP scenario is the possibility  to find long lived squarks and gluinos.

The main characteristic of the  neutralino LSP scenarios is associated with the nature of the LSP, we found usually a dominanty wino or higgsino LSP.  
This implies that the chargino is often nearly degenerate with the LSP hence leads to invisible decay products or that the chargino is stable at the collider scale, hence the importance of the searches for long-lived/stable particles.   This is not a unique feature of the model as long-lived charginos can also be found in the MSSM. We also found a number of topologies that could not be constrained by current SMS results, including 2, 3 or 4 jets + MET topologies with either light jets or third generation quarks.    The main reason  is that these topologies are associated with asymmetric decays, that is  each of the pair produced particles has a  different decay chain whereas most SMS results assume symmetric decays. The most striking example is a $bt$ + MET topology that comes from stop pair production where one of the stop decays into a quark and  the LSP and the other to a quark and  an invisible chargino~\cite{Graesser:2012qy}.
Note however that a  recent result by CMS~\cite{CMS:2014wsa}  gives limits on stop pair production  for this topology independently of the relative branching fractions of each stop. It can be expected that this result will exclude most points with the $bt$ + MET missing topology in scenarios where the neutralino is lighter than $\approx 200$~GeV and the stop mass is below $\approx 700$~GeV.
Other examples include decays of squarks through a heavier neutralino.
 Furthermore we have stressed that there is a nice complementarity between the searches for stable charginos at colliders and the direct searches for DM. A significant number of the points which are below the reach of ton-scale detectors have a chargino which is  either  collider stable  or lead to displaced vertices. 
Similarly the singlino LSP, which occurs for a small fraction of the points,  can hardly be probed by DM direct detection.

Another feature of the UMSM is the split u-type/d-type RH squark masses which are found for specific choices of the $U(1)'$ charge. Despite conventional decays, these squarks are harder to detect. They do not benefit from contributions of all flavours of squarks and therefore the production cross section is lower.

As concerns the Higgs sector, the model predicts mostly a SM-like Higgs, although deviations up to 25\% can be observed in the signal-strengths for either $\gamma\gamma$ or fermionic final states. 
Moreover an invisible branching fraction of the Higgs up to 25\% can be found. Over some of the parameter space the second Higgs lies below the TeV scale and can be probed at LHC13TeV  in the standard $\tau^+\tau^-$ mode relevant at large $\tan\beta$ but also in the $t\bar{t}$ or $hh$ mode at small values of $\tan\beta$. It remains to be seen to which extent the SUSY decay modes can be exploited.
Finally we should stress that the lowest values of $\tan\beta$ lead to quite enhanced rates for $\Delta  M_s$. Refining the constraints on the CKM elements  which is one of the important source of uncertainties implies that this observable would strongly constrain the low $\tan\beta$ scenarios.

\section{Acknowledgements}
We are grateful to Sabine Kraml for useful comments on the manuscript and to Suchita Kulkarni, Andre Lessa, Veronika Magerl, Wolfgang Magerl, Michael Traub and Wolfgang Waltenberger for useful discussions. We thank   
Vincent Tisserand and Diego Guadagnoli for useful discussions on the current determination of the CKM matrix elements. 
We also thank  Tomaso Lari, Jamie Boyd and Shlomit Tarem  for exchanges on  ATLAS results for  stable particles.
G.B. thanks the LPSC where part of this work was done for its hospitality. This work was  supported in part by the LIA-TCAP of CNRS,  by the French ANR, Project DMAstro-LHC, ANR-12-BS05-0006, by the {\it Investissements d'avenir}, Labex ENIGMASS and by the European Union as part of the FP7 Marie Curie Initial Training Network MCnetITN (PITN-GA-2012-315877).
This research was supported in part by the Research Executive Agency (REA) of the European Union under the Grant Agreement PITN-GA2012-316704 (``HiggsTools'').
The work of AP  was  also supported by the Russian foundation for Basic Research, grant
RFBR-15-52-16021-CNRS-a.

\begin{appendices}

\renewcommand{\thesection}{\Alph{section}}
\setcounter{section}{0}
\numberwithin{equation}{section}

\section{Radiative corrections in the Higgs sector}
\label{sec:app}

To introduce radiative corrections in a gauge invariant manner we use an effective Lagrangian,
\beq \begin{split}
-\mathscr{L}_{eff} = & \; \l_1 |H_d|^4/2 + \l_2 |H_u|^4/2
+ \l_3 |H_d|^2 |H_u|^2 \\
& + \l_4 |H_u \cdot H_d|^2
+ \l_5 ((H_u \cdot H_d)^2+(H_u \cdot H_d)^{* 2})/2 \\
& + (\l_6 |H_d|^2 + \l_7 |H_u|^2) ((H_u \cdot H_d) + (H_u \cdot H_d)^*) \\
& + \l_8 |H_d|^2 |S|^2 + \l_9 |H_u|^2 |S|^2 \\
& + \l_s (S H_u \cdot H_d + S^* (H_u \cdot H_d)^*),
\end{split} \eeq
where  $\l_s$ is the only dimensionful parameter. 
To compute the $\l$'s we adapt the \NTools\cite{Ellwanger:2005dv} code to the UMSSM. Here we do not consider pure UMSSM corrections from  gauge contributions since the $U(1)'$ gauge coupling is small compared to the Yukawa coupling of the top quark~\cite{Barger:2006dh}. 

The minimization conditions for the loop improved Higgs potential become
\beq \begin{split}
\left(m^c_{H_d}\right)^2 = & \left(m^\mathrm{tree}_{H_d}\right)^2 + \l_s \frac{v_s v_u}{\sqrt{2} v_d} -\l_1 \frac{v_d^2}{2} -(\l_3+\l_4+\l_5) \frac{v_u^2}{2} + 3 \l_6 \frac{v_u v_d}{2} + \l_7 \frac{v_u^3}{2 v_d} -\l_8 \frac{v_s^2}{2}\\
\left(m^c_{H_u}\right)^2 = & \left(m^\mathrm{tree}_{H_u}\right)^2 + \l_s \frac{v_s v_d}{\sqrt{2} v_u} -\l_2 \frac{v_u^2}{2} -(\l_3+\l_4+\l_5) \frac{v_d^2}{2} + \l_6 \frac{v_d^3}{2 v_u} + 3 \l_7 \frac{v_u v_d}{2} -\l_9 \frac{v_s^2}{2}\\
\left(m^c_S\right)^2 = & \left(m^\mathrm{tree}_S\right)^2 + \l_s \frac{v_u v_d}{\sqrt{2} v_s} -\l_8 \frac{v_d^2}{2} -\l_9 \frac{v_u^2}{2}.
\end{split} \eeq \\

The corrected CP-even mass-squared matrix elements $(\mathcal{M}_+^c)_{ij}$ are
\beq \begin{split}
\left({\mathcal{M}_{+}^c}\right)_{11} & = \left({\mathcal{M}_{+}^0}\right)_{11} + \l_1 v_d^2 + \left(\l_s \frac{v_s}{\sqrt{2}} -3 \l_6 \frac{v_d^2}{2} + \l_7 \frac{v_u^2}{2}\right)\frac{v_u}{v_d}\\
\left({\mathcal{M}_{+}^c}\right)_{12} & = \left({\mathcal{M}_{+}^0}\right)_{12} +  (\l_3+\l_4+\l_5) v_u v_d -\frac{3}{2}(\l_6 v_d^2 + \l_7 v_u^2) -\l_s \frac{v_s}{\sqrt{2}}\\
\left({\mathcal{M}_{+}^c}\right)_{13} & = \left({\mathcal{M}_{+}^0}\right)_{13} + \l_8 v_s v_d -\l_s \frac{v_u}{\sqrt{2}}\\
\left({\mathcal{M}_{+}^c}\right)_{22} & = \left({\mathcal{M}_{+}^0}\right)_{22} + \l_2 v_u^2 + \left(\l_s \frac{v_s}{\sqrt{2}} + \l_6 \frac{v_d^2}{2} -3 \l_7 \frac{v_u^2}{2}\right)\frac{v_d}{v_u}\\
\left({\mathcal{M}_{+}^c}\right)_{23} & = \left({\mathcal{M}_{+}^0}\right)_{23} + \l_9 v_s v_u -\l_s \frac{v_d}{\sqrt{2}}\\
\left({\mathcal{M}_{+}^c}\right)_{33} & = \left({\mathcal{M}_{+}^0}\right)_{33} + \l_s \frac{v_u v_d}{\sqrt{2} v_s}.
\end{split} \eeq

The corrected CP-odd mass-squared matrix elements read
\beq \begin{split}\label{eq:oddcorr}
\left({\mathcal{M}_{-}^c}\right)_{11} & = \left({\mathcal{M}_{-}^0}\right)_{11} + \left(\l_6 \frac{v_d^2}{2} + \l_7 \frac{v_u^2}{2} - \l_5 v_u v_d + \l_s \frac{v_s}{\sqrt{2}}\right)\frac{v_u}{v_d}\\
\left({\mathcal{M}_{-}^c}\right)_{12} & = \left({\mathcal{M}_{-}^0}\right)_{12} + \l_6 \frac{v_d^2}{2} + \l_7 \frac{v_u^2}{2} - \l_5 v_u v_d + \l_s \frac{v_s}{\sqrt{2}}\\
\left({\mathcal{M}_{-}^c}\right)_{13} & = \left({\mathcal{M}_{-}^0}\right)_{13} + \l_s \frac{v_u}{\sqrt{2}}\\
\left({\mathcal{M}_{-}^c}\right)_{22} & = \left({\mathcal{M}_{-}^0}\right)_{22} + \left(\l_6 \frac{v_d^2}{2} + \l_7 \frac{v_u^2}{2} - \l_5 v_u v_d + \l_s \frac{v_s}{\sqrt{2}}\right)\frac{v_d}{v_u}\\
\left({\mathcal{M}_{-}^c}\right)_{23} & = \left({\mathcal{M}_{-}^0}\right)_{23} + \l_s \frac{v_d}{\sqrt{2}}\\
\left({\mathcal{M}_{-}^c}\right)_{33} & = \left({\mathcal{M}_{-}^0}\right)_{33} + \l_s \frac{v_u v_d}{\sqrt{2} v_s},
\end{split} \eeq
which leads to the one-loop corrected pseudoscalar mass

\beq \begin{split}
\left(m^c_{A^0}\right)^2 = & \left(m_{A^0}^\mathrm{tree}\right)^2 + \frac{2}{ \sin 2\beta} \left\{ \sqrt{2} \l_s v_s\left(1+x^2\right) +\lambda_{567} v^2 \right.\\
& \left. + \left( 
 \left[\sqrt{2} \l_s v_s \left(1+x^2\right) +\lambda_{567} v^2\right]^2
 - 4\sqrt{2} \l_s v_s x^2 \lambda_{567} v^2 \right)^{1/2} \ \right\},
\end{split} \eeq
where $\lambda_{567}=-\l_5 \sin 2\beta + \l_6 \cb^2 + \l_7 \cb^2$ and $x= v \sin 2\beta / 2 v_s$.

Note that in 
\NTools~the CP-odd matrix $\mathcal{M}_{-}$ is defined in the basis of the two CP-odd bosons $\{A^{0}, S_I\}$, we must therefore perform the 
 transformation $T^T \mathcal{M}_{-} T$ where 
\beq T = \left(\begin{array}{ccc} \cb & -\sb & 0 \\ \sb & \cb & 0 \\ 0 & 0 & 1 \end{array}\right) \eeq
as in~\cite{Belanger:2005kh}.

Finally,  the charged Higgs mass is corrected as
\beq
\left(m^c_{H^\pm}\right)^2 = \left(m^\mathrm{tree}_{H^\pm}\right)^2 + \left(\l_s \frac{v_s}{\sqrt{2}} + \l_6 \frac{v_d^2}{2} + \l_7 \frac{v_u^2}{2}\right)\frac{2}{\sin 2\beta} -(\l_4+\l_5) \frac{v^2}{2}.  
\eeq

\end{appendices}

\bibliography{umssm_Run1-DM}{}

\end{document}